\documentclass[12pt]{JHEP3}

\usepackage{amsmath,epsfig}
\usepackage{amssymb,amsfonts}
\usepackage{latexsym}
\usepackage[latin1]{inputenc}
\usepackage{slashed}
\usepackage{empheq}
\numberwithin{equation}{section}
\usepackage{subcaption}
\usepackage{xcolor}

\usepackage{graphicx}
\usepackage{longtable}
\usepackage{color}

\newcommand{\exclude}[1]{}
\def\a#1{\alpha_{#1}}

\def\beq{\begin{equation}}
\def\eeq{\end{equation}}
\def\be{\begin{equation}}
\def\ee{\end{equation}}
\def\bea{\begin{eqnarray}}
\def\eea{\end{eqnarray}}
\def\bal{\begin{align}}
\def\eal{\end{align}}

\def\2b2[#1,#2][#3,#4]{\left( \begin{array}{cc} #1 & #2 \\ #3 & #4 \end{array}
\right)}
\def\3b3[#1,#2,#3][#4,#5,#6][#7,#8,#9]{\left( \begin{array}{ccc} #1 & #2 #3 \\
#4 & #5 & #6\\#7&#8&#9\end{array} \right)}

\newcommand\fverb{\setbox\pippobox=\hbox\bgroup\verb}
\newcommand\fverbdo{\egroup\medskip\noindent%
                        \fbox{\unhbox\pippobox}\ }
\newcommand\fverbit{\egroup\item[\fbox{\unhbox\pippobox}]}

\newcommand{\bear}{\begin{eqnarray}}

\newcommand{\eear}{\end{eqnarray}}
\newcommand{\de}{\partial}
\newcommand{\bsea}{\begin{subeqnarray}}
\newcommand{\esea}{\end{subeqnarray}}
\newbox\pippobox

\def\f{\varphi}

\def\g{\gamma}

\def\6{d}
\def\a{\alpha}

\def\half{\frac12}

\def\pa{\partial}

\def\e{\epsilon}
\def\m{\mu}
\def\n{\nu}

\def\sp{\;\;\;,\;\;\;}

\def\sq
\def\a{\alpha}

\def\hri#1#2{\href{http://arxiv.org/abs/#1}{[ArXiv:#1]#2}}
\def\hre#1#2{\href{http://arxiv.org/abs/#1/#2}{[ArXiv:#1/#2]}}

\def\e{\epsilon}

\title{Brane cosmology and the self-tuning of the cosmological constant}
\author{A. Amariti$^\dagger$, C. Charmousis$^\sharp$, D. Forcella$^\natural$,
  E. Kiritsis$^\natural$$^\flat$, F. Nitti$^\natural$,
~\\
$^\dagger$  \href{https://www.mi.infn.it/en/}
{INFN, Sezione di Milano}, Via Celoria 16, I-20133 Milan, Italy.
\\
~\\
 $^\sharp$
Laboratoire de Physique Th\'eorique, CNRS, Univ. Paris-Sud, \\ Universit\'e Paris-Saclay, 91405 Orsay, France\\
~\\
$^\natural$ \href{http://www.apc.univ-paris7.fr}{APC, AstroParticule et Cosmologie}, Universit\'e Paris Diderot, CNRS/IN2P3, CEA/IRFU,
Observatoire de Paris, Sorbonne Paris Cit\'e,\\
 10, rue Alice Domon et L\'eonie Duquet, 75205 Paris
Cedex 13, France\\
~\\
$^\flat$ \href{http://hep.physics.uoc.gr}{Crete Center for Theoretical Physics}, Institute for Theoretical and Computational Physics,
Department of Physics,  P.O. Box 2208,\\
University of Crete, 70013, Heraklion, Greece
}

\preprint{CCTP-2018-16\\
ITCP-IPP 2018/12}

\abstract{The cosmology of branes undergoing the self-tuning mechanism of the cosmological constant is considered. The equations and matching conditions are derived in several coordinate systems, and an exploration of possible solution strategies is performed. The ensuing equations are solved analytically in the probe brane limit. We classify the distinct behavior for the brane cosmology and we correlate them with properties of the bulk (static) solutions. Their matching to the actual universe cosmology is addressed.
}


\begin{document}
\maketitle

\section{Introduction and summary}

The cosmological constant problem\footnote{For recent reviews, see for
example \cite{Padilla:2015aaa,Burgess}.} can be seen as  a clash
between two frameworks which, each on its own, have been widely
successful in describing physical phenomena. On one side,  Effective
(quantum) Field Theory (EFT) has established itself as the correct
description of micro-physics of non-gravitational interactions; on the
other hand, General Relativity (GR) gives an accurate description of
the  observed gravitational phenomena from  macroscopic  down to sub-millimeter scales.

The clash between these frameworks can be phrased as the fact that any
EFT calculation of the vacuum energy density receives large
contributions from all short distance (UV) modes, and at the same time
it is a source for the gravitational fields on very large scales
(IR). This however is at odds with the currently observed (tiny) value
of the space-time curvature on large scales (as inferred from the acceleration of the expansion of the visible universe).

One possible resolution of this clash is the introduction of new
degrees of freedom which implement a dynamical mechanism for the
relaxation  of the cosmological constant to a small value.  A
mechanism such that, regardless of the value of vacuum energy, flat
four-dimensional space-time is a solution of the gravitational field
equations -without fine tuning the coupling constants of the theory- is called {\em self-tuning}.

Within the context of local 4d field theory coupled to 4d general
relativity , this is very hard to achieve, as explained long
ago by Weinberg \cite{Weinberg:1988cp}. Examples of 4d
theories evading his argument require specific non-minimal couplings
between gravity and the extra sector \cite{Charmousis:2011bf,Charmousis:2011ea} or a more
drastic violation of the IR-UV decoupling.

In \cite{selftuning}, a self-tuning theory was proposed based on the holographic
AdS/CFT duality. The theory is  formulated in the
language of braneworld scenarios\footnote{It has also a four-dimensional incarnation along the lines of \cite{smgrav}.} \cite{rs}, and consists of a
five-dimensional scalar-tensor ({\em bulk}) theory (5d Einstein gravity
minimally  coupled to a scalar field) coupled to a four-dimensional
theory localized on a codimension-one defect ({\em brane}) and including the
Standard Model fields. In the dual, field theoretical language, the
bulk theory is interpreted as a strongly interacting, UV complete
quantum field theory coupled to the weakly interacting brane
fields, along the lines of the well-established connection between
holography and brane-world phenomenology \cite{Verlinde,RZ,APR,smgrav}.
 The  interaction between the two sectors can be thought of arising from a heavy
messenger sector, which at scales below their mass $\Lambda$ (which
effectively acts as a UV cut-off) can be replaced by effective
couplings between the brane and the bulk, which take the form of
induced {\em brane
potentials} multiplying the standard four-dimensional Einstein-Hilbert,
cosmological  and scalar kinetic term  on the brane.

In the class of models in \cite{selftuning}, a working self-tuning
mechanism is in place, due to the higher-dimensional nature of gravity
and the  interplay between  the brane and the
bulk. Solutions are determined by solving the system of
 bulk Einstein equations plus Israel's matching conditions across the
 brane. For generic values of the brane vacuum energy, solutions in
which the brane geometry is flat Minkowski space can generically exist.  These
solutions correspond to the Poincar\'e-invariant vacua of the
theory.  The brane is static and its location in the bulk is
stabilized at a fixed radial position\footnote{This is unlike earlier unsuccessful attempts to establish a well-defined self-tuning theory, \cite{ArkaniHamed,Kachru,csaki}.}.  At the
same time, a mechanism for gravity quasi-localization analogous to the
DGP mechanism \cite{DGP} in curved space-time  allows
gravitational interactions to behave as four-dimensional in a
range of scales. As it was shown in \cite{selftuning},  under some
mild assumptions on the values of the brane potentials at the location
of the brane, the vacuum solutions are stable under small
fluctuations. The embedding of this model in a consistent holographic
framework, as well as the introduction of  general bulk-brane
couplings (including an induced Einstein-Hilbert term) allows this
model to bypass the problems of previous proposals along the same
lines \cite{ArkaniHamed,Kachru,csaki}. In particular,
solutions with no bulk singularities, or with holographically acceptable ones, can be found.

The work \cite{selftuning} analyzed in detail the
Poincar\'e-invariant  vacuum solutions that lead to a brane embedding with a flat world-volume. The next logical step is to
explore  curved-brane  solutions,  in particular the ones which
can be related to cosmological evolution as seen by brane observers.

A first  step in this direction, taken in
\cite{dS}, was to consider solutions in which
the  brane position is still time-independent  but now its intrinsic
geometry  is allowed to be curved and maximally symmetric
(four-dimensional de Sitter or anti-de
Sitter space-time).  It was shown
that such solutions do not generically exist, if the bulk is dual to
the ground-state of a holographic QFT on a {\em flat} space-time.
For  these solutions to exist,  instead, a modification of
the bulk metric  is required all the way to the boundary of
five-dimensional AdS: essentially, the slicing of the bulk must be
adapted to the brane geometry at all values of the radial coordinate.  In the holographic, dual
QFT  language, this means that these solutions can only exist if the
metric on which the dual 4d QFT lives (and determines the bulk solutions)  is itself curved (in this case,
dS or AdS). This has two important consequences: on the one hand, it
shows that  the only vacuum solutions with a Poincar\'e-invariant UV metric  are the self-tuning, stabilized  flat brane-worlds
previously  found in \cite{selftuning}, and that to have static\footnote{i.e. with a time indepedent brane position} curved solutions one  must
 introduce a hard modification of the UV theory (i.e. a change in the
 background metric seen by the UV QFT). In other words, there is no
 competition in the same theory between static solutions with different
 curvatures. On the other hand, the analysis of \cite{dS} showed also
 an interesting  way of obtaining four-dimensional de Sitter space,
 which evades the ``swampland'' constraints. Such constraints, if valid, seem to rule out
 de Sitter as a solution of the effective supergravity equations
 arising from string theory.

The purpose of this paper is to initiate a detailed investigation of
the cosmology  of the self-tuning theories. Our approach
here is complementary to the one undertaken in \cite{dS}: rather than
looking for static curved solutions, here we
focus on non-vacuum, time-dependent solutions, where the brane  is
moving in the bulk, but the leading (radial) boundary conditions on the metric and
scalar field near the AdS boundary are static. In the holographic dictionary,
these  solutions correspond to time-dependent states in the same
theory that contains the vacuum self-tuning solutions of
\cite{selftuning}, rather than states in a different theory with a
deformed background metric (as was the case studied in
\cite{dS}.

The time-dependence  of the brane position in a
curved bulk as well as, generically, a time-dependent  bulk
metric,  result in an  FRW-like induced metric on the brane.  i.e. in
cosmological solutions. Our goal here will be to understand what are
the general features of the cosmological evolutions in self-tuning
models, what types of cosmological histories are possible, and what is
the relation between the cosmology and the self-tuning mechanism (in
particular whether at late times the brane can relax into a self-tuning
vacuum or possibly  in a weakly curved de Sitter geometry).

 The general problem we set out to solve involves non-linear PDEs in
time and the radial direction, and  the most
effective approach will likely be a numerical analysis. In order to
have an analytic handle on the dynamics,  in most of the paper we make
the further simplifying assumption that the brane can be treated as a
probe in a static bulk. Even in this case, the brane motion in a
curved bulk results in a cosmological brane metric, as was found earlier in \cite{mirage}.

The probe limit has the advantage that the equations governing the brane
dynamics can be reduced to a single, ordinary differential equation
equivalent to the one describing the relativistic dynamics of a point
particle in one dimension. This will allow us to understand analytically the motion of the brane, and
the resulting cosmology, especially in scaling regions of the bulk and
close to the stable self-tuning vacua. These,  in particular, will take
on a very simple interpretation in the probe limit, as extrema of the
effective potential for the one-dimensional brane motion.

Another virtue of the probe-brane approximation is that, since the
bulk solution is static, one can manifestly read-off the regularity
in the IR of the bulk geometry. This is not so in time-dependent bulk
geometries, for which the question of IR-regularity requires
identifying the presence of apparent horizons and switching  to  in-falling
coordinates. This however makes it harder to apply the holographic
dictionary close to the UV boundary.

The probe-brane approximation  requires constraining assumptions on the
size of the brane potentials in relation to the bulk curvature
scale. Although  this is a limitation of this approach, it
nevertheless gives a qualitative understanding of what kind of
cosmological evolutions are possible, and we expect that many of these
qualitative features will carry on to the general (fully backreacted)
system. Moreover, as we shall see,  there are certain regimes (namely,
when the brane approaches the asymptotic UV region of the bulk) where
the probe approximation becomes universally accurate, and does not depend on
particular assumptions for the brane potentials. Interestingly, the universal behavior of the UV bulk region is one of the key points of the self tuning mechanism \cite{selftuning}.
Below we summarize our main results.

Our first result is a derivation of   the  fully backreacted
equations for time-dependent backgrounds  of general Einstein-scalar theory coupled to a  general co-dimension one defect,
in the presence of brane induced cosmological and kinetic terms.  We
derive the set of bulk equations and Israel matching conditions, in different
coordinate systems. This is a  generalization of previous work on
brane-world cosmology  in the absence of
bulk scalars, and/or  brane-induced induced terms, \cite{Bin}-\cite{CGP}.

We then proceed to study the system in the probe-brane
approximation. In this case, the only degree of freedom is the brane
radial  position as a function of time, $u(\tau)$. The dynamics can be
recast in terms of the  (non-canonical)  Lagrangian dynamics of a
point-particle in one dimension. The  induced  metric on the brane has
the  cosmological FRW form, and it is completely determined by the
brane trajectory $u(\tau)$ and the bulk scale factor.

In this context, several analytic results are  obtained:
\begin{enumerate}
\item As a general statement, a brane moving in the radial direction
  towards the IR (small bulk scale factor) region  of the bulk corresponds
  to a contracting FRW universe; similarly, a brane moving towards
  the UV (large bulk scale factor) corresponds to an expanding universe. This is because the
  induced brane time-dependent scale factor is essentially the same as the bulk
  scale factor, which is monotonically decreasing from the UV to the
  IR. This setup provides an interesting answer to the question posed
  in \cite{afim} where a holographic view of cosmology was postulated.
  In this view, an expanding universe cosmology is seen as an inverse
  RG flow. In the probe brane setup, we can see clearly that expansion
  can only occur when the brane has enough intrinsic energy
  (i.e. kinetic energy in the bulk), so it can move opposite to the
  gravitational force. 
  
\item Stationary points of the  one-dimensional effective potential
  felt by the brane correspond to equilibrium points. The stationarity
  condition is shown to arise as the probe approximation of the fully
  backreacted Israel condition for self-tuning vacua. Similarly, the
  stability conditions for fluctuations around a self-tuning solution,
  discussed in \cite{selftuning}, imply the stability in the
  Lagrangian system (positive kinetic term and second derivative of
  the effective potential at the minimum). The converse is not true: stability in the probe approximation is a weaker
  condition than full stability, since some of the bulk modes decouple
  in this limit.

\item In the extreme UV, corresponding to the  near-AdS boundary region,
  with a  diverging bulk scale factor, we find a universal behavior
  for an expanding brane: in the expanding  regime the cosmology
  approximates a 4d de Sitter geometry at late times. The effective
  Hubble constant is determined by the values of
  the UV limits  of the brane cosmological constant  and induced Planck
  scale (as  these  quantities are functions of the brane
  position in the bulk). This regime, once reached, lasts forever, and
  the geometry approaches de Sitter better and better as the brane
  moves towards the boundary.  Remarkably, the probe brane
  approximation is always a good one in the UV expanding regime,
  regardless of the details of the brane potentials.  In
  general, however, it is not guaranteed that a brane coming from
  the IR region will always reach the extreme UV, because depending
  on the signs of the brane potentials  this may
  lie in  a classically forbidden region (see point 5 below).

\item The extreme IR, corresponding to a vanishing bulk scale factor,  is perceived on the brane as a big-bang or big
  crunch  singularity, depending whether the system is going away from
  or falling into the IR. This may coincide with a
  (good) bulk singularity, or
  with a Cauchy horizon if the IR corresponds to a regular AdS-like fixed
  point of the dual field theory. Depending on the behavior of the
  bulk  and brane scalar potentials in the extreme IR, as well as on
  the total ``energy''  (in the analog Lagrangian language) of the
  brane, the cosmology can mimic the one driven by perfect fluids with various
  equations of state parameter $w$, ranging from a cosmological constant ($w=-1$) up to radiation
  ($w=1/3$).

\item As in any Lagrangian system, the potential may contain a classically
  forbidden region where the brane velocity becomes imaginary. When reaching
  the boundary of a classically allowed  region, the brane motion is
  inverted and turns (for example) from contracting to
  expanding. In cosmological terms, this corresponds to a regular
  bounce, which is forbidden in purely 4d general relativity with
  sources obeying the null energy condition, but is allowed in mirage cosmology, \cite{mirage,GGK}.

\end{enumerate}

In this paper we have paved the way for a comprehensive study of the
cosmology of general (self-tuning or not) Einstein-scalar theories
coupled to a general codimension-one defect with induced gravity.
 Several directions and improvements beyond the present work can be
 foreseen.

The most important ingredient we have omitted in this paper, beside
the brane backreaction,  is the
presence of  four-dimensional, cosmologically active matter on the
brane. This of course has to be included if we want the system to
undergo a phase   where  cosmological
history is the standard one, with a period driven by
sources situated on the brane. Four-dimensional matter can be
included as a brane-localized perfect fluid source in the field
equations\footnote{This was studied already in \cite{mirage} but in the absence of an induced brane curvature term.} , and the interplay between brane and bulk will depend on the
coupling between the fluid and the holographic (bulk)
sector. Particularly intriguing is the possibility that the
``energy'' of the probe brane one-dimensional motion may be dissipated
and converted into ordinary matter on the brane, allowing for the
brane to become trapped by a self-tuning solution along the way.

Another interesting development would be the analysis of  the mirage
cosmology over the dS and AdS brane solutions (ie with curved sliced
bulk) studied in \cite{dS}. This would allow  broader possibilities due
to the more general boundary conditions in the UV. Other
generalizations include for example the probe-brane motion in a more
general geometry like an dilatonic five-dimensional black
hole\footnote{In the Randall-Sundrum context this has been studied for example in \cite{kraus},\cite{Bowcock}, and \cite{CGP} including the DGP term.}: although this does not correspond to a Poincar\'e-invariant
vacuum state, it may be still relevant for cosmology as it possesses
the same symmetries of a four-dimensional FRW slice.

Finally, we mention the possibility of looking for exact solutions of
the bulk/brane system \cite{Davis}, by taking as a starting point special classes
of known exact solutions of five-dimensional Einstein-dilaton
theories, for example the ones obtained in \cite{exact},\cite{axi}.\\

The paper is organized as follows.

In section 2 we establish the setup, review the self-tuning vacuum
solutions and discuss on general grounds what type of ansatz will lead
to cosmological solutions.

In section 3 we establish the probe-brane approximation and  derive the
corresponding equations of motion in the language of relativistic Lagrangian
mechanics. We  then derive the corresponding  FRW cosmology on the brane
and the  associated cosmological   parameters. Finally, we  discuss the
non-relativistic limit and  stationary points of the system.

In section 4 we consider the probe-brane motion in the asymptotic
regions (IR and UV) of the bulk geometry. Introducing  a general
parametrization of  the bulk and brane potentials, we find approximate
analytical scaling solutions in these regions, and analyze the
corresponding cosmology induced on the brane.

From the results obtained in the previous sections, in Section
5 we draw general conclusions about the possible cosmological
histories of a probe brane universe in these models.

The Appendix contains several technical details. In
Appendix A we collect the full set of  bulk field equations and Israel
matching conditions for time-dependent ansatz, in various coordinate
systems. In Appendix B we investigate a simplified ansatz for a
time-dependent background and discuss why it generically does not
lead to a solution  of the full system. In Appendix C and D we provide
technical details of the probe-brane action and equation of motion.
Appendix E contains  computational  details of the solutions in the
asymptotic regimes.

\section{Self-tuning setup and time-dependent solutions}

We consider a scalar-tensor Einstein theory in a five-dimensional bulk space-time parametrized by coordinates $x^a\equiv (u, x^\mu)$ along the lines described in \cite{selftuning}.  This theory describes  a holographic CFT, and the scalar is dual to the operator driving the RG flow in that QFT\footnote{The generic holographic theory has many scalars but the scalar-tensor theory we will use has only one. It can be shown that to describe the relevant physics, we should keep only the effective scalar that flows and neglect the others.}.
 We consider a four-dimensional brane embedded in the bulk parametrized by coordinates $x^\mu$ . The most general 2-derivative  action to consider\footnote{There are higher order derivatives both on the brane and in the bulk, that are neglected here. The higher derivatives in the bulk are suppressed at strong coupling in the dual QFT. The higher derivatives on the brane are suppressed by the cutoff on the brane that as argued in \cite{smgrav} is of the order of the four-dimensional Planck scale.}  reads,
\be \label{A1}
S = S_{bulk} + S_{brane }
\ee
where,
\be\label{A2}
S_{bulk} = M^{3} \int d^{5}x \sqrt{-g} \left[R - {1\over 2}g^{ab}\de_a\f\de_b \f - V(\f)\right] + S_{GH},
\ee
\be\label{A3}
S_{brane} = M^{3} \int d^4 \xi \sqrt{-\g}\left[- W_B(\f) - {1\over 2} Z_B(\f) \gamma^{\mu\nu}\de_\mu\f\de_\nu \f + U_B(\f) R^{(\gamma)} \right]+\cdots,
\ee
where $M$ is the bulk Planck mass, $g_{ab}$ is the bulk metric, $R$ is its associated Ricci scalar,
$\xi^\mu$ are world-volume coordinates,  $\gamma_{\mu\nu}$, $R^{(\g)}$ are respectively the induced metric and intrinsic curvature of the brane, while $V(\f)$ is the bulk scalar potential. $S_{GH}$ is the Gibbons-Hawking term at the space-time boundary (e.g. the UV boundary if the bulk is asymptotically $AdS$).

 The ellipsis in the brane action involves higher-derivative terms of the gravitational sector fields ($\varphi,\gamma_{\m\n}$) as well as the action of the brane-localized fields (the ``Standard Model'' (SM), in the case of interest to us).
 $W_B(\f), Z_B(\f)$ and $U_B(\f)$ are scalar potentials which are generated by the quantum corrections of the brane-localized fields (that couple to the bulk fields, see \cite{smgrav}). As such, they are localized on the brane. In particular, $W_B(\f)$ contains  the brane vacuum energy, which takes contributions from the brane matter fields.  All of $W_B(\f), Z_B(\f)$ and $U_B(\f)$ are cutoff dependent and generically, $W_B(\f)\sim \Lambda^4$, $Z_B(\f)\sim U_B(\f)\sim \Lambda^2$ where $\Lambda$ is the
UV cutoff of the brane physics as described here.

\subsection{Field equations and matching conditions}

The  bulk field equations depend only on $V(\f)$ and are given by:
\be
R_{ab} -{1\over 2} g_{ab} R = {1\over 2}\de_a\f\de_b \f -  {1\over 2}g_{ab}\left( {1\over 2}g^{cd}\de_c\f\de_d \f + V(\f) \right),  \label{FE1}\ee
\be
\de_a \left(\sqrt{-g} g^{ab}\de_b \f \right)- {\de V \over \de \f} =0 \label{FE2}
\ee
The brane, being codimension-1, separates the bulk in two parts, denoted by ``$UV$'' (which contains the conformal $AdS$ boundary region or more generally, in non-asymptotically  $AdS$ solutions,  the region where the volume form becomes infinite ) and ``$IR$'' (where the volume form eventually vanishes, and may contain the $AdS$ Poincar\'e horizon, or a (good) singularity, or a black hole horizon, \cite{exact,gubser,thermo,cgkkm}). We  take the coordinate $u$ to increase towards the IR region.

Denoting $g^{UV}_{ab}, g^{IR}_{ab}$ and $\f^{UV}, \f^{IR}$  the solutions for the metric and scalar field on each side of the brane, and by $\Big[ X\Big]^{IR}_{UV}$ the jump of a quantity $X$ across the brane, Israel's junction conditions are:
\begin{enumerate}
\item Continuity of the metric and scalar field:
\be\label{FE3}
\Big[g_{ab}\Big]^{UV}_{IR} = 0,   \qquad \Big[\f\Big]^{IR}_{UV} =0
\ee
\item Discontinuity of the extrinsic curvature and normal derivative of $\f$:
\be\label{FE4}
\Big[K_{\mu\nu} - \gamma_{\mu\nu} K \Big]^{IR}_{UV} =   {1\over \sqrt{-\gamma}}{\delta S_{brane} \over \delta \gamma^{\mu\nu}}  ,  \qquad \Big[n^a\de_a \f\Big]^{IR}_{UV} =- {1\over \sqrt{-\gamma}}{\delta S_{brane} \over \delta \f} ,
\ee
where $K_{\mu\nu}$ is the extrinsic curvature of the brane, $K = \gamma^{\mu\nu}K_{\mu\nu}$  its trace, and $n^a$ a unit normal vector to the brane,  oriented towards the $IR$.
\end{enumerate}
Using the  form of the brane action, equations (\ref{FE4}) are given explicitly by:
\be
\Big[K_{\mu\nu} - \gamma_{\mu\nu} K \Big]^{IR}_{UV} = \Bigg[\half W_B(\f) \gamma_{\mu\nu} + U_B(\f) G^{(\gamma)}_{\mu\nu} - Z_B(\f) \left(\de_\mu \f \de_\nu \f - \half \g_{\mu\nu} (\de \f)^2 \right)+ \label{FE5}
\ee
$$ \qquad + \left(\gamma_{\mu\nu}\gamma^{\rho\sigma}\nabla^{(\g)}_\rho \nabla^{(\g)}_\sigma - \nabla^{(\g)}_\mu\nabla^{(\g)}_\nu\right) U_B(\f) \Bigg]_{\f_0(x)},
$$
\be
\Big[n^a\de_a \f\Big]^{IR}_{UV}  = \left[{d W_B \over d \f} - {d U_B\over d \f} R^{(\g)} + \half {d Z_B \over d \varphi}(\de \f)^2 - {1\over \sqrt{\g}}\de_\mu \left( Z_B \sqrt{\g} \g^{\mu\nu}\de_\nu \f \right) \right]_{\f_0(x)},  \label{FE6}
\ee
where $\f_0(x^\mu)\equiv \f(x^{\m},u_0)$ is the scalar field evaluated at the brane position $u_0$.

\subsection{Review of vacuum solutions and self-tuning}
\label{subsec:VacSol}
Before discussing the brane cosmology, in this section we review the
vacuum solutions of the model introduced in the previous section, and
the corresponding self-tuning mechanism for the cosmological
constant. By {\em vacuum} here we mean time-independent bulk solutions
displaying four-dimensional Poincar\'e invariance. They are dual to the
ground-state of the corresponding dual QFT. More details can be found in previous work, \cite{selftuning}.

The most general solution of the bulk equations (\ref{FE1}-\ref{FE2})
and junction conditions  (\ref{FE3}-\ref{FE4}) enjoying  full 4d Poincar\'e
invariance consists in two halves of time-independent five-dimensional geometries (which we
call UV and IR, to use the holographic terminology, as we explain below) separated by  a flat
static brane sitting at a fixed radial position $u_*$:
\be\label{rev1}
ds^2 = \left\{\begin{array}{ll} du^2 + e^{2A_{UV}(u)}
    \eta_{\mu\nu}dx^\mu dx^\nu & \:\: u< u_*\\ & \\
du^2 + e^{2A_{IR}(u)}
    \eta_{\mu\nu}dx^\mu dx^\nu  &\:\: u>u_* \end{array}\right. ,  \qquad \varphi(u) =
\left\{\begin{array}{ll}\f_{UV}(u) & \:\: u< u_* \\ & \\ \f_{IR}(u) &
    \:\: u>u_* \end{array}\right.
\ee
with the brane embedding given simply by $\xi^\mu = x^\mu$,
$u=u_*$. The scale factor  and scalar field are continuous across the
brane, while their first derivatives have a jump, determined by Israel's junction
conditions at $u=u_*$, as we shall discuss in a moment. It is
convenient to rewrite the bulk solution in terms of a
first order formulation by introducing UV and IR superpotentials \cite{Wolfe}
$W_{UV}(\f)$ and $W_{IR}(\f)$, i.e. scalar functions of $\f$ such that
\bea
&& {dA_{UV} \over du} = - {W_{UV}(\varphi_{UV}(u)) \over 6}, \qquad {dA_{IR} \over
  du} = - {W_{IR}(\varphi_{IR}(u)) \over 6},  \label{rev2a}\\ && {d\f_{UV} \over du} =
- {d W_{UV} \over d\f}(\f_{UV}(u)),  \qquad {d\f_{IR} \over du} =
- {d W_{UV} \over d\f}(\f_{IR}(u)). \label{rev2b}
\eea
Both $W_{UV,IR}$ satisfy the superpotential equation,
\be\label{rev3}
-{W^2\over 3} + {1\over 2} \left(d W\over d\f \right)^2 = V(\f).
\ee
Israel's jump conditions (\ref{FE4}) are most easily written in terms of the
superpotentials as
\be\label{rev4}
W_{IR}(\f_*)- W_{UV}(\f_*) = W_B(\f_*), \quad {d W_{IR}\over
  d\f}(\f_*)- {d W_{UV} \over d\f} (\f_*) = {d W_B\over d\f}(\f_*),
\ee
where $\f_* \equiv \f_{UV}(u_*) = \f_{IR}(u_*)$ is the value of the scalar at the position of the brane.  Below, we review the
main features of the UV and IR geometry and of the solution of the
junction conditions.

\subsubsection{The UV geometry} We call the ``UV geometry'' the half which connects to an asymptotic AdS
boundary. To guarantee the presence of the UV  half we have to take a bulk scalar
potential which admits at least one (local) maximum (which we set at $\f=0$ without loss of generality) , and
which therefore allows for $AdS$ solutions with $\f=0$, and
\be
A(u) = -{u\over \ell}\sp V(0) = -{12\over \ell^2}
\label{ec1}\ee
where  $\ell$ is the AdS length determined by the relevant maximum of the potential.

Close to  the maximum,
\be \label{UV1}
V(\f) = -{12\over \ell^2} + {m^2 \over 2} \f^2 + O(\f^3)
\ee
where $m$ is restricted by the BF bound, which in 5d reads
\be\label{UV2}
m^2 \geq -{4\over \ell^2}.
\ee
Domain-wall solutions with non-trivial $\f(u)$ connecting to the
maximum are dual to RG flows driven by a relevant operator with
dimension $\Delta = \Delta_+$ where
\be \label{UV3}
\Delta_{\pm} = 2 \pm \sqrt{4 + {m^2 \ell^2}}
\ee
where  we have  assumed the  standard holographic
dictionary\footnote{For $-4 < m^2 \ell^2 < -3$ there exists an
  alternative dictionary where the operator dimension is
  $\Delta_-$. We shall not use this alternative here.}. Since
$\f=0$  is a maximum of the bulk potential, we have $m^2 <0$, $0<
\Delta_-<2$, and $2< \Delta_+<4$ (the dual operator is relevant).

Close to $\f=0$, the metric and scalar field profile behave as
\be\label{UV4}
A_{UV}(u) = -{u\over \ell}+ \ldots, \quad \f_{UV}(u) =
\f_-\ell^{\Delta_-}e^{\Delta_- u/\ell} + \ldots  +
\f_+\ell^{\Delta_+}e^{\Delta_+ u/\ell} + \ldots, \quad u \to -\infty
\ee
where $\f_-$ and $\f_+$ are the two independent integration constants of the bulk
Klein-Gordon equation and in the dual QFT they correspond respectively to the source and
the vev of the relevant operator deforming the CFT in the UV. The
ellipses indicate subleading terms which are completely fixed by the
integration constants. Since the  scale factor diverges as
$u\to\-\infty$, the UV region can be thought of as  the large-volume
half of the geometry.

The superpotential $W_{UV}$ corresponding to the solution (\ref{UV4}) takes  two possible asymptotic forms:
\be\label{UV5}
W_-(\f) = {6\over \ell} + {\Delta_- \over 2\ell} \f^2 + \ldots + C
\f^{4\over \Delta_-} + \ldots  \qquad \f \to 0
\ee
or
\be \label{UV6}
W_+(\f) = {6\over \ell} + {\Delta_+ \over 2\ell} \f^2 + \ldots \qquad
\f\to 0
\ee
The $W_-$-type solution  corresponds to the generic case  $\f_-\neq 0$
and describes in the dual language a relevant deformation of the CFT
obtained by giving a source to  the relevant operator dual to $\f$ (this source is $\phi_-$). In
the subleading terms, we have highlighted a particular non-analytic
term, proportional to a free constant $C$, which plays the role of the
single integration constant of the first order differential equation
(\ref{rev3}). It determines the ratio between the vev and the source
of the dual operator. All other subleading terms we omitted, are either
$C$-independent or they are fixed in terms of $C$.
Notice that, since  $C$ enters at subleading
order, the leading UV behavior of the superpotential as $\f\to 0$ is
universal. Notice also that $W_-$ does not contain $\f_-$, which is
to be  thought of as a boundary condition for the first order flow of
$\f_{UV}$ (equation \ref{rev2b}) for a given $W_{UV}$. Therefore, a choice
of $W_-$ corresponds to a  family of holographic RG flows,
differing from each other by the value of the UV source. The
subleading coefficient  $\f_+$ in equation (\ref{UV4}) is proportional
to $C \f_-^{\Delta_+/\Delta_-}$.

The $W_+$-type superpotential (\ref{UV6})  corresponds to solutions
with $\f_-=0$, which are dual to a pure vev deformation with no source.  Notice that this superpotential does not contain any free
parameter, so it is a single point in the space of solutions of the
superpotential  equation (\ref{rev3}). As it  will be clear when we
discuss the IR below, solutions of this type are generically singular
in the IR, unless the bulk potential is appropriately tuned.

We always refer as ``UV limit''  a  $u\to -\infty$ asymptotic
solution of the form (\ref{UV4}) or equivalently as the $\f \to 0$ limit
of a superpotential of the form (\ref{UV5}-\ref{UV6}).

\subsubsection{The IR Geometry} \label{sss:IR}

The  IR name indicates the far interior of the geometry in analogy with standard holography. For generic solutions
of the form (\ref{rev1}), the interior contains a singularity (which
is typically at a finite coordinate position $u_0 > u_*$) where the
scale factor vanishes and the scalar field diverges ( a full
classification of IR geometries can be found e.g. in \cite{exotic}).  The only
exception is the case when the geometry is asymptotically
$AdS$ in the IR too, in which case the singularity is replaced by a
Poincar\'e horizon. Certain special types of
singularities are acceptable in holography (for example, they  describe color
confinement of the dual gauge theory and the presence a mass gap), but they have to satisfy
certain criteria (which we  loosely refer to as ``regularity'')
which strongly constrains the IR superpotential, to the point of
selecting  a {\em single} solution of the superpotential equation
(\ref{rev3}). Moreover, certain classes of  bulk potentials do not
admit a ``regular'' solution at all and therefore they belong to a
holographic swampland, \cite{gubser,thermo,cgkkm}.

In all cases discussed above, the scale factor vanishes in the IR
limit, therefore the IR  is the small-volume part of the solution. The type of geometry in the IR  depends on the bulk potential. We shall
consider two classes of IR solutions: mildly singular ones (with constraints
on the type of singularity) and asymptotically AdS in the IR. We start
from the latter.

\paragraph*{IR-AdS.}
 These are solutions where the scalar
field reaches a second extremum of the bulk potential (a
minimum) at $\f = \bar{\f}$, corresponding to an IR conformal fixed
point of the dual QFT. Close to the minimum,
\be \label{IR1}
V(\f)  = -{12\over \ell_{IR}^2} + {m^2_{IR} \over 2}(\f-\bar{\f})^2 + \ldots
, \qquad \f \to \bar{\f}
\ee
with $m^2_{IR} >0$.  Now we have $\Delta^{IR}_- <0$
and $\Delta^{IR}_+>4$ (these quantities are defined as in  (\ref{UV3}) with $m$ replaced
by $m_{IR}$). The scale factor and scalar field profile are given by
\be\label{IR2}
A_{IR}(u) = -{u\over \ell_{IR}}+ \ldots, \quad \f_{IR}(u) = \bar{\f} +
\f_-\ell^{\Delta_-^{IR}}e^{\Delta_-^{IR} u/\ell} + \ldots, \quad u \to +\infty
\ee
and there is no place for the $\f_+$-type contribution since this would
diverge as $u\to +\infty$  and drive the solution away from the fixed
point (i.e. the dual operator is irrelevant at the IR fixed point).  Correspondingly, there is a single solution of the
superpotential equation which reaches the fixed point, and it is  given by
\be\label{IR3}
W_{IR}  = {6\over \ell_{IR}} + {\Delta_-^{IR} \over 2\ell_{IR}} (\f-\bar{\f})^2 + \ldots \qquad
\f\to \bar{\f}.
\ee
Since this contains no free integration constant, requiring the
solution to reach the IR fixed point completely fixes $W_{IR}$.
All other solutions to the superpotential equation  miss the IR fixed
point, and what happens to them depends on the behavior of the
potential for $\f > \bar{\f}$. If the solution misses all possible IR
fixed points, this behavior is captured by the  the general
classification of the large-$\f$ asymptotics given below.

\paragraph*{Exponential IR.}\label{ExpSuperpot}
If  the solution misses all finite-$\f$ fixed points, then it must
reach the large-field region $\f\to \infty$. We can classify the
solutions assuming the potential $V(\f)$  is dominated in this limit
by a single exponential,
\be\label{IR4}
V(\f) \simeq - V_{\infty} e^{2\kappa \f} \qquad \f \to +\infty
\ee
where $V_\infty$ and $\kappa$ are  positive  constants\footnote{The general static solution with an exponential potential has been found in \cite{exact}}.
This allows  a general classification\footnote{To be more precise we define $2\kappa=\lim_{\f\to\infty}{V'\over V}$. In string theory $\kappa<\infty$.}
, as the cases of constant or
power-law asymptotics are essentially captured by setting $\kappa=0$
(as we shall see in section 4).

In this case we always find a curvature singularity at a finite
coordinate value $u_0$. For generic solutions, the singularity is
unacceptable according to various criteria\footnote{for example, it is {\em
  bad}  in Gubser's classification \cite{gubser}.}. As it turns out,
the only solution which has an acceptable (or ``good'' in the
holographic sense) IR singularity (which we
 call henceforth ``regular'' by an abuse of terminology) are those
associated to IR superpotentials with asymptotics
\be\label{IR5}
W_{IR} = W_{\infty} e^{\kappa\f} \quad W_\infty = \sqrt{2V_{\infty}
  \over 2/3 - \kappa^2}\qquad \f \to +\infty
\ee
These asymptotics are special in the sense that (as it was the case for
the IR AdS geometry) they do not admit any free parameter, and the
solution behaving as in (\ref{IR5})  is
an isolated point in the space of solutions of the superpotential
equation (\ref{rev3}). One key point is that this special solution
exists only if
\be\label{gubser}
\kappa < \sqrt{2\over 3},
\ee
which is usually referred to as the {\em Gubser bound}.

The other solutions of (\ref{rev3}), which do depend on a free
integration constant, have an exponential divergence independent
of $\kappa$, given by
\be
W \sim \exp\left[{\sqrt{2\over 3}\f}\right]\;.
\label{ec2}\ee
 Due to the
Gubser bound (\ref{gubser}) this is a stronger divergence than the one
in (\ref{IR5}) and one can show that it always leads to a bad singularity.
For more details we
refer the reader to the discussion in \cite{thermo,exotic} and references therein.

As a consequence,  if (\ref{gubser}) is not satisfied, the potential does not admit
any ``regular'' solution reaching $\f\to \infty$. Therefore, we
always assume that the bulk potential asymptotics satisfy Gubser's
bound at large $\f$.

With a superpotential of the form (\ref{IR5}), the scale factor
vanishes and the
scalar field diverges  close to the singularity at $u_0$ as
\be \label{IR6}
A_{IR}(u) \sim (u_0-u)^{1\over 6\kappa^2}, \qquad \f_{IR}(u) \sim
- {1\over \kappa} \log \left[W_{\infty}\kappa^2(u_0 - u)\right],
\ee
where we assumed the singularity is reached from below, i.e. $u$ is a
monotonic coordinate going from the UV at $u=-\infty$ to the IR at
$u=u_0$.

In either the AdS or exponential asymptotic case, we found that the IR
superpotential $W_{IR}$ is uniquely fixed by ``regularity''. This
means that, generically (forgetting the brane for a moment) it will
match in the UV with one of the solutions of the type $W_-$ in
(\ref{UV5}) , with a specific
value of the integration constant. It is only in very special cases
(which require a tuning of the bulk potential) that the regular
solution will match the $W_+$ solution in the UV. Examples of
this kind were recently discussed in \cite{exotic}.

\subsubsection{The junction conditions and the self-tuning mechanism}

From our discussion of the UV and IR halves of the bulk we can derive
the following conclusions:
\begin{enumerate}
\item The UV behavior is universal and common to a continuous family of
  solutions, parametrized by the integration constant $C$ appearing in
  $W_{UV}$ as in equation (\ref{UV5}).
\item In contrast, the IR behavior is restricted by the regularity
  requirement, which completely fixes\footnote{In rare cases there may be a discrete finite set of regular solutions signaling distinct saddle points of the holographic theory, differing only in the vev of the perturbing operator.}  the IR superpotential
  $W_{IR}$.
\end{enumerate}
With these considerations in mind, we can now look at the junction
conditions across the brane (\ref{rev4}).  Since $W_{IR}$ is
completely fixed by regularity,  we can look at (\ref{rev4}) as a
system of  two
non-linear equations in two unknowns: the scalar field value at the
brane $\f_*$, and the value of the integration constant $C$ hidden in $W_{UV}$ (or
alternatively, the value  of $W_{UV}$  at the brane, which serves as
an initial condition for the superpotential equation in the
UV). Generically, theses equations will admit a solution $(\f_*, C_*)$
which fixes both the brane position (in $\f$-space) {\em and} the UV
superpotential $W_{UV}(\f)$. Finally, the position of the brane in the
$u$-coordinate is determined by integrating the flow equation for
$\f_{UV}(u)$ by giving as extra input the UV boundary condition $\f_-$
in (\ref{UV4}) (which is one of the boundary data defining the
holographic theory).

The essence of the self-tuning mechanism is that the procedure above
 leads to  solutions with Minkowski brane geometry, for generic
 brane cosmological term $W_B(\f)$. Moreover, the brane is stabilized
 at $u_*$ since solutions of this kind typically come in discrete
 sets, as the system of equation is, generically, non-degenerate.

\subsection{Searching for cosmological solutions}

Having discussed the vacua of the bulk brane system, we now turn to states describing a
cosmology. We emphasize that we are looking for time-dependent
solutions in the {\em same holographic theory} (i.e. with the same  UV boundary
conditions) that gives  us the Minkowski vacuum state discussed in the
previous section. In other words, here we do not want the
time-dependence of the solution to arise because we  turn on
time-dependent sources (couplings) in the dual QFT. In  this sense, the approach we follow here is
orthogonal to the one pursued in
\cite{dS}, where de-Sitter brane geometries were found by
deforming the QFT boundary metric to be de Sitter, and which could be
extended to more general FRW metrics. The reason is that in that case it was shown that the only way to generically obtain such solutions was to consider the bulk QFT to be defined (at the boundary) on a similar curved manifold, dS$_4$ or AdS$_4$.

More specifically, we look for time-dependent solutions of the bulk
equations plus junction conditions such that all the {\em source
terms} in the UV (i.e. the leading terms in the metric and scalar
field asymptotics close to the AdS boundary) are unmodified with
respect to the vacuum solution (\ref{UV4})
\be \label{src0}
ds^2_{UV} \to  du^2 + e^{-2u/\ell}\left(\eta_{\mu\nu} + \ldots\right)
dx^\mu dx^\nu ,
\quad \f_{UV} \to \f_- \ell^{\Delta_-}e^{\Delta_-u} + \ldots, \quad u\to -\infty
\ee
and all the time dependence is buried in the subleading terms
(e.g. the vev term  $\f_+$ in (\ref{UV4}), as well as subleading
corrections to the metric starting at order $e^{4u/\ell}$, are allowed
to be time-dependent).

The general form of a cosmological solution with the features sketched above
consists of a   bulk metric and dilaton profile which, on each side of the brane, can be
put (up to bulk diffeomorphisms) in  the general form
\be\label{src1}
ds^2_{bulk} =  b_\alpha(t,z)^2 dz^2 -n_{\alpha}(t,z)^2 dt^2 +
a_{\alpha}(t,z) \delta_{ij} dx^i dx^j, \quad \f = \f_{\alpha}(z,t), \qquad \alpha = \{UV, IR\}
\ee
where we assume a spatially-flat universe\footnote{In the general
  ansatz (\ref{src1}) we have denoted by $z$ the holographic radial
  coordinate, and we reserve the notation $u$ for the special case
  when the metric takes the {\it domain-wall} form as in equation
  (\ref{rev1}) or (\ref{src0})} .  The brane is described, up
to world-volume reparametrizations,
  by the  world-volume coordinates  $(t,x_i)$ and an embedding
function  $z= z_0(t)$. Although there are three unknown functions in
the ansatz (\ref{src1}), they can be reduced to two by a further
gauge-fixing, but we leave equation (\ref{src1}) general for
convenience. Examples of gauged-fixed metrics can be found in
Appendix \ref{equations}.

Close to the UV boundary, the functions
$b_{UV},n_{UV},a_{UV}$ and $\f_{UV}$  must be such that equation
(\ref{src1}) takes the asymptotic form (\ref{src0}). One then has to
solve Einstein's equations (\ref{FE1}-\ref{FE2}) and Israel's junction
conditions (\ref{FE3}-\ref{FE4}), subject to the UV  boundary
conditions, and to IR regularity. The detailed form of the field
equations and junction conditions in  the ansatz (\ref{src1}) can be
found in Appendix \ref{app:diag}. In appendices \ref{app:khaler},
\ref{app:ch-y} and \ref{app:xristos} we also report the bulk equations
and junction conditions for different, gauge-fixed forms of the bulk
ansatz. Each of them may be more useful for specific purposes
(e.g. the ansatz discussed in appendix \ref{app:ch-y} is convenient
for discussing the presence of apparent horizons in the IR, and for possible
numerical implementations).

A quick look at the form of Einstein's equations and junction
conditions is enough to convince one-self that it is not easy, to obtain some insight from these equations, let alone solve
them for general enough time-dependent backgrounds. Therefore, it is
natural to look for simplifying assumptions which could make the
system more tractable. In the rest of this section, we discuss some
possible attempts at a simplification, and whether or not one expects
them to work.

We insist on looking for solutions in {\em generic}
theories, i.e. which do not need special relations to be imposed on
the bulk and brane potentials.

\paragraph*{Static bulk, moving brane.} The first thing one might try
is to look for solutions in which the time-dependence is completely
encoded in the brane embedding, which describes a trajectory in the
$(t,z)$ plane in a static bulk. If such a solution were possible, it
would be  very simple to impose the same UV boundary conditions and IR
regularity requirement as for a static brane, and one would be left
with the problem of solving for the brane embedding in a vacuum bulk
solution of the form (\ref{rev1}), but with a time-dependent $u_*$.

In the absence of a bulk scalar field, or if the latter is constant,
such an ansatz is actually the most general one: the bulk theory
is pure gravity with a (negative) cosmological constant, and due to a careful application of Birkhoff 's theorem including braneworld boundaries \cite{Bowcock}, the
solution can be always be put in a static
form, which is the AdS-Schwarzschild metric.  This
ansatz is widely used when discussing brane-world cosmology in the
context of e.g. the Randall-Sundrum model with no bulk scalars \cite{kraus,Bowcock,CGP}.

On the other hand, if we look for solutions with a  non-trivial
scalar-field profile, such a simplification is generically impossible.  In the
presence of a bulk scalar,  it is
generically inconsistent to impose Israel's junction conditions for a
moving brane in a static bulk, unless the brane-potentials are tuned
to the bulk solution. A proof of this fact can be found in appendix  A
of \cite{dS}.  The  argument relies on the fact that,
 in a static,  fixed bulk, Israel's junction conditions for the metric
 (plus eventually the brane matter content) completely determine the
 brane-induced cosmology, and this in turn is generically inconsistent
 with the junction conditions for the jump in the derivative of the
 scalar field. 

The above argument can be evaded in
special situations in which  the  brane potentials are
  specifically adjusted   to the bulk solutions, in such a way that  the bulk metric
  and scalar field are smooth across the brane, and the brane does not
  backreact. This results in  an  {\em evanescent brane}
  \cite{dS} which does not affect the bulk geometry\footnote{The existence of such branes may be forbidden by swampland criteria.}. As
  mentioned above, such solutions do not arise in generic theories and
  require special fine-tunings. We shall not pursue this direction
  here. 
  The constraints arising from the junction conditions are relaxed when we consider a 
  {\em probe brane}, which will be analysed in detail in the
  following sections. Another possibility is to look for moving branes
  in more general static bulk solutions (e.g. dilatonic black holes, which
  preserve the same symmetries as the cosmological solutions in which
  we are interested), and investigate if this allows to relax some
  constraints. This  will not be pursued here, and we leave it for
  future work.

\paragraph*{Time-dependent domain-wall ansatz} Since generically one has to abandon the idea of a static
bulk, one may  look for special ansatze which may simplify the
analysis. For example, instead of working with the general bulk metric
(\ref{src1}), one may try and look for solutions  generalising the
static ansatz (\ref{rev1}) by promoting, on each side of the brane,  the scale factor to a function of
$u$ and $t$,
\be \label{src2}
ds^2 = du^2 + e^{A(u,t)}\eta_{\mu\nu}dx^\mu dx^\nu , \quad \f = \f(u,t).
\ee
This is a special case of (\ref{src1}) with $b=1$ and $n=a$.  From the
counting of degrees of freedom and gauge-invariance, it is already quite clear
that this generically cannot work, as  general coordinate invariance usually
allows to impose only one condition (and not two) on the metric
coefficients. In Appendix \ref{simplifiedApp}, we perform a  linearized
analysis (around the static case)  of the solutions with the
simplified  bulk ansatz (\ref{src2})  and the corresponding junction
conditions. There, it is shown that, in the linear regime,  the existence of
a solution of this form is only possible if a certain (non-generic)
condition on the brane potentials is satisfied, in agreement with
the above expectation. Although obtained explicitly only in the linear
regime, this result strongly suggests that a similar non-generic
condition will arise also in the full non-linear system.

\paragraph*{Half-static bulk.} A third possibility to find a somewhat
simplified cosmological solution is to assume a generic ansatz
(\ref{src1})  on one side of the brane, while allowing the other side to be
static. In this case, a solution to the junction conditions is
possible if the moving brane ``cancels out'' the time-dependence of the
time-varying half of the bulk. This situation is particularly
convenient if we assume the static half to be the IR, because in this
case  the regularity condition is unchanged with respect to the
discussion in subsection  \ref{sss:IR}. On the other hand, the motion
of the brane will result in a general UV geometry with time-dependent
vevs.

Although this possibility is interesting, there are  indications
that it cannot generically lead to a solution, based once again on a
linearized analysis around an equilibrium position. Indeed, based on
small fluctuations analysis performed in \cite{selftuning}, one arrives at the
conclusion that there are no  (harmonic) perturbations
of the static solutions such that one side of the brane is
time-independent. Below we sketch the argument in a quantum-mechanical
framework, for the details of which we refer the reader to \cite{selftuning}.

Choosing $z$ as the conformal radial coordinate, the bulk fluctuations are first decomposed  in spatially homogeneous harmonic modes
on both the UV and IR side, $\psi_{{uv}}(z,t;\omega)$,
$\psi_{{ir}}(z,t;\omega)$,  satisfying $\de_t^2 \psi_{{uv,ir}} =
-\omega^2 \psi_{{uv,ir}}$. As shown in \cite{selftuning}, the
problem of  scalar perturbations of the brane/bulk system around an
equilibrium position can be put in the form of an equivalent
Schr\"odinger equation for the two-component vector $\Psi$ with
components  $(\psi_{{uv}}, \psi_{{ir}})$,
\be \label{src3}
{\cal H}  \Psi = \omega^2 \Psi , \qquad {\cal H } \equiv\left(\begin{array}{cc}
    H_{{uv}} & 0  \\ 0 & H_{{ir}}\end{array}\right),   \qquad
\Psi \equiv \left(\begin{array}{c}
    \psi_{{uv}}  \\ \psi_{{ir}}\end{array}\right)
\ee
where the ${uv}$ and ${ir}$  Hamiltonians  are completely
determined by the background bulk solution in the UV and IR. One must impose a
normalizability condition both at the  UV and IR boundaries of the problem for
$\psi_{{uv}}$ and  $\psi_{{ir}}$, respectively. Furthermore,
the matching conditions across
the brane  at $z_*$ translate into a linear relation between $\Psi$ and
its normal derivative at the interface,
\be \label{src4}
\de_z \Psi = \left[\Gamma^{(1)} + \omega^2 \Gamma^{(2)}\right] \Psi
\ee
where $\Gamma^{(1)}$ and $\Gamma^{(2)}$ are constant, $\omega$-independent
two-by-two matrices which depend on the bulk data and the brane potentials
evaluated on the background static solution, and are given explicitly
in Appendix D of \cite{selftuning}.  The fluctuation in the brane
position is not an independent dynamical variable, but it is
completely determined by $\psi_{{uv}}$ and  $\psi_{{ir}}$ at
the interface.

We now search for a solution which is static in the
IR. For  $\omega \neq 0$, this  implies that $\psi_{{ir}}(r,t)=0$
identically. This  in turn requires that the two-by-two matrix on the
right hand side of (\ref{src4}) must be upper-triangular, i.e. its
$\tiny{ir,uv}$ component must vanish,
\be
\Gamma^{(1)}_{{ir,uv}} + \omega^2 \Gamma^{(2)}_{{ir,uv}} = 0
\label{ec3}\ee
This is possible  in either one of the following two cases:
\begin{enumerate}
\item Either $\Gamma^{(1)}_{{ir,uv}} =
  \Gamma^{(2)}_{{ir,uv}} = 0$;
\item Or, if  the matrix components do not vanish,  the frequency must
  be fixed to be
\be \label{src5}
\omega^2_*=   -
  {\Gamma^{(1)}_{{ir,uv}}\over \Gamma^{(2)}_{{ir,uv}}}.
\ee
\end{enumerate}
The first case  requires a special relation
  to be imposed at the interface on the value of the bulk data  and brane potentials, and
  is therefore non-generic.  In the second case,
 we must proceed to solve the Schrodinger problem
  for $\psi_{{uv}}$ for $z<z_*$, with Hamiltonian $H_{uv}$,
  energy $\omega_*^2$,
  normalizability condition at the UV boundary, plus the
  condition at the interface
\be \label{src6}
\de_z \psi_{{uv}}  = \left[\Gamma^{(1)}_{{uv,uv}} + \omega^2_*
  \Gamma^{(2)}_{{uv,uv}}\right] \psi_{{uv}}
\ee
This problem is clearly over-determined: it is equivalent to a
one-dimensional Schr\"odinger problem in a box with linear boundary
conditions at both ends and specified energy. The only way this
can admit a non-trivial solution is if the frequency $\omega_*$ (which
is purely fixed by the data at the interface, by equation
(\ref{src5}))  is {\em also}
part of the spectrum of the bulk Scrh\"odinger operator in the
UV. This is not generically the case, but rather it requires special
relations to be imposed between the bulk and brane data.

According to the  argument above, in a  generic theory, there are no solutions in which one side
of the bulk is  static and the other is  time-dependent,  at least at
the linearized level. This does not
exclude the existence of non-perturbative solutions of this kind
(i.e. far from the static regime), but the argument we gave makes it  unlikely,
because it ultimately amounts to counting  equations vs. degrees of
freedom. Therefore, it should be used as a warning that looking for
such simplified solutions in the full system may not be a fruitful
enterprise,  and it is another avenue we shall not follow. \\

The conclusion of the  analysis performed in this subsection is that some easy shortcuts do not seem to work: to determine a  cosmological solution in a
generic theory, one must start with the  general ansatz
(\ref{src1}), possibly in one of the fully gauge-fixed forms  we give
in Appendix A. As it is apparent from the explicit form of the
junction conditions, this is a non-trivial problem in general, and
little can be said of generic solutions, including the discussion of
the IR regularity conditions. Although one may hope to find concrete
examples, by looking at very specific exact solutions e.g. with purely
exponential potentials \cite{exact}, we shall not pursue this here.

Our principal aim here is to build a general picture of possible cosmologies, even approximate, in order to assess the cosmological potential of this setup and how cosmology intertwines with the mechanism of the self-tuning of the cosmological constant that was presented in \cite{selftuning}.
To do this generically, we are led to make a different simplifying assumption: in the rest of
the paper we  study the brane motion in the {\em probe limit}, in
which the size of the  brane-induced terms is small compared to the
bulk curvature scale, and one can neglect the brane backreaction. This
situation is similar, but  more generic, to the one of an evanescent brane
described at the beginning of this section: in  the latter case, a
special relation must be imposed between the brane potentials, so that
similar sized terms cancel each other in the junction conditions. The
probe limit that we follow here,  on the other hand, is an
approximation which requires order of magnitude hierarchies but no
specific tunings between different brane-induced terms. In this sense,
it is more generic. As we shall see, the   gain is important, since the
probe brane motion is, essentially, exactly solvable in any given bulk
solution.

The probe limit has been used in the past to define {\em mirage cosmology} \cite{mirage}, essentially driven by the motion of (universe)-branes in non-trivial bulk fields. Such a motion is perceived as a cosmological evolution on the brane, because the induced metric becomes time-dependent due to the brane motion.

We therefore proceed, in the rest of this paper, to a full analysis of
the probe limit in a generic bulk background with the asymptotic
features discussed in section \ref{subsec:VacSol}.

\section{The probe brane limit}




The probe approximation consists of  studying the brane
motion in a given geometry satisfying the bulk Einstein's equations
(\ref{FE1}-\ref{FE2}) while ignoring the brane backreaction on the bulk solution.
This means that all terms in the
brane-induced action are small compared to the corresponding bulk
terms, i.e. that when we construct the bulk solution, we can
effectively set to zero  anything appearing on the  right hand side of
the matching conditions (\ref{FE4}). Strictly speaking this implies that we are outside
of the self-tuning framework reviewed in section 2, i.e. we cannot
assume e.g. the brane cosmological term to be arbitrarily large.
Although this is a caveat of the probe approximation, the cosmological solutions we shall find will be useful, 
among other things, for studying the cosmology of theories that self-tune the cosmological constant.
There may be also special cases like eg. evanescent branes \cite{dS}, where, although
each individual term on the right hand side of (\ref{FE4}) is large,
they effectively cancel and the probe approximation is still valid. 
Finally, as we shall see explicitly in section \ref{asym}, the probe
approximation is {\em always} valid if the brane reaches the UV
region, no matter  the details or the magnitude of the induced
terms. Interestingly, we will find that in this regime the induced
brane geometry is close to de Sitter.

In the approximation in which the right hand sides of the matching
conditions (\ref{FE4}) can be neglected, the bulk geometry is smooth
across the brane. We must just solve the bulk equations unperturbed by the presence of the brane. Moreover, since we are looking for a solution with
the same sources as the vacuum, the simplest possibility is to
consider the whole of the bulk to be in a Poincar\'e-invariant,
time-independent  vacuum state, and the cosmological evolution to
arise solely due to the motion of the brane\footnote{The case where the bulk solutions is sliced by dS or AdS geometries discussed in \cite{F,dS} can also be considered, but we shall deal with in in a future publication.}. The bulk geometry  is
therefore specified by a single time-independent scale-factor  and scalar field
profile,
\be
ds^2\equiv g_{MN}dx^Mdx^N=du^2+e^{2A(u)}(-dt^2+dx^idx^i)\sp \varphi(u)\;.
\label{c1}\ee
This solution is described in the first order formalisms in terms of a single
superpotential $W = W_{UV}= W_{IR}$, which is completely fixed by
regularity in the IR, and satisfies the relations
\be
{dA \over du} =-{W\over 6}\sp {d\varphi \over du} =\frac{d W}{d \varphi} \sp -{W^2\over 3}+{1 \over 2} \left(\frac{d W}{d \varphi}\right)^2=V
\label{c7}\ee

We now turn to the brane dynamics in the probe limit, in the
fixed bulk geometry described above. Using  world-volume coordinates $\xi^{\m}$, the world-volume action in full gauge invariant form is
\be
S_b= M^3 \int d^4\xi\sqrt{-\hat g}\left(-W_B(\varphi)+U_B(\varphi)\hat R-{Z_B\over 2}(\pa\varphi)^2+\cdots\right)\;,
\label{c2}\ee
where hats indicate induced quantities.
Considering the static gauge, $\xi^{\m}=x^{\mu}$, the only dynamical
 variable describing  the brane dynamics is $u(x^{\m})$ that we
 take to be only function of time, therefore the probe brane limit is
 described by a single degree of freedom $u(t)$.

The induced metric is given by
\be
d\hat s^2\equiv \hat g_{\m\n}d\xi^{\m}d\xi_{\n}=
-(e^{2A}-\dot u^2)dt^2+e^{2A}dx^idx^i
\label{c3}\ee
where a dot denotes a derivative with respect to $t$.

Using the induced  metric and scalar field in the brane action
(\ref{c2}), the problem of probe brane motion translates into a
Lagrangian mechanics problem governed by the action\footnote{The case without the scalar coupling has been addressed in \cite{rou}.} (see appendix
\ref{approbe} ):
\be
S_b[u(t)]=M^3 V_3\int dt~e^{4A}\left[{F\over \sqrt{ 1-e^{-2A}\dot u^2}}-\sqrt{ 1-e^{-2A}\dot u^2}\left(W_B+F\right)
\right]\equiv \int dt~L_b
\label{c12}\ee
where $V_3$ is the spatial volume (measured in the bulk coordinates)
and  
\be
F(\varphi)=-{U_BW^2\over 6}+W {d W\over d\f} {d U_B \over d\f}+{1\over
  2}Z_B\left({d W \over d\f}\right)^2 ~.
\label{c13}\ee

\subsection{The general solution}

The action in (\ref{c12}) is explicitly time-independent and therefore
the Hamiltonian is conserved. The momentum conjugate to $u$ and the
Hamiltonian are given by
\be
p_u={\delta L_b\over \delta \dot u}={e^{2A}\dot u\over ( 1-e^{-2A}\dot u^2)^{3\over 2}}\left[F+ ( 1-e^{-2A}\dot u^2)(W_B+F)\right]
\label{c14}\ee
and
\be
H=\dot u p_u-L_b={e^{2A} [e^{2A}W_B+(F-W_B)\dot u^2]\over ( 1-e^{-2A}\dot u^2)^{3\over 2}}=E\;,
\label{c15}\ee
where $E$ above is a real constant, and we have omitted the overall factors
$M^3V_3$ for simplicity.  

For a given value of the integration constant $E$, we may now solve equation (\ref{c15}) to find $\dot u$:
\be
E( 1-e^{-2A}\dot u^2)^{3/2}=e^{2A}[e^{2A}W_B+(F-W_B)\dot u^2]
\label{c29}\ee
This equation expresses $\dot u^2$ as a function of $e^{2A(u)}$ and of
$\varphi(u)$, hidden inside $W_B$ and $F$. Solving for the trajectory
is then reduced to  a single integration to obtain $u(t)$. 

Equation (\ref{c29}) can be cast in  a simpler form, by expressing it in terms of
 the brane proper time coordinate $\tau$, defined by a world-volume
 coordinate transformation that brings the induced metric to the form
\be
d\hat s^2\equiv -d\tau^2+e^{2A(\tau)}dx^idx^i
\label{c32}\ee
From (\ref{c3}) we obtain
\be
\sqrt{e^{2A}-\dot u^2}~dt=d\tau
\quad
\rightarrow
\quad
\frac{d u}{ d t}
=
\frac{e^A}{\sqrt{1+\left(\frac{d u}{d \tau}\right)^2}}
\frac{d u}{ d \tau}\;.
\label{cc1}
\ee
We define the new variable $y$ as
\begin{equation} \label{y}
y \equiv  \sqrt{1+\left(\frac{d u}{d \tau}\right)^2}={1\over \sqrt{1-e^{-2A}\dot u^2}}\;.
\end{equation}
By its definition $y$ satisfies
\be
y\geq 1
\label{d3}\ee
Then equation (\ref{c29}) becomes
\begin{equation}
E  = e^{4 A} y [F(y^2-1)+W_B]
\label{cc3}
\end{equation}
This is a cubic equation in $y$, whose solutions are discussed in
appendix \ref{AppCubic}. Therefore, the problem of probe brane motion for a
given  ``energy'' $E$ reduces to finding solutions $y(u)$   to the
algebraic equation (\ref{cc3}) and restricting to  $y(u)\geq 1$.
The energy $E$ is the only relevant initial condition for the brane motion.
It can be scaled to 1, if non-zero,  by scaling the  boundary coordinates.
The other initial  condition to the brane motion corresponds to a  trivial shift of the initial time point.

\subsection{The Mirage Cosmology}

Once a  solution  of the cubic equation (\ref{cc3})  has been obtained,
it is possible to find the brane motions by solving the differential equation for the brane trajectory $u(\tau)$, in terms of proper time:
\begin{equation}
\frac{d u}{d \tau} =\pm \sqrt{y^2(u)-1},
\label{cc5}
\end{equation}
which can be immediately integrated to give
\be \label{cc6}
\int_{u_0}^{u(\tau)} {d u \over  \sqrt{y^2(u) -1}}  = \pm (\tau-\tau_0)
\ee
After inverting  equation (\ref{cc6}) for $u(\tau)$,  the cosmological scale factor  is then determined by
the bulk geometry plus the brane trajectory $u(\tau)$,
\be
a(\tau) = e^{A(u(\tau))},
\label{ac1}\ee
and the  effective brane Hubble parameter is given by
\be
H \equiv{ dA\over d\tau} = {dA \over du} {du\over d\tau} =
  \pm \frac{W(\varphi(u(\tau)))}{6} \sqrt{y^2(u(\tau))-1}
\label{ec4}\ee
Since the bulk scale factor $e^{A(u)}$ is monotonically decreasing
with $u$  from the UV to the IR, expansion or contraction depends on the sign of
the velocity in equation (\ref{ec4}): a brane going towards the UV describes an expanding
universe, and a brane going towards the IR describes a contracting
one, \cite{mirage}.

Integrating equation (\ref{cc5}) directly may sometimes be
impractical, because it requires using the explicit form  of the bulk
solution $\f(u)$ to write $y(u)$. Alternatively, we will often find it more
convenient to solve (\ref{cc3}) for  $y$ in terms of $\f$ (on which
$F$ and $W_B$ depend), then integrate the equation for
$\f(\tau)$.
\begin{equation}
\label{dynphi}
\frac{d \varphi}{d \tau}
=
\frac{d \varphi}{d u}
\frac{d u}{d \tau}
=
\pm
\frac{d W}{d \varphi}  \sqrt{y^2(\f)-1}.
\end{equation}
where we have used equation (\ref{c7}) to rewrite $d\f/du$ in terms of
$W$.  Then,  knowing the form of the bulk solution  $\f(u)$ and inverting
 the relation between $\f$ and $u$, will give us the trajectory
 $u(\tau)$. But this is unnecessary, if we are interested only in the scale factor
 $a(\tau)$: the latter  we can obtain directly by solving from
 $A(\f)$   the equation 
\be
{dA \over d\f} = - {6 \over W} {d W\over d\f}, 
\ee
which follows from (\ref{c7}), and substituting the solution $\f(\tau)$
of (\ref{dynphi}).

\subsection{The non-relativistic limit}

Although we have formally solved the problem in general, it is not
easy to read-off directly  from the probe-brane action (\ref{c12}) what the brane
motion will be, since the Lagrangian is highly non-canonical. Things
however simplify, and a certain degree of intuition can be gained by
looking at the action, in the limit when the brane motion is
non-relativistic, i.e. when
\be
e^{-A}|\dot{u}(t)|\ll 1.
\label{ec5}\ee
In terms of proper time $u(\tau)$, from equation (\ref{cc1}-\ref{y})
it is clear that  this condition corresponds to solutions near unity, $y
\simeq 1$.

By expanding the action (\ref{c12}) for small $\dot u^2$ we can approximate it as
\be
S_b
\label{smllvelact}
\simeq V_3\int dt~\left[-W_B~e^{4A}+{1\over 2}(W_B+2F)e^{2A}\dot
  u^2+{\cal O}(\dot u^4)\right]. 
\ee
The effective potential for slow brane motion is
\be
V_{eff}(u)=e^{4A}~W_B\;.
\label{c19}\ee
Note however, that the effective ``mass'' is $u$ dependent.
Consider the small velocity action from (\ref{c12})
\be
S_{\rm linear}= V_3\int dt~\left[-V_{eff}+{1\over 2}M(\varphi)\dot u^2\right]\sp M(\varphi)=(W_B+2F)e^{2A}
\label{c24}\ee
with $V_{eff}$ given in (\ref{c19}).
The equations of motion stemming from this action are
\be
{d\over dt}\left({1\over 2}M(\varphi)\dot u^2+V_{eff}\right)=0
\label{c25}\ee
so that
\be
{1\over 2}M(\varphi)\dot u^2+V_{eff}=E~~~\to~~~\dot u=\pm\sqrt{2(E-V_{eff})\over M}
\label{c26}\ee
This equation can be written in terms of the proper time $\tau$ giving
\be
\label{nonrel}
\frac{d u}{d \tau} = \pm \sqrt\frac{2(E-e^{4A} W_B)}{(2F+3W_B)e^{4A}-E}
\simeq
\pm \sqrt\frac{2(E-e^{4A} W_B)}{(2F+W_B)e^{4A}}
\ee
where we used the fact that in the non relativistic regime $E \simeq V_{eff}$.

We can obtain this equation also by expanding the cubic equation (\ref{cc3})
in the non-relativistic regime. It corresponds to $\frac{d u}{d \tau} \rightarrow 0$ ($y^2 \rightarrow 1^+$).
In this case the cubic equation (\ref{cc3})  further simplifies as
\begin{equation}
E  = e^{4 A}  (F(y^2-1)+W_B)
=e^{4A}  \left(F \left(\frac{d u}{d \tau}\right)^2+W_B)\right)
\label{cubicc3}
\end{equation}
Solving  for
 $\frac{d u}{d \tau}$ gives back equation (\ref{nonrel}).

\bigskip
\subsection{Self-tuning extrema in the probe limit}
\bigskip

In the previous section we derived the equations describing a generic
probe-brane trajectory. Before analyzing the resulting cosmology, it
is interesting to revisit the static self-tuning solutions discussed in
\cite{selftuning}. They correspond to bulk metrics of the form
(\ref{c1}) but different scale factors $A_{UV}(u)$, $A_{IR}(u)$ and
scalar field profiles $\varphi_{UV}(u)$, $\varphi_{IR}(u)$. Self-tuning
solutions at a fixed brane position $u_*$ are found by imposing the
junction conditions (\ref{FE5}-\ref{FE6}), which in this case simplify to
\be \label{s-t}
W_{IR}(\varphi_*) - W_{UV}(\varphi_*) = W_{B}(\varphi_*), \quad
W_{IR}'(\varphi_*) - W_{UV}'(\varphi_*) = W_{B}'(\varphi_*) ,
\ee
where $\varphi_* \equiv \varphi(u_*)$ and $W_{UV,IR}(\varphi)$ are the
superpotentials in the UV and IR, respectively. As shown in
\cite{selftuning}, these equations fix a discrete set of values for the brane position
$\varphi_0$, which in turn determine the coordinate of the brane $u_0$ once
the UV boundary conditions on $A$ and $\varphi$ are fixed.

These static solutions lead to a Minkowski induced metric on the
brane, and can be considered  as the ``vacuum states'' of the
theory. As we now show,  in the probe-brane
limit,  these solutions correspond to  extrema of the effective potential for
non-relativistic brane motion, (\ref{c19}) and therefore to static (flat) brane solutions.

A static solution in the probe brane limit  is found by setting
$\ddot{u}=\dot{u}=0$ in equation (\ref{c16}), which results in the condition
\be \label{ext1}
\de_u\Big( {e^{4A(u)} W_B(\varphi(u))}\Big)\Big|_{u=u_*} = 0
\ee
This is the same as  the extremality condition  for the effective
potential $V_{eff}(u)$ defined in (\ref{c19}), as it can be understood
intuitively in the non-relativistic Lagrangian description
(\ref{smllvelact}).

Using  equation  (\ref{c7}), the left hand side of equation (\ref{ext1})   can be rewritten as
\be
\de_u \Big( {e^{4A(u)} W_B(\varphi(u))}\Big)=  W_B'W'- {2\over 3}WW_B\;,
\label{c18}\ee
where we have used the first order equations in (\ref{c7}).
The vanishing of the right hand side  should be thought of as an equation for the equilibrium position $\varphi_*$:
\be
W_B'(\varphi_*)W'(\varphi_*)={2\over 3}W(\varphi_*)W_{B}(\varphi_*), \qquad \varphi_*
= \varphi(u_*)
\label{c20}\ee

We  now show that  this condition comes from equations (\ref{s-t})
in the probe limit. The probe approximation is effective if the
right-hand sides of those  equations above are small. We call the IR
superpotential $W(\varphi)$. This is fixed by regularity, as discussed in
section 2.2. In the probe approximation, the $W_{UV}$
solution will  be very close to $W(\varphi)$.
We can therefore write
\be
W_{IR}=W(\varphi)\sp W_{UV}=W(\varphi)-\delta W,
\label{c22}\ee
with $\delta W\ll W$. $\delta W$ satisfies the linearized perturbation of equation (\ref{c7}) namely \be
-{2\over 3}W\delta W+W'\delta W'=0
\label{c23}\ee
We may now write the Israel matching conditions in (\ref{s-t}) as
\be
\delta W(\varphi_*)=W_{B}(\varphi_*)\sp  \delta W'(\varphi_*)=W'_{B}(\varphi_*)
\label{c37}\ee
Therefore (\ref{c37}) and (\ref{c23}) are equivalent to (\ref{c20}).
We conclude that the extremality condition for equilibrium we found,
(\ref{s-t}), is  the probe limit of the Israel conditions, (\ref{s-t}).

We now discuss the stability of such an equilibrium position, by
looking at small fluctuations $u(t) = u_* + \delta u(t)$. For small
$\delta u$, we can expand the probe action (\ref{c12}) to quadratic
order,  and we obtain
\be\label{squad}
S^{(2)}[\delta u(t)] = V_3\int dt~\left[-V_{eff}''(u_*) (\delta u)^2
  +{1\over 2}M(\varphi(u_*))\left(\dot {\delta u}\right)^2\right]
\ee
where
\be\label{ext2}
M(\varphi)=(W_B+2F)e^{2A} , \sp V_{eff} = e^{4A}W_B.
\ee
This is  the same action (\ref{smllvelact}) we found in the
non-relativistic limit,  expanded to quadratic order around the extremum\footnote{Notice that
  we obtained equation (\ref{squad}) in a different approximation
   from (\ref{smllvelact}) (small fluctuations vs. small
   velocity). Therefore, although similar looking, the quadratic
   action can also describe relativistic  fluctuations.}.

Stability of the extremum at $u_*$ requires
\be\label{ext3}
{d^2 V_{eff}\over du^2} (u_*) >0, \quad M(\varphi(u_*) )\geq 0
\ee

These requirements follow  from the probe limit of
a set of  {\em sufficient} conditions for the absence of ghostlike and
tachyonic modes around a static solution, which were formulated in
\cite{selftuning} for the fully backreacted system, namely
\be\label{notach}
W_B''(\varphi_*) > W_{IR}''(\varphi_*) -W_{UV}''(\varphi_*) \quad \Rightarrow \quad \text{no tachyons}
\ee
and
\be\label{noghost}
\left\{\begin{array}{l} \left[{W_B\over W_{IR}W_{UV}} - {U_B\over
        3}\right]_{\varphi_*} > 0 \\
\\
\quad \text{and} \\
\\
 Z_B(\varphi_*)\left[{W_B\over W_{IR}W_{UV}} - {U_B\over
        3}\right]_{\varphi_*} > \left(U_B'(\varphi_*)\right)^2\end{array}\right.\quad \Rightarrow \quad \text{no ghosts}
\ee
Notice that these two conditions also require $Z_B(\varphi_*) >0$.

We first consider the condition (\ref{notach}). In the probe
limit (\ref{c22}), this becomes
\be \label{notach-pr}
W_B''(\varphi_*) > \delta W''(\varphi_*)
\ee
To compute the left hand side, we can take a derivative with respect to $\varphi$
of equation (\ref{c23}) to obtain
\be\label{notach2}
\delta W'' = \left({4\over 9} \left({W\over W'}\right)^2 - {2\over 3}
    {W W'' \over (W')^2} + {2\over3} \right)\delta W
\ee
On the other hand, the second derivative with respect to $u$ of the
effective potential is given by
\bea\label{notach3}
{d^2 V_{eff}\over du^2}  = &&\left[(W')^2 W_B'' + \left(W''W' - {2\over 3}
  W'W\right) W_B' - {2\over 3}(W')^2 W_B +\right. \nonumber \\
&& \left.-  {2\over 3}W \left(W'W_B' -
  {2\over 3} W W_B\right)\right] e^{4A}.
\eea
Evaluating the left hand side at $u_*$,  i.e. where the extremality
condition (\ref{c20}) is satisfied, this expression simplifies to
\be\label{notach4}
{d^2 V_{eff}\over du^2} =  (W'(\varphi_*))^2 \left\{W_B''(\varphi_*) -\left[
  {2\over 3} + {4\over 9}\left({W\over W'}\right)^2 - {2\over 3} {W''W
    \over (W')^2} \right]_{\varphi_*}W_B(\varphi_*) \right\}.
\ee
 Recalling that $ W_B(\varphi_*) = \delta W(\varphi_*)$ and using (\ref{notach2}), the condition
 (\ref{notach-pr}) is equivalent to the positivity of the left hand
 side of (\ref{notach4}).

We now turn to the positivity of the probe brane fluctuation
kinetic term around equilibrium,  $M(\varphi_*)>0$ in equation
(\ref{ext2}). From the definition of $F$ in equation (\ref{c13}), we have
\be \label{ghost1}
M(\varphi_*) =  e^{2A(u_*)} \left[\left(W_B- {U_B\over 3} W^2\right) + 2 W' W U_B' + Z_B( W')^2
\right]_{\varphi_*}
\ee
Now suppose both  exact no-ghost sufficient conditions (\ref{noghost}) are
satisfied. Approximating $W_{IR}W_{UV} \simeq W^2$, it follows that
\bea\label{ghost2}
M(\varphi_*) > && e^{2A(u_*)}\left[{(U'_B)^2 W^2\over Z_B} + 2 W' W U_B' + Z_B( W')^2
\right]_{\varphi_*} = \nonumber \\
&& =  {e^{2A(u_*)} \over Z_B(\varphi_*)}\Big( U_B' W +
  Z_B W'\Big)_{\varphi_*}^2 \geq 0.
\eea
Therefore the exact no-ghost conditions imply the positivity of the probe
brane fluctuations. Notice however that the converse is not true: in
the probe limit, we have only the one condition $M(\varphi_*) >0$, which
does not guarantee that {\em both} conditions (\ref{noghost}) are
satisfied. This is because  half of the full
spectrum is lost in the probe limit, roughly those corresponding to
the bulk scalar perturbations, which decouple from the brane
fluctuations.

\section{Asymptotic cosmologies\label{asym}}

In this section we provide a  discussion of the probe brane cosmology
when the brane probes the asymptotic UV and IR regions.
Namely,  we look at the cosmology on the brane, defined by the parameter $a(\tau)$, in the asymptotic regime of its motion in the five-dimensional space-time.

One such asymptotic region is the UV region, specified by values of
$u$ near $-\infty$. The other  asymptotic region is the IR. If the
bulk interior is regular (in which case it must be asymptotically
AdS),  such region is all the way down to values of $u$ near
$+\infty$. Otherwise, the interior singularity is reached at a finite
coordinate value $u_0$, and the asymptotic region is reached for $u
\simeq u_0$.  

From the point of view of the cosmology on the brane, a motion of the
brane toward the IR  corresponds to a contracting space-time, while a
motion of the brane towards the UV boundary  corresponds to an
expanding space-time. Recall that $u$ increases monotonically from the
UV to the IR. Therefore,  
what distinguishes the two cases is the sign of the velocity
$\dot{u}$: if it is positive,  the brane is moving towards the IR
(contracting cosmology);  if it is negative the brane is moving
towards the UV (expanding cosmology). 

Because in this section we restrict the analysis only to these asymptotic regions,  all the expressions will be valid only in the (leading) asymptotic limit. We indicate this with the symbol $"\simeq"$ instead of $"="$. The regime in which this approximation is valid should be clear by the context and it will be specified in each subsection.

The induced cosmology is controlled by the brane scale factor
$a(\tau)$  defined in (\ref{ac1}), with  the brane cosmic time
coordinate $\tau$  defined in (\ref{c32}). The corresponding Hubble
parameter is defined as 
\be
H = {1\over a}{d a \over d\tau}.
\ee 

We will be particularly interested in the emergence of (approximate) scaling
behavior, of the kind
$a(\tau) \sim \tau^\alpha$
with $\alpha$  a real parameter. In the present setup, this behavior can only
occur near a scaling region of the bulk solution, where $e^{A(u)}$ is
a power-law in $u$ or, in the limiting case, a simple exponential of
the form $e^{\pm u/\ell}$.  As we will see, for very general bulk
potentials 
we will find scaling solutions of the scaling type   in the asymptotic IR and UV regions discussed above.

It is convenient to parametrize a scaling solution 
by an equation of state parameter $w$,  as is
introduced in  standard FRW cosmology:
 \be
 a(\tau) \simeq a_0 \tau^{\frac{2}{3(1+w)}}
 \label{Atau}
 \ee
Then, the mirage cosmology mimics the ordinary 4d cosmology driven by
a single fluid with pressure $p$ and energy density $\rho$ related by
the equation of state $p= w \rho$. For example,   
 The case $w=0$ describes a matter dominated universe while $w=\frac{1}{3}$
  describes a radiation dominated universe. For  the case $w=-1$ the relation
  (\ref{Atau}) becomes exponential, $a  = a_0 \exp (H_0 t)$ and describes a de Sitter
  universe,  driven by a cosmological constant. 

We also introduce the deceleration parameter, 
  \be
  \label{decpar}
  q \equiv -{ a \,\, \de_\tau^2{a} \over (\de_\tau a)^2} 
  =  - 1-{1\over H^2}{d H \over d\tau} =
   \frac{1}{2}(1+3w).
   \ee
The universe is decelerating for $q>0$ and accelerating for $q<0$. From the above expression we see that the value $\omega= -
\frac{1}{3}$ (which in 4d corresponds to curvature-dominated evolution) separates the accelerating case ($w < -\frac{1}{3}$) from the decelerating
  one   ($w  > -\frac{1}{3}$).

Using the tools introduced above, in the next subsections we will  proceed to discuss the approximate solution for the brane
motion, and the resulting cosmology, in  asymptotic regions of the
bulk space-time.  In appendix \ref{appdet} we derive all the relevant calculations that are not explicitly reported in this section.

\subsection{The near-AdS case}

We start in this sub-section with the study the dynamics for the case of a near-AdS bulk superpotential.
The form of $W$ here is approximated close to the fixed point
(see  (\ref{UV5}), (\ref{UV6}) and (\ref{IR3}) and the discussion
in section \ref{subsec:VacSol} for details)
as
 \begin{equation}
W_{UV}(\varphi ) =
 \frac{6}{ \ell} +  \frac{\Delta_{\pm}  }{2 \ell}\varphi ^2
+\mathcal{O}(\varphi^3),
\quad
W_{IR}(\varphi ) =
 \frac{6}{ \ell} +  \frac{\Delta_{-}  }{2 \ell}
 (\varphi-\overline\varphi)^2
+\mathcal{O}((\varphi - \overline\varphi)^3)
\label{ec6}\end{equation}
$W_{UV}$ describes the near-AdS boundary  near $\f=0$.
$W_{IR}$ describes the IR AdS asymptotics, in the neighbourhood of $\f=\bar \f$.
Note that the AdS length $\ell$ is different in the UV and the IR expansions but we will not keep track of this.
We have kept the leading quadratic, non-constant contributions\footnote{In the UV we allowed also the plus solution related to $\Delta_+$ for the case the bulk theory is driven by a scalar vev rather than a scalar source.}, although they do not contribute to leading order to several of the formulae below, as they are useful for estimating the asymptotics in appendix \ref{appd1} .

Close to a fixed point, we  can take the brane potentials to be
approximately constant (i.e.  approximately equal to either their UV or IR fixed-point
values), and a convenient parametrization is:
\begin{equation}
W_B \simeq \frac{h_W}{\ell^2}, \quad
U_B \simeq \frac{h_U}{6}, \quad
Z_B \simeq h_Z
\label{ec7}\end{equation}
where $h_W$, $h_U$ and $h_Z$ are  constants with the dimension of
inverse mass. The constants $h_U$ and $h_V$ must be positive,  for the interactions to have good properties on the brane \cite{selftuning}. On the other hand, $h_W$ can have either sign.
Small, smooth variations of the brane potentials around the fixed
point values  do not change the leading scaling solutions of the brane
cosmology, and will be neglected in what follows.


In either asymptotic AdS region described by superpotentials (\ref{ec6}) and
(\ref{ec7}),  the function $F$ defined in (\ref{c13}) becomes
\be
F_{UV} \simeq \frac{\Delta_\pm^2 \varphi ^2 h_Z}{2 \ell^2}-\frac{\left(\Delta_\pm  \varphi ^2+12\right)^2 h_U}{144 \ell^2}
\label{ec8}
\ee
\be
F_{IR} \simeq \frac{\Delta_-^2 (\varphi -\bar \varphi )^2 h_Z}{2 \ell^2}-\frac{(\Delta_-  (\varphi -\bar \varphi )^2+12)^2 h_U}{144 \ell^2}
\label{ec9}\ee
The equations
\be
A'=-\frac{W}{6}\sp \partial_\varphi W=\varphi'
\ee
  are solved in the UV and in the IR respectively by
\begin{equation}
\label{phieq}
A_{UV} \simeq A_0 -\frac{1}{\Delta_\pm} \log |\varphi| - \frac{\varphi^2}{24}
, \quad
\varphi_{UV} \simeq \varphi_\pm e^{\frac{\Delta_\pm}{\ell}u}
, \quad
u \rightarrow -  \infty
\end{equation}

\begin{equation}
\label{phieq2}
A_{IR} \simeq A_0 -\frac{1}{\Delta_-} \log |\varphi-\bar \varphi| - \frac{ (\varphi-\bar \varphi)^2}{24}
, \quad
\varphi_{IR} \simeq \bar \varphi +\varphi_- e^{\frac{\Delta_-}{\ell}u}
, \quad
u \rightarrow + \infty
\end{equation}
where the integration constants have  been fixed as $A_0$ and $\varphi_{\pm}$.
The cubic equation  in the UV is
\be
\label{cubicUV}
E-\frac{ y e^{4A_0}
\left(h_W+\frac{1}{2} \Delta _\pm^2 \left(y^2-1\right) \varphi ^2 h_Z-\left(y^2-1\right) \left(\frac{\Delta _\pm \varphi ^2}{12}+1\right)^2 h_U\right)}{\ell^2 |\varphi|^{\frac{4}{\Delta _\pm}}} \simeq 0
\ee
while in the IR  it is
\be
\label{cubicIR}
E-\frac{y e^{4A_0}
\left(h_W+\frac{1}{2} \Delta _-^2 \left(y^2-1\right)
(\varphi -\bar \varphi) ^2 h_Z-\left(y^2-1\right) \left(\frac{\Delta _- (\varphi -\bar \varphi)^2}{12}+1\right)^2 h_U\right)}{\ell^2 |\varphi -\bar \varphi|^{\frac{4}{\Delta _-}}} \simeq 0
\ee
These equations can be further  approximated
by expanding $\varphi$ close to the fixed point.
In the UV we expand  $|\varphi |\simeq \delta \varphi$
while in the IR we expand  $|\varphi - \varphi| \simeq \delta \varphi$, where in both cases we consider $\delta \varphi \rightarrow 0^+$.
In this limit the cubic equation simplifies to
\be
\label{cubicUVIR}
E-\frac{y e^{4A_0}
\left(h_W-\left(y^2-1\right)  h_U\right)}{\ell^2 \delta \varphi ^{\frac{4}{\Delta _\pm}}} \simeq 0
\ee
in both the UV and in the IR case.
In the following we study the IR and the UV cases separately.

%
%
\subsubsection{The motion near the UV\label{mUV}}
%
%
In the UV, 
at small $\varphi$, the solution of (\ref{cubicUVIR})
can be approximated as
\begin{equation}
\label{xUVAdS}
y \simeq \sqrt{\frac{h_W}{h_U}+1}
\end{equation}
The derivation is detailed  in appendix \ref{appd1}, but it can be
simply understood by noticing that in the UV both $\Delta_\pm$ are
positive, and to satisfy equation (\ref{cubicUVIR}) as $\delta\f \to
0$ the numerator  of the second term on the left hand side
must necessarily  vanish in this limit. This leads directly to the
solution (\ref{xUVAdS}), the other possible solution $y=0$ being
unphysical because of the constraint $y >1$.    

Then, for $h_U \ll h_W$
 the solution is in the ultra-relativistic regime $y \rightarrow \infty$,
while for $0<h_W \ll h_U$ the solution is in the non-relativistic
regime $y \rightarrow 1^+$. In both cases we need to require that $h_W>0$.
If $h_W<0$ there is no solution to the cubic equation, which implies
that the brane is in a classically forbidden region. $h_W<0$  is
equivalent to having  a negative cosmological constant on the brane,
which indicates  that the brane universe cannot continue to expand, as
it will be doing in this regime otherwise.

Assuming $h_W,h_U>0$,  with $y$ approximately constant and given in
equation (\ref{xUVAdS}), we can immediately integrate the equation for
the trajectory, (\ref{cc5}), to find 
\begin{equation}
u(\tau) = \eta (\tau-\tau_0) \sqrt{\frac{h_W}{h_U}}
\label{ec11}\end{equation}
where $\eta = \pm 1$ is the same sign appearing in equation
(\ref{cc5}). As discussed at the beginning of this section,
the space-time on the brane is expanding for $\eta=-1$ and contracting
for $\eta=+1$. 

In the asymptotic UV region the bulk warp factor is approximately
$A(u) \simeq - {u\over \ell}$, therefore   the cosmological scale
factor induced on the brane is 
\be \label{ec111} 
a(\tau) \simeq a_0 e^{-\eta \tau \sqrt{{h_W\over h_U} }}
\ee
where $a_0$ is set by initial conditions. Equation (\ref{ec111}) tells
us that  the
space-time on  brane is approximately de Sitter.  The associated Hubble scale of the de Sitter expansion/contraction is given by
\be \label{ec10-i}
H_{eff}\equiv {1\over a}{d  a \over d\tau}={1\over \ell}\sqrt{\frac{h_W}{h_U}}
\ee
(where  $\ell$ and $h_{W,U}$ are the constants defined in the UV
region).





Remarkably,  the brane geometry is an approximate dS universe, with the  Hubble
parameter (\ref{ec10-i})  determined by the
values of the   brane Planck scale and cosmological constant close to
the UV fixed point, which are controlled by $h_W$ and $h_U$
respectively. Indeed, from the brane action (\ref{c2})  and from the
definitions (\ref{ec7}) close to the fixed point, we may identify the
effective four-dimensional Planck scale $M_4$ and cosmological
constant $\Lambda_{eff}$  by writing equation (\ref{c2})  in the
standard four-dimensional form (after setting $\f$ to its UV value),
\be \label{4d}
S_b = {M^2_4 \over2} \int (R - 2\Lambda_{eff}), \qquad M_4^2 = M^3 {h_U
  \over 3}, \quad M_4^2 \Lambda_{eff} =  M^3 {h_W \over \ell^2}.
\ee
Then, the Hubble parameter (\ref{ec10-i}) can be rewritten:
\be \label{ec12-i}
H_{eff}^2 = {\Lambda_{eff} \over 3}
\ee
which is the same one would obtain from the purely 4d action  (\ref{4d}).

The results (\ref{ec111}-\ref{ec10-i})  also hold when the brane spends time
close to an intermediate fixed point which is neither the IR nor the
ultimate UV, as we discuss more at length in the final section of
this paper.

\subsubsection{IR case}
%
%

In the IR,  the solution of the cubic equation
(\ref{cubicUVIR})
can be approximated  as
\begin{equation}
\label{AdSIR}
y \simeq
\left( \frac{E \ell^2 \delta \varphi ^{4/\Delta_- }}{ e^{4A_0}  h_U} \right)^{1/3}
\end{equation}
(see  appendix \ref{appd1}). In this case, $\Delta_-<0$ and  $y$ diverges as
$\delta\f \to 0$. In particular,  the non-relativistic
regime (which requires $y \simeq 1$ cannot be reached,  and the solution (\ref{AdSIR})  of equation (\ref{cubicUVIR}) is necessarily in the ultra relativistic regime.

To find the brane trajectory, we first solve  equation (\ref{dynphi})
for the deviation $\delta\f(\tau)$
  of the scalar field
 from the IR value. 
Using the IR expansion for $W$ from equation (\ref{ec6}), as well as
(\ref{AdSIR}),  one arrives at
\begin{equation}
\varphi(\tau) \simeq \bar \varphi+
(\mu |\tau -\tau _0|)^{-\frac{3 \Delta_- }{4}}   
\label{ec13}\end{equation}
where $\mu$ is a constant with dimension of mass given explicitly by
$$
\mu \equiv \left(\frac{8 E}{27  e^{4A_0} \ell h_U}\right)^{1\over 3}. 
$$ 
Equation (\ref{phieq2}) then leads  to
\begin{equation}
a(\tau) = e^{A(\tau)} \simeq
e^{-\frac{1}{\Delta_-}\log|\varphi- \bar \varphi|} =a_0 |\tau -\tau _0|{}^{3/4} 
\label{ec14}\end{equation}
The brane cosmology here is controlled by a parameter $w$ defined
in (\ref{Atau}), that is here  $w  = -\frac{1}{9}$, a value between the domination of the
matter and the curvature cosmology.
Moreover $w > -\frac{1}{3}$ signaling that
we are in a decelerated case.

From the asymptotic behavior of $\f$ in equation (\ref{phieq2}) we can
read-off the brane trajectory close to the singularity, 
\begin{equation}
u(\tau) \simeq - {3 \ell \over 4} \log (\mu |\tau-\tau_0|)
\label{ec15}\end{equation}
and as discussed at the beginning of this section,
the space-time on the brane is contracting if $\tau<\tau_0$
and expanding if $\tau>\tau_0$.
This signals the fact that from the brane perspective $u_0 \equiv u(\tau_0)$ corresponds to a big bang or big crunch singularity respectively.

\subsection{Exponential bulk superpotential}

We now turn to the second alternative where the infrared is signalled by a bulk singularity, and
the scalar field runs to infinity in the interior of bulk space. 
This case differs from the one discussed in the previous section (IR
AdS)  because here there are no fixed points at finite $\varphi$.
We assume the potential is  dominated by an exponential (see
(\ref{IR4})) and the only acceptable solution of the superpotential
equation  is then  given by   (\ref{IR5}), where the requirement of regularity corresponds to (\ref{gubser}).

Thus, in  this subsection we study the probe cosmology near the IR
with a  bulk superpotential with exponential asymptotics in the large
$\f$ region: 
\begin{equation}
W  \simeq W_\infty  e^{\kappa \varphi} \qquad  \f \to +\infty
\label{ec16}\end{equation}
with $W_\infty$ a positive real constant, and $\kappa$ a real constant that
needs to satisfy the Gubser bound (\ref{gubser}), $\kappa <
\sqrt{2/3}$ in order for the singularity to be qualified as a "good" singularity. As in the previous section,  the expression (\ref{ec16}) for the superpotential is approximate and valid only asymptotically for $\varphi \rightarrow \infty$.

We have also studied
 the power-law superpotential $W  \simeq  \varphi^p$.
As expected we find a result between  between AdS and the limit $\kappa=0$.
The details are in appendix \ref{PL}.

It is convenient to define the new dimensionless variable:
\begin{equation}
\epsilon(\tau) \equiv  W_\infty \kappa^2 (u_0 - u(\tau)),
\label{ec17}\end{equation}
with $u_0$ a positive real constant corresponding to the  location
of the bulk (good) IR singularity. Hence $\epsilon$
corresponds to the coordinate  distance between the position of the brane and the position of the IR singularity.
In all of this section,  we only focus on the asymptotic region near $u_0$, hence all the expressions will be valid only in the limit $\epsilon \rightarrow 0 $, corresponding to $\varphi \rightarrow \infty$.
By solving the equations $A'=-\frac{W}{6}$ and $\partial_\varphi W=\varphi'$, in terms of this new variable, we obtain
the approximate solutions, valid in the regime $\epsilon \ll 1$,
\begin{equation}
\varphi  \simeq-\frac{\log (\epsilon )}{\kappa }, 
\qquad
e^A \simeq e^ {A_0} \epsilon^{\frac{1}{6 \kappa ^2}}
\label{ec18}\end{equation}
where $A_0$ is an integration constant.
The equation of motion for $\epsilon(\tau)$, which follows from
(\ref{cc5}), is 
\begin{equation}
\label{differential}
\frac{d \epsilon}{d \tau}
=
\frac{d \epsilon}{d u}
\frac{d u}{d \tau}
  \simeq \eta \; W_\infty  \; \kappa^2 \sqrt{y^2-1}
\end{equation}
with $ \eta = \pm $ corresponding to a motion of the brane towards the IR singularity ("+"), and hence a contracting space-time on the brane, or outgoing from the IR singularity ("-"), and hence an expanding space-time on the brane.
We need to solve the cubic equation (\ref{cc3}) in order to obtain $y$ and substitute  it in equation (\ref{differential}).

We assume that each of the brane potentials is dominated, for large
$\varphi$ by a single exponential, and we parametrize them as
\begin{equation}
W_B  \simeq h_W e^{\gamma_W \varphi}, \quad
U_B  \simeq h_U e^{\gamma_U \varphi}, \quad
Z_B \simeq h_Z e^{\gamma_Z \varphi} \qquad \f \to \infty.
\label{ec19}\end{equation}
with $\gamma_i$ real constants, $h_W$ has dimension of mass, and
$h_{U,Z}$ have dimension of length. Such potentials are only approximate expressions valid in the limit $\varphi  \rightarrow  \infty$, near the position of the IR singularity.
We also require $\gamma_i < \kappa$ to ensure the validity of the
probe brane approximation.
The parametrization of the IR behavior  (\ref{ec19})  is very
general, and possible deviations from this behavior (e.g. exponentials
times power-laws) will only affect the details of the solution at
subleading orders, except for very specific critical values of the
constants $\gamma_i$. It includes the case of constant potentials, in the limit
of vanishing $\gamma_i$.

Substituting  equations (\ref{ec19})  into  (\ref{c13}) we obtain:
\begin{equation}
\label{EG}
F \simeq \left(\frac{W_\infty}{\epsilon}\right)^2
\left(
\frac{h_U}{\epsilon^{\frac{\gamma _U}{\kappa }}}\left( \kappa   \gamma _U - \frac{1}{6}\right)
+
\frac{\kappa ^2 h_Z}{2 \epsilon ^{\frac{\gamma _Z}{\kappa}}}  \right)
\end{equation}
Inserting the above expression for $F$ into  the cubic equation
(\ref{cc3}), the latter becomes approximately
\be
\label{cubicFNG}
E \simeq e^{4 A_0} y \epsilon ^{\frac{2}{3 \kappa ^2}-2} \left(\frac{1}{6} W_\infty^2 \left(y^2-1\right) \left(\frac{h_U \left(6 \kappa  \gamma _U-1\right)}{\epsilon ^{\frac{\gamma _U}{\kappa }}}+\frac{3 \kappa ^2 h_Z}{\epsilon ^{\frac{\gamma _Z}{\kappa }}}\right)+\frac{\epsilon ^2 h_W}{\epsilon ^{\frac{\gamma _W}{\kappa }}}\right)\simeq
\ee
$$
\simeq
e^{4 A_0} y \epsilon ^{\frac{2}{3 \kappa ^2}-2} \left(\frac{1}{6} W_\infty^2 \left(y^2-1\right) \left(\frac{h_U \left(6 \kappa  \gamma _U-1\right)}{\epsilon ^{\frac{\gamma _U}{\kappa }}}+\frac{3 \kappa ^2 h_Z}{\epsilon ^{\frac{\gamma _Z}{\kappa }}}\right)\right)
\label{ec20}
$$
where in the second line of (\ref{cubicFNG})
we expanded in small $\epsilon$.
Observe that the inequalities
\be
\frac{\gamma_W}{ \kappa}<1\sp
\gamma_Z>0\sp \gamma_U>0
\ee
imply that
\be
\frac{\gamma_U}{\kappa} > \frac{\gamma_W}{\kappa} -2
\quad
\text{and}
\quad
\frac{\gamma_Z}{\kappa} > \frac{\gamma_W}{\kappa} -2
\label{ec21}\ee
As a consequence of these inequalities,  the last term in the first line
of (\ref{cubicFNG}) is always negligible in the limit of small  $\epsilon$.

There are two relevant regimes, depending on the
relative values of $\gamma_U$ and $\gamma_Z$, each one  leading  to a further
simplification of equation (\ref{ec20}) : 
\be
\label{EG1}
\text{If} \,\,
\gamma_U < \gamma_Z
\quad
 \rightarrow
\quad
E\simeq \frac{1}{2} W_\infty^2 e^{4 A_0} \kappa ^2 y \left(y^2-1\right) h_Z \epsilon ^{\frac{2}{3 \kappa ^2}-\frac{\gamma _Z}{\kappa }-2} + \dots
\ee
\be
\label{FG}
\text{If} \,\,
\gamma_U > \gamma_Z
\quad
\rightarrow
\quad
E \simeq \frac{1}{6} W_\infty^2 e^{4 A_0} y \left(y^2-1\right) h_U \left(6 \kappa  \gamma _U-1\right) \epsilon ^{\frac{2}{3 \kappa ^2}-\frac{\gamma _U}{\kappa }-2} +
\dots
\ee
The ellipsis in the above equations refers  to higher orders in $\epsilon$.
The regime $\gamma_U =\gamma_Z$
 can be studied in a similar manner.

If both exponents of $\epsilon$ in (\ref{EG1}) and (\ref{FG})
 are positive,  we are in the ultra-relativistic regime as $\epsilon \rightarrow 0$.
This is because in this case the solution can only exist
for large values of $y$.
On the other hand, if at least one of the exponents of $\epsilon$ in (\ref{EG1}) and (\ref{FG})  is negative,
the cubic equation can
be solved for finite $y$.
We  observe that in this case it is possible to find a
solution in the non-relativistic regime, that corresponds to $y\rightarrow 1^+$.
We shall study the ultra-relativistic and the non-relativistic
regimes of the solutions separately.

\subsubsection{The ultra-relativistic regime}

As discussed above, the ultra-relativistic regime at small $\epsilon$ requires that both the exponents of $\epsilon$ in the second line of (\ref{cubicFNG}) are positive.

\begin{equation}
\label{relns2}
\frac{2}{3 \kappa^2}-\frac{\gamma_Z}{\kappa}-2>0
\quad
{\rm and}
\quad
\frac{2}{3 \kappa^2}-\frac{\gamma_U}{\kappa}-2>0
\end{equation}
In the following, we define the quantity $\gamma$
as
\begin{equation}
\label{gamma}
\gamma \equiv \max\left( \gamma_U,\gamma_{Z}\right)
\end{equation}
This will allow us to express,
at least at the qualitative level,
the various results,  in terms of a single parameter
$\gamma$.
Observe that in this case  (\ref{relns2}) together with $\gamma_U,\gamma_Z >0$
implies
\be
\label{ckappa}
\kappa < \sqrt \frac{1}{3}
\ee
The full solutions of the cubic equations and of the
asymptotic cosmology are given in appendix \ref{appIRexpUR}.
Here we provide the qualitative behavior.

The solution of the cubic equation behaves as
\be
y \propto \frac{1}{\epsilon^{\frac{1}{3} \left(\frac{2}{3 \kappa^2}-\frac{\gamma}{\kappa}-2 \right)}}
\label{ec23}\ee
The equation for the brane dynamics in term of the variable $\epsilon(\tau)$ behaves as
\be
\frac{d \epsilon}{d \tau} \propto \pm
 \epsilon ^{\frac{1}{3} \left(2-\frac{2}{3 \kappa ^2}+\frac{\gamma }{\kappa }\right)}
 \label{ec24}\ee
Solving for $\epsilon(\tau)$ and inserting the solution in equation
(\ref{ec18}) we obtain the cosmological scale factor  $a(\tau)  = e^{A(\tau)}$:
\be
a(\tau) \propto
|\tau -\tau _0|^\frac{3}{6 \kappa ^2-6 \kappa  \gamma +4}
\label{ec25}\ee
The Hubble parameter $H$ is therefore
\begin{equation}
H=
\frac{3}{2 \tau (3 \kappa (\kappa-\gamma)+2)}, 
\qquad
\frac{\dot H}{H^2}
 =
2 \kappa  (\gamma-\kappa)-\frac{4}{3}, \qquad - 2 < \frac{\dot H}{H^2} < - \frac{4}{3}
\label{ec26}\end{equation}
The equation of state parameter $w$ (defined
in (\ref{Atau})) is given by
\begin{equation}
w =-\frac{1}{9} +\frac{4}{3} \kappa (\kappa-\gamma)\sp -\frac{1}{9} < w < \frac{1}{3}
\label{ec27} \end{equation}
  This value is between the
radiation value ($w = \frac{1}{3}$) and the curvature dominated cosmology
 ($w = -\frac{1}{3}$). The matter dominated case, $w=0$ is included in this region
 and it corresponds to $\gamma = \kappa- \frac{1}{12 \kappa}$.
The deceleration parameter defined in  (\ref{decpar}) in this case is
\begin{equation}
\label{decexp}
q = \frac{1}{3} + 2 \kappa (\kappa-\gamma)> 0
\end{equation}
implying a decelerated expansion/contraction\footnote{To avoid
  semantic confusion, we make this point more explicit: if $\ddot{a}<0$, then
  the  expansion speed decreases but the contraction speed {\em
    increases} in absolute value, i.e. the approach to the big crunch
  is faster and faster.}
The brane trajectory in this case is given by
 \begin{equation}
u(\tau) -u_0 = -\frac{\epsilon(\tau)}{a_1 \kappa^2} \propto -\left(\pm(\tau -\tau _0)\right)^{\frac{9 \kappa ^2}{3 \kappa ^2-3 \gamma  \kappa +2}}
 \label{ec28}  \end{equation}
 Then $\partial_\tau u(\tau) > 0 $ for $\tau<\tau_0$, corresponding to
 the fact that the space-time on the brane is contacting
  while  $\de_\tau u(\tau) < 0 $ for $\tau>\tau_0$,
  corresponding to
 the fact that the space-time on the brane is expanding.

\subsubsection{The  non-relativistic regime}

The non-relativistic regime corresponds to a solution
of  (\ref{cubicFNG}) that is near one:  $y \rightarrow 1^+$.
In the regime where $\e\to 0$ and if
one of the following inequalities hold,
\begin{equation}
\label{relns1}
\frac{2}{3 \kappa^2}-\frac{\gamma_Z}{\kappa}-2<0
\quad
\text{or}
\quad
\frac{2}{3 \kappa^2}-\frac{\gamma_U}{\kappa}-2<0
\end{equation}
we deduce from (\ref{cubicFNG}) that $y\to 1^+$ and we are therefore in the non-relativistic regime.

Then we study the cubic  equation by expanding $y$ in  (\ref{cubicFNG}) as $y = 1 + \delta_x$,  for small $\delta_x$.
While the details of the derivation are in appendix \ref{appIRexpNR}, here we just present the relevant formulae.

We simplify the exposition by  ignoring the various constants and by considering the single exponent $\gamma$
defined in (\ref{gamma}). It is enough to consider a single exponent dominating in (\ref{cubicFNG}) (say $\gamma_U$ or $\gamma_Z$) as the alternative case is qualitatively similar.
At lowest order, the equation is solved by
\be
y \equiv 1+ \delta_y
\quad
\text{with}
\quad
\delta_y \propto
\epsilon^{2-\frac{2}{3 \kappa ^2}+\frac{\gamma }{\kappa }}
\label{ec29}\ee
and the expansion is consistent only if $\delta_y \rightarrow 0^+$,
i.e. if the exponent of $\epsilon$ in (\ref{ec29}) is positive, in
agreement with (\ref{relns1}).  The combined requirements
\be
\frac{2}{3 \kappa^2}-\frac{\gamma}{\kappa}-2<0\;\;~~~ {\rm and}~~~\;\;
\gamma < \kappa, 
\ee
together with the Gubser bound (\ref{gubser}), imply
\be
\label{constr22}
\kappa > \frac{\sqrt 2}{3}
\ee
If this relation is not satisfied then there is no non-relativistic regime.

The solution $\epsilon(\tau)$ of equation (\ref{differential}) is
\be
\epsilon(\tau)
\propto
|\tau -\tau _0|^\frac{6 \kappa ^2}{2-3 \kappa  \gamma }
\label{ec35}\ee

From equation  (\ref{ec18}) we obtain the cosmological scale factor on
the brane, 
\be
a(\tau) \equiv e^{A(\epsilon(\tau))} \propto
|\tau -\tau _0|^\frac{1}{2-3 \kappa  \gamma }
\label{ec31}\ee
as well as   the Hubble  parameter $H$ and its derivative, 
\begin{equation}
H=
\frac{1}{\tau  (2-3 \gamma  \kappa )}, 
\qquad
\frac{\dot H}{H^2} =3 \gamma  \kappa -2, \qquad -\frac{4}{3}<\frac{\dot H}{H^2} < 0. 
\label{ec32}\end{equation}
The equation of  state parameter $w$ defined
in (\ref{Atau}) is given here by
\begin{equation}
w = \frac{1}{3}-2 \kappa  \gamma\sp
-1<
w <\frac{1}{9}
\label{ec33} \end{equation}
This range includes a matter-like equation ($w=0$, corresponding to
 $\gamma = \frac{1}{6 \kappa}$), but not radiation ($w=1/3$).

We evaluate the deceleration parameter to obtain
\be
\label{decnrel}
 q = 1-3 \kappa \gamma
\ee
This is positive in the region
\be
\label{regg}
\frac{2}{3 \kappa} - 2 \kappa
< \gamma< \frac{1}{3 \kappa}
\ee
In this regime the contraction and the expansion are decelerated.
Observe that (\ref{regg}) is a non empty region,
compatible with the constraint
(\ref{constr22}), because
\be
\frac{2}{3 \kappa} - 2 \kappa
<\frac{1}{3 \kappa}
\quad
\rightarrow
\quad
k^2>\frac{1}{6}
\label{ec34}\ee
On the other hand in the region
\be
\label{regg2}
 \frac{1}{3 \kappa}
<
\gamma
< \kappa
\ee
the deceleration parameter in (\ref{decnrel})  is negative  and
the contraction and the expansion are accelerated.

We conclude this analysis by summarizing the results that we have just obtained
in the  case of an exponential bulk superpotential, both in the ultra-relativistic and in the
non-relativistic regime.
\begin{itemize}
\item {\bf Ultra-relativistic regime}:
this regime is allowed if both
\be
\frac{2}{3 \kappa^2}-\frac{\gamma_Z}{\kappa}-2>0~~~
{\rm and}~~~
\frac{2}{3 \kappa^2}-\frac{\gamma_U}{\kappa}-2>0
\ee
This requires in particular that  $0<\kappa < \sqrt \frac{1}{3}$.
The  cosmology is determined by the exponent
$w=\frac{4}{3}\kappa(\kappa-\gamma) - \frac{1}{9}$ where $\gamma$ is defined in (\ref{gamma}).

The deceleration parameter $q = \frac{1}{3} + 2 \kappa (\kappa-\gamma)$ is always positive.
The (good) singularity is  at $u_0 = u(\tau_0)$.

\item {\bf Non-relativistic regime}:
this regime is allowed if either
$\frac{2}{3 \kappa^2}-\frac{\gamma_Z}{\kappa}-2$
or
$\frac{2}{3 \kappa^2}-\frac{\gamma_U}{\kappa}-2$
are negative.
Furthermore, we have to require $ \frac{\sqrt{2}}{3}<\kappa < \sqrt \frac{2}{3}$.
The  cosmology is determined by the exponent
$w=\frac{1}{3}-2 \kappa \gamma$ where $\gamma$
is as in (\ref{gamma}).
The deceleration parameter $q=1-3 \kappa \gamma$ is  positive
if $\frac{2}{3 \kappa} - 2 \kappa
< \gamma< \frac{1}{3 \kappa}$, while it is negative if
$\frac{1}{3 \kappa}<\gamma< \kappa$.

\end{itemize}
In the two extreme regimes described above the analysis is
particularly simple. However they do not exhaust all the
possibilities,  since intermediate  situations are also allowed. 
 
\subsection{Brane cosmology in  scaling regions:  a summary}

In this subsection we quickly summarize the cosmology that we have found in this section exploring the probe brane dynamics.
The table below describes the various cosmologies, by displaying the functional behaviour of $a(\tau)$, whether the cosmology found is expanding or contracting, and whether the cosmology found is accelerating or decelerating.

\begin{center}
\begin{tabular}{|c|c|c|c|}
\hline
  & $ a(\tau)$& w & q 
  \\
  \hline
\begin{tabular}{c}
UV AdS
\end{tabular}
 &
 $ \sim e^{ - \frac{|\tau -\tau _0|}{\ell}\sqrt{h_W\over h_U}}$
  &
$ -1 $
&
 $<0$ 
 \\
\hline
\begin{tabular}{c}
IR AdS \\
Ultra-relativistic
\end{tabular}
 & $\sim |\tau -\tau _0|{}^{3/4} $ & $-{1\over 9}$ & $>0$ 
 \\
\hline
\begin{tabular}{c}
IR Exponential\\
Ultra-relativistic \\
$0 < \kappa < \sqrt \frac{1}{3}$
\end{tabular}
&$\sim |\tau -\tau _0|^\frac{3}{6 \kappa ^2-6 \kappa  \gamma +4}$&
$ -{1\over 9} +{4\over 3} \kappa(\kappa-\gamma) \in \left(-{1\over 9},
  {1\over 3}\right) $ & 
$>0$
\\
\hline
\begin{tabular}{c}
IR Exponential \\
Non-relativistic \\
$ \frac{\sqrt 2}{3} < \kappa < \sqrt \frac{2}{3}$
\end{tabular}
&
$\sim |\tau -\tau _0|^\frac{1}{2-3 \kappa  \gamma }$
&$ {1\over 3} - 2\gamma\kappa \in \left(-1,{1\over 9}\right)$ & 
\begin{tabular}{c}
$<0$ if $  \frac{2}{3 \kappa} - 2 \kappa
< \gamma< \frac{1}{3 \kappa}$ \\
\hline
$>0$ if $
 \frac{1}{3 \kappa}< \gamma < \kappa
  $
\end{tabular}
\\
\hline
\end{tabular}
\end{center}

\section{Brane cosmological evolution}

In the previous section we have analyzed the probe brane motion and
the corresponding cosmological evolution in the asymptotic regions (UV
and IR) of the bulk geometry. We are now ready to give a qualitative
picture of the possible histories that the brane universe
can follow.

\begin{figure}[h]
\begin{center}
\includegraphics[width=15cm]{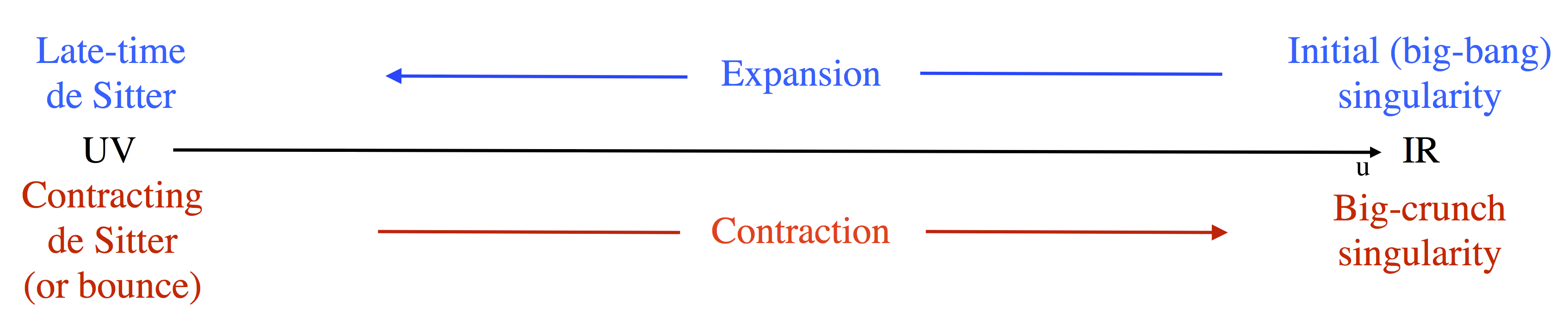}
\caption{Mirage cosmological evolution on the brane. The black line
  denotes the bulk radial direction $u$, from the AdS boundary (UV) to the deep
interior (IR). The upper blue line and lower red line indicate the  two
possible directions of the probe brane motion along the radial
direction, and correspond to an
expanding and contracting universe, respectively. The deep IR  is
perceived on the brane as a big bang or big crunch singularity,
regardless of the nature of the bulk interior (IR singularity or
Poincaré horizon).}  \label{fig-history}
\end{center}
\end{figure}

First, we collect  a few general properties which emerge from the
previous sections.
\begin{enumerate}

\item A brane moving towards the UV=AdS-like boundary undergoes cosmological expansion,
  while a brane moving towards the IR corresponds to a contracting
  universe.

\item For a brane emerging from the deep IR, the origin of AdS or the (good) IR singularity
  maps to the big-bang (initial) singularity\footnote{Note that if the
    bulk is asymptotically AdS in this regime, this is only an
    apparent 4d singularity, and it appears because of the brane embedding and the coordinate system chosen.} for the brane observer. The
  expansion is in a scaling regime, with an effective equation of
  state  which may result in either  deceleration or
  acceleration, depending on the large-$\f$ behavior of the brane and
  bulk potentials.

\item As the brane moves towards an AdS boundary, the brane evolution
  approaches a de Sitter expansion, whose parameters are set by the
  induced data (brane cosmological constant and Planck scale). If the
  brane reaches  this region, then  backreaction is automatically
  negligible and the probe approximation becomes accurate
  independently of the model details. 
  The UV fixed point can be the near AdS boundary, or can be an
  intermediate, quasi-fixed point, as it happens in theories with
  intermediate walking behavior, \cite{matti,afim}. 

\end{enumerate}
These general features are summarized in Figure \ref{fig-history}\\

The details of the cosmological history  depends of course on the
features of the bulk geometry and the initial conditions. However, using  the
above building blocks, we can sketch  a few general possible
histories which can emerge, independently of many of these details. In
doing this, we have to remember that we have neglected the effect of
brane matter (as well as  bulk backreaction). At some point  along the evolution of the brane-universe the brane energy densities 
must dominate, if we want to describe a universe close to our own.
In the ``normal'' phase of cosmology, four-dimensional matter accurately
accounts for observation. We have also assumed that the bulk dual QFT is living on Minkowski space. Other options, \cite{dS}, change the brane-world cosmology.

Therefore,  we may use
five-dimensional effects for triggering ``exotic'' behavior, but we have to make
sure there is room for normal textbook brane-energy driven FRW evolution along the way.

\paragraph*{From an IR Big Bang to a UV de Sitter.}

The simplest possible probe brane history is one which starts in a big
bang in the IR, with the brane emerging from the IR fixed-point or
singularity, reaches the UV near-boundary region,  and there it undergoes a late de Sitter expansion. 
Note, that if there is an IR standard fixed-point, and the
geometry there is that of AdS, then the big-bang singularity on the
brane is an ``apparent singularity". The singularity is  due to the embedding and
the fact that there is a Poincar\'e horizon at that point in the bulk. Such brane
singularities were analyzed  in \cite{su}, \cite{mirage} and found to
be coordinate/embedding singularities.

The later de Sitter phase may look
appealing at first sight for realising either  an  early-time inflationary phase
(which would erase all memory of the IR big bang) or late-time
acceleration after an intermediate cosmological phase driven by brane
matter. However this picture  is too simplistic, as we discuss below.

First, one may be tempted to identify  the asymptotic UV de Sitter phase with the
current late-time accelerating phase. Recall that the
induced de Sitter scale $H_{dS}$ is, by equation (\ref{ec12-i})
\be \label{his1}
H_{eff}^2 \sim  \Lambda_{eff},
\ee
where $\Lambda_{eff}$  is  set by the values of the  brane
cosmological constant and induced Planck scale in the UV as can be
seen in equation (\ref{4d}).

 Naively one may think that the
 brane cosmological constant  is set by the brane vacuum energy $\Lambda^4$
 (including the SM loop contributions), in which case the value of
 $\Lambda_{eff}$ would simply be $\Lambda^4/M_4^2$, and we would get
 the old cosmological constant problem back.  However, things may be
 subtler. First, it is not obvious what the value of the brane
 cosmological term $W_B$ should be in the far UV, because this
 crucially depends on how the dilaton couples to the brane fields.
Second,  the brane-world description is supposed to be valid
 only up to a UV cut-off, the scale of the messengers, coupling the
 standard model fields to the  strongly interacting holographic sector
 \cite{selftuning}. What happens in the far UV beyond this scale
 requires integrating back the messengers and understanding the full
 dynamics.  Without further investigation, what can be said here is
 that the description of the
self-tuning mechanism studied in \cite{selftuning} will not be
applicable, and  whether the far UV mirage cosmology   is compatible
with the observed late-time acceleration depends on a deeper
understanding of the UV of the theory.

Another possibility is that the de Sitter phase approached close to
the  UV fixed point may represent early universe inflation, which
should be followed by reheating and radiation/matter
domination. However this cannot work in the simple ``vanilla'' picture
described in this paragraph. In fact, the more the brane approaches
the UV the more the cosmology approaches de Sitter, and there is no
mechanism to make de Sitter cosmological constant decay and produce
Standard Model particles: there is no inflaton which can leave the
slow-roll regime in this picture. A way to realise inflation in a
slightly more complicated setup is discussed next.

\paragraph*{Intermediate inflationary epoch.}

A different scenario, which can give rise to a
period of inflation in the middle of the bulk,  is  one in which  the background RG flow geometry
approaches an extremum of the potential from the IR,  misses it and
continues towards another, further extremum in the ultimate UV.

This
situation can be realised in a multi-field setup, in which an extremum
of the potential is attractive in some directions but repulsive in
others \cite{multifield}. In this case, if the background flow starts
close to the attractive direction, it will spend a long time close to
the fixed point (and give rise to an induced dS brane geometry) but
eventually deviate away along the repulsive direction, which will
effectively put an end to  inflation. This can happen automatically, without fine-tuning, near a potential extremum that just violates the BF bound, \cite{matti}.

There is also the possibility of realising this scenario in a
single-field flow, in which the potential has multiple extrema along
the $\phi$ direction, as in \cite{exotic}. In such a case,  the flow can
approach from the IR a minimum of the bulk potential but continue on
to a maximum further away. In the region close to the minimum the
geometry will be approximately AdS, and the solution will be similar,
for a period of time,
to the one described in section \ref{mUV} (since the minimum will be seen
as a UV quasi-fixed point).


In order to have a long enough inflation, the solution must linger
 around the skipped fixed point (with the scalar field approximately
 constant) long enough. When  the brane leaves the vicinity of the
 intermediate fixed point, inflation ends.

\paragraph*{Back to self-tuning.}

In order to obtain a  realistic scenario, after a period of inflation
(which may be realised as sketched above), one must be able to produce
matter on the brane, and end up in an ordinary cosmology. For example,
after the intermediate period of inflation, the brane may get caught
in a self-tuning minimum and start oscillating, producing matter. If
the effective potential around the minimum is steep enough, in a
regime where the  quasi-localised  gravity is 4d, the
subsequent evolution will be  driven by brane-matter  in an
essentially four-dimensional way, as previous experience with these
models suggests \cite{Deffayet}.

Deviations from standard 4d gravity may  also give interesting effects
in this regime. As one can show in simple  models,  corrections to the
Friedman equation  arising from the higher-dimensional nature of the
setup may produce a ``geometric''  late-time acceleration driven by
ordinary matter \cite{CGP,next_paper}.

To realise this scenario however one must go beyond the setup presented
in this work.  For
one, brane matter and its couplings to the holographic CFT sector have
to be explicitly included. Also, in order to be trapped around the
self-tuning solution, a mechanism to dissipate the brane kinetic
energy must be in place. This could result from energy transfer from
the brane  to ordinary matter or some other sector (e.g. production of
light
particles such as axions)\footnote{Such brane-bulk energy-exchange mechanisms and their effect on brane cosmology have been studied in \cite{irs,z,bb}.}. In the simple setup we  have been studying
so far (probe limit), on the other hand, the brane energy is conserved.  Even if a
self-tuning solution is present,  the brane will typically overshoot it,
(unless its initial energy is very close to the energy corresponding to
the bottom of the effective potential). The question of how a
self-tuning solution may be reached starting from a non-equilibrium
initial condition  will be the subject of future work.

\paragraph*{Bouncing universe.} In the absence of a mechanism to
dissipate energy and be trapped in a self-tuning extremum of the
effective potential, the brane will continue towards the UV of the
geometry.
 There are two possible outcomes for the cosmological
history. One, which was discussed  at the beginning of this section, is that the brane continues
indefinitely towards the ultimate UV and is trapped in a de Sitter
phase forever. The alternative is that the UV is actually {\em not}
accessible due to the features of the brane potentials: recall from
section 4.1  that, in order for the probe brane equations to have a
solution  in the asymptotic UV, the brane potentials must be such that
$W_B/U_B>0$ in the far UV. If this is not the case, the UV belongs to
a ``forbidden region'', (the relativistic analog of having $E<V$ in
classical mechanics) and the trajectory turns around at a finite location $u_b$
where $\dot{u}=0$.

From the point of view of the one-dimensional
problem, the UV is hidden behind a potential barrier. If this is the
case, the cosmology undergoes a {\em bounce}, changes  from expanding
to contracting, and the brane goes back towards the IR eventually
ending in a big crunch singularity. Whether the cosmology is bouncing or not may depend not
only on the geometry and on the brane potentials, but also on the
initial conditions, i.e the initial value of the energy, since the
barrier in the UV may not be infinite and it may be overcome if the
energy is large enough. More specifically, a bounce occurs whenever
the variable $y$ in equation (\ref{cc3}) crosses unity\footnote{The fact that regular bounces can occur in mirage (brane) cosmology has been explored in the past, \cite{mirage,GGK}. This can happen in the absence of scalars but in the presence of at least two transverse dimensions to the brane.}.

To summarise, we sketch some of the possible brane histories which may
arise in this scenario:
\begin{enumerate}
\item Big bang $\longrightarrow$ de Sitter.
\item Big bang $\longrightarrow$ bounce $\longrightarrow$ big crunch.
\item Big bang $\longrightarrow$ Inflation period $\longrightarrow$
  matter production and oscillations around a self-tuning extremum (if
  a suitable mechanism for dissipating energy is at work).
\end{enumerate}



\section*{Acknowledgements}
\addcontentsline{toc}{section}{Acknowledgements}

We would like to thank Pau Figueras for discussions.
Some of the material of the probe brane cosmology in the presence of induced curvature terms were worked out in the master's thesis of Konstantinos Roubedakis at the University of Crete, \cite{rou}.

This work was supported in part by  the Advanced
ERC grant SM-grav, No 669288.


\newpage
\appendix
\renewcommand{\theequation}{\thesection.\arabic{equation}}
\addcontentsline{toc}{section}{Appendix\label{app}}
\section*{Appendix}

\section{Einstein equations and junction conditions in various
  coordinates\label{equations}}

In this appendix we provide the non-linear bulk equations and junction conditions for a set of useful coordinates to solve the time dependent cosmological problem.

\subsection{The general diagonal ansatz} \label{app:diag}
In this subsection we introduce the following ansatz for the metric:
\be
ds^2 = b(t,z)^2 dz^2 -n(t,z)^2 dt^2 + a(t,z) \delta_{ij} dx^i dx^j
\ee
where now $t$ is the time coordinate, $z$ the radial coordinate, and $x^i$ are three dimensional space-like coordinates.
The bulk equations of motion for the metric and the scalar field are (using
the convention $d_t = \dot{}$ and $d_z = {}'$ ):
\begin{equation}
\begin{aligned}
\frac{1}{3}
\Big(\Big(\frac{\dot \varphi}{2}  \frac{b}{n}\Big)^2
+ \Big( \frac{\varphi'}{2}\Big)^2 -\frac{V(\varphi) b^2}{4} \Big)
+ \frac{a'}{a} \Big(\frac{a'}{a}+\frac{n'}{n}\Big)
-\frac{b^2}{n^2} \Big( \frac{\dot a}{a} \Big(
 \frac{\dot a}{a} - \frac{\dot n}{n}
\Big)
+\frac{\ddot{a}}{a}
\Big)
 &= 0
\\
\frac{1}{3}
\Big(\Big(\frac{\varphi'}{2}  \frac{n}{b}\Big)^2
+ \Big( \frac{\dot \varphi}{2}\Big)^2 +\frac{V(\varphi) n^2}{4}  \Big)
 + \frac{\dot{a}}{a}  \Big(\frac{\dot{a}}{a}
  + \frac{\dot b }{b }\Big)
- \frac{n^2}{b^2} \Big(
\frac{a'}{a} \Big( \frac{a'}{a}  - \frac{b'}{b}\Big)
+   \frac{a''}{a} \Big)
&= 0
\\
\frac{ n'   }{ n} \frac{  \dot a  }{a }
+  \frac{a'}{a} \frac{\dot b}{b}
+\frac{1}{6} \varphi' \dot \varphi
- \frac{\dot{a'}}{a}
&= 0
\\
  \Big(\frac{a\dot \varphi}{2n}\Big)^2
-\Big(\frac{a \varphi'}{2 b} \Big)^2
-\frac{1}{4} a^2 V(\varphi)
-2 \frac{a}{b}
\Big(\frac{a' }{b} \frac{b'}{b}
- \frac{a'}{b} \frac{n'}{n}
-\frac{a''}{b}
+
\frac{1}{2}\frac{a }{b}
\Big(
\frac{b'}{b} \frac{n'}{n}
-\frac{n''}{n} \Big)
\Big)
&~+
\\
+2 \frac{a }{n} \Big(
\frac{ \dot a}{ n} \frac{\dot n}{ n}
 -\frac{\dot a}{ n} \frac{\dot b}{ b }
- \frac{\ddot a}{n}
+\frac{1}{2}\frac{a}{ n}
\Big(
\frac{\dot b}{ b } \frac{\dot n}{ n}
- \frac{\ddot b}{b }\Big)\Big)
  +\Big(\frac{a'}{b}\Big)^2-\Big(\frac{ \dot a}{ n}\Big)^2
&=0
\end{aligned} \label{eom-app}
\end{equation}
The normal vector to the brane is:
\begin{equation}
n_{a} = \frac{n b }{\sqrt{n^2-\dot z_0^2 b^2}} (1,-\dot z_0,0,0,0)
\end{equation}
The  non vanishing components of the brane extrinsic curvature are
\begin{equation}
K_{zz} = \frac{\dot z_0^2 b^3 }{n}X,\quad
K_{zt} = -\dot z_0 b n X, \quad
K_{tt} = \frac{n^3  }{b} X,\quad
K_{ij} = \frac{a(n^2 a' + \dot z_0 b^2 \dot a )}{b n \sqrt{n^2-b^2\dot z_0^2}} \delta_{ij}
\end{equation}
with
\begin{equation}
X =
\frac{ \dot z_0^3 b^3 \dot b -n^3 n'-\dot z_0  b n^2 (\dot z_0  b'+2 \dot b ) +b^2 n (\dot z_0
(2 \dot z_0  n'+\dot n )-\ddot z_0  n )}{ (n^2-\dot z_0^2 b^2)^{5/2}}
\end{equation}
The pullback metric is given by
\begin{equation}
ds^2 = -\Big(\hat n(z_0,t)^2-\hat b(z_0,t)^2 \dot r_0 \Big)   dt^2   + \hat a(z_0,t)^2 \delta_{ij} dx^i dx^j
\end{equation}
Observe that this metric can be obtained from the original one from the equation
\begin{equation}
\hat {g}_{\mu \nu} = M^a_\mu M^b_\nu g_{a b}
\end{equation}
where
\begin{equation}
M=\left(
\begin{array}{ccccc}
\dot z_0& 1 & 0 & 0 & 0 \\
 0 & 0 & 1 & 0 & 0 \\
 0 & 0 & 0 & 1 & 0 \\
 0 & 0 & 0 & 0 & 1 \\
\end{array}
\right)
\end{equation}
The non vanishing components of $K_{\mu \nu}$ are
\begin{equation}
\begin{aligned}
K_{tt} =
\frac
{\dot z_0^3 b^3 \dot b-n^3 n'
-\dot z_0 b n^2 (\dot z_0 b'+2 \dot b)
+b^2 n (\dot z_0 (2 \dot z_0 n'+\dot n)-\ddot z_0 n)}
{bn  \sqrt{n^2-\dot z_0^2 b^2}},
\quad
K_{ij} = \frac{a(n^2 a' + \dot z_0 b^2 \dot a )}{b n \sqrt{n^2-b^2\dot z_0^2}} \delta_{ij}
\end{aligned}
\end{equation}
One can verify that
\begin{equation}
\begin{aligned}
K = &K_{ab} g^{ab} = K_{\mu \nu} \hat{g}^{\mu \nu}=
\\
=&
\frac{3 (n^2 \!\!-\!\!\dot z_0^2 b^2) (\dot z_0 \dot a b^2+a' n^2)+a (\dot z_0 b n^2 (\dot z_0 b'+2 \dot b )\!\!-\!\!\dot z_0^3 b^3 \dot b+b^2 n (\ddot z_0 n\!\!-\!\!\dot z_0 (2 \dot z_0 n'+\dot n ) )+n' n^3)}{a b n (n^2-\dot z_0^2 b^2)^{3/2}}
\end{aligned}
\end{equation}
The non-vanishing components of $G_{\mu \nu}$ are
\begin{equation}
G_{tt} = \frac{3 \dot a^2}{a^2}
\quad
G_{ij} = \Big(\frac{\dot a^2+2 a \ddot a}{\dot z_0^2 b^2-n^2}-\frac{\dot a (\dot z_0 b (\dot z_0 \dot b+\ddot z_0 b )-n \dot n))}{(\dot z_0^2 b^2-n^2)^2}\Big) \delta_{ij}
\end{equation}
The conditions given by the discontinuity of the extrinsic curvature and normal derivative of $\varphi$ are
\begin{equation}
\begin{aligned}
&
\Big[ K_{\tau \tau} - \hat{g}_{\tau \tau} K\Big]_{UV}^{IR} =
\Big[ \frac{3 b \dot r_0 \sqrt{n^2-\dot z_0^2 b^2} (\dot z_0 \dot a+a' )}{a  n}\Big]_{UV}^{IR}
=
&
\\
&
\Big(
\frac{3 \dot a  \dot \varphi}{a} \big(d_{\varphi}U_B + \frac{\dot a}{a} U_B\big)+\frac{1}{2}  (b^2 \dot z_0^2-n^2)W_B-\frac{1}{2} \dot \varphi^2 Z_B
\Big)_{\varphi_0(x)}
&
\end{aligned}
\end{equation}

\begin{equation}
\begin{aligned}
&
\Big[K_{ij} - \hat{g}_{ij} K \Big]_{UV}^{IR}  = \Big[
\frac{a(
a (b'+\dot b \dot z_0) ( b^2  \dot z_0^2-2  n^2)
+a b n (n'+\dot n \dot z_0-n  \ddot z_0)
)
}{n (n^2 - b^2 \dot z_0^2)^{\frac{3}{2}}}
&
\\
&
+\frac{
2 a b (b^2 z_0^2-n^2 )(a'+\dot a \dot z_0))
}{n (n^2 - b^2 \dot z_0^2)^{\frac{3}{2}}}
\Big]_{UV}^{IR} 
=
\Big(
\Big(\frac{2 a \dot a (n \dot n-b \dot z_0 (\dot b \dot z_0+b \ddot z_0))}{(n^2-\dot z_0^2 b^2)^2}-\frac{2 a \ddot a+\dot a^2}{n^2-\dot z_0^2 b^2}\Big) U_B
&
\\
&
+
\Big(\frac{a^2 \dot \varphi  (n \dot n-b \dot z_0 (\dot b \dot z_0+b \ddot z_0))}{(n^2-\dot z_0^2 b^2)^2}
-\frac{a (2 \dot a \dot \varphi +a \ddot \varphi )}{n^2-\dot z_0^2 b^2}
\Big)
d_{\varphi}U_B
-\frac{a^2  \dot \varphi ^2 d^2_{\varphi}U_B}{n^2-\dot z_0^2 b^2}
-\frac{a^2  \dot \varphi ^2 Z_B}{2 (n^2-\dot z_0^2 b^2)}
+\frac{a^2 W_B}{2}  \Big)_{\varphi_0(x)}
&
\end{aligned}
\end{equation}

\begin{equation}
\begin{aligned}
&
\Big[n^a d_{a} \varphi \Big]_{UV}^{IR} =
\Big[
\frac{\dot z_0 b^2 \dot \varphi+n^2 \varphi'}{b n \sqrt{n^2-\dot z_0 b^2}}
 \Big]_{UV}^{IR}
=
\Big(
\frac{6  \big(\frac{\dot a (n \dot n-b \dot z_0 (\dot b \dot z_0+b \ddot z_0))}{n^2-b^2 \dot z_0^2}-\frac{a \ddot a+\dot a^2}{a}\big)}{a (n^2-b^2 \dot z_0^2)} d_\varphi U_B
+d_\varphi W_B+
&
\\
&
+ \Big(\frac{\dot \varphi (2 (b \dot z_0 (\dot b \dot z_0+b \ddot z_0)-n \dot n)-3 a \dot a)}{(n^2-b^2 \dot z_0^2)^2}+\frac{3 \dot a \dot \varphi+a \ddot \varphi}{a (n^2-b^2 \dot z_0^2)}\Big) Z_B
+\frac{ \dot \varphi^2}{2 n^2-2 b^2 \dot z_0^2}d_\varphi Z_B
\Big)_{\varphi_0(x)}
&
\end{aligned}
\end{equation}

\subsection{The K\"ahler ansatz} \label{app:khaler}
We start with the K\"ahler ansatz for the metric:
\be
ds^2 = \omega(\rho,\tau)^2 (d\rho^2 -  d\tau^2 )  + \gamma(\rho,\tau)^2 \delta_{ij} dx^i dx^j
\label{1}\ee
where $\tau$ is the time coordinate, $\rho$ the radius coordinate, and $x^i$ are three dimensional space-like coordinates.
The equations of motion for the metric and the scalar field are:
\begin{equation}
\begin{aligned}
 \frac{1}{4} \big(\dot \varphi^2+\varphi'^2
 -V(\varphi )\omega^2 \big)
  -\frac{3 \dot \gamma^2}{\gamma^2}
+\frac{3 \dot \gamma\dot\omega}{\gamma \omega}
- \frac{3 \ddot \gamma}{\gamma}
+\frac{3 \gamma'^2}{\gamma^2}
+ \frac{3 \gamma' \omega'}{\gamma \omega}
&
= 0
&\label{2}
\\
\frac{3 \dot\omega \gamma'}{\gamma \omega}
+\frac{1}{2} \dot \varphi \varphi'
+\frac{3\dot \gamma \omega'}{\gamma \omega}
-\frac{3 \dot\gamma'}{\gamma}
&
= 0
&
 \\
  \frac{1}{4} \big(\dot \varphi^2+\varphi'^2
 +V(\varphi )\omega^2 \big)
 + \frac{3 \dot \gamma^2}{\gamma^2}
+\frac{3 \dot \gamma\dot\omega}{\gamma \omega}
-\frac{3 \gamma'^2}{\gamma^2}
+\frac{3 \gamma' \omega'}{\gamma \omega}
-\frac{3 \gamma''}{\gamma}
&
= 0
&
\\
\frac{\gamma^2}{4 \omega^2} \Big( (\dot \varphi^2-\varphi'^2 -  V(\varphi) {\omega^2} \Big)
-\frac{\dot \gamma^2}{\omega^2}
+\frac{\gamma^2 \dot\omega^2}{\omega^4}
-\frac{2 \gamma \ddot \gamma}{\omega^2}
-\frac{\gamma^2 \ddot \omega}{\omega^3}
&
+
~~~
&
\\
+\frac{\gamma'^2}{\omega^2}
-\frac{\gamma^2 \omega'^2}{\omega^4}
+\frac{2 \gamma \gamma''}{\omega^2}
+\frac{\gamma^2 \omega''}{\omega^3}
&
=0
&
\end{aligned}
\end{equation}
The normal vector to the brane is:
\begin{equation}
n_{a} = \frac{\omega}{\sqrt{(1- \dot \rho_0^2)}} (1,-\dot \rho_0,0,0,0)
\end{equation}
The  non vanishing components of the brane extrinsic curvature are:
\begin{equation}
K_{ab} = \nabla_a n_{b} - n^c n_{a} \nabla_c n_{b}
= \frac{1}{2} (\nabla_a n_{b}+\nabla_b n_{a}- n^c \nabla_c (n_{a} n_{b}))
\end{equation}
 with:
 \begin{equation}
K_{\rho\rho} = - \dot \rho_0^2 X,
\quad
K_{\rho\tau} = \dot \rho_0 X,
\quad
K_{\tau\tau} = -X,
\quad
K_{ij} =   \frac{\gamma(\dot \gamma \dot \rho_0  + \gamma')}{\omega\sqrt{1- \dot \rho_0^2}} \delta_{ij}
\end{equation}
where
\begin{equation}
X =
\frac{
(1- \dot \rho_{0}^2) (\dot\omega  \dot\rho_{0}  +\omega')+\omega \ddot \rho_0 }
{(1- \dot \rho_0^2) \sqrt{1- \dot \rho_0^2}},
\end{equation}
The induced metric $\hat g_{\mu \nu}$ is given by
\begin{equation}
ds^2 = -\Big(\hat \omega(\rho_0(\tau),\tau)^2-\dot \rho_0(\tau)\Big)   d\tau^2   + \hat \gamma(\rho_0(\tau),\tau)^2 \delta_{ij} dx^i dx^j
\end{equation}
Observe that the brane induced metric can be obtained from the bulk metric using the matrix $M$ as in the equation
\begin{equation}\label{M}
\hat g_{\mu \nu } = M_\mu^a M_\nu^b g_{a b}
\end{equation}
where
\begin{equation}
M=\left(
\begin{array}{ccccc}
\dot \rho_0(\tau)& 1 & 0 & 0 & 0 \\
 0 & 0 & 1 & 0 & 0 \\
 0 & 0 & 0 & 1 & 0 \\
 0 & 0 & 0 & 0 & 1 \\
\end{array}
\right)
\end{equation}
The non vanishing components of $K_{\mu \nu}$ are
\begin{equation}
K_{tt} = \frac{(\dot \rho_0-1) ( \dot \omega \dot \rho_0+\omega')-\omega \ddot \rho_0 }{\sqrt{1-\dot \rho_0^2}},
\quad
K_{ij} = \frac{\gamma (\dot \gamma \dot \rho_0 + \gamma')}{\omega  \sqrt{1-\dot \rho_0^2}} \delta_{ij}
\end{equation}
One can verify that
\be
K = K_{ab} g^{ab} = K_{\mu \nu} \hat{g}^{\mu \nu}=
-
\frac{3 \omega  (\dot \rho_0^2-1) (  \dot \gamma \dot \rho_0 +\gamma')
+
\gamma  ((\dot \rho_0^2-1) (\dot \omega  \dot \rho_0  +\omega')
- \omega  \ddot \rho_0)}
{\gamma  \omega ^2 (1-\dot \rho_0^2)^{3/2}}
\ee
The non-vanishing components of the Einstein tensor $G_{\mu \nu}$ are
\begin{equation}
G_{\tau\tau} =
\frac{3 \dot \gamma^2}{\gamma^2},
\quad
G_{ij} =
\Big(
\frac{ \dot \gamma^2}{\omega^2}  +  \frac{ 2  \gamma (\ddot \gamma \omega -\dot \gamma \dot \omega) }{\omega^3}
-\frac{2 \gamma \dot \gamma   } {\omega^2}
\frac{ \dot \rho_0 \ddot \rho_0 } {(\dot \rho_0^2-1)^2 }
\Big)
\delta_{ij}
\end{equation}
The conditions given by the discontinuity of the extrinsic curvature and normal derivative of $\varphi$ are
\begin{equation}
\begin{aligned}
&
\Big[ K_{\tau \tau} - \hat{g}_{\tau \tau} K\Big]_{UV}^{IR} =
\Big[ \frac{ 3 \omega }{\gamma}(\dot\rho \dot \gamma+\gamma')\sqrt{1-\dot \rho_0^2}\Big]_{UV}^{IR}
=
&
\\
&
=\big(\frac{\gamma ^2 (\omega ^2 W_B (\dot \rho_0^2-1)- \dot \varphi^2Z_B)+6\gamma  \dot \gamma \dot \varphi  d_{\varphi} U_B +6  \dot \gamma^2U_B}{2 \gamma ^2}
\Big)_{\varphi_0(x)}
&
\end{aligned}
\end{equation}

\begin{equation}
\begin{aligned}
&
\Big[K_{ij} -\hat{g}_{ij} K \Big]_{UV}^{IR}  =
\Big[-\frac{\gamma (\gamma (\ddot \rho_0 \omega-(\dot \rho_0^2-1) (\dot \rho_0 \dot \omega+\omega' ))-2 (\dot \rho_0^2-1) \omega(\dot \rho_0 \dot \gamma+\gamma'))}{(1-\dot \rho_0^2 )^{3/2} \omega^2} \delta_{ij}\Big]_{UV}^{IR}
&
\\
&
=
\Big(
-\Big(\frac{2 \gamma ( \omega  \ddot \gamma -  \dot \gamma  \dot \omega) +\omega \dot \gamma^2}
{\omega^3 (1-\dot \rho_0^2)}
\frac{2 \gamma \dot \gamma  \dot \rho_0 \ddot \rho_0 }{\omega^2 (1-\dot\rho_0^2)^2}\Big)U_B
-\Big(
\frac{\gamma (\dot \varphi  (2 \omega \dot \gamma-\gamma \dot \omega)+\gamma \omega \ddot \varphi )}{\omega^3 (1-\dot\rho_0^2)}
 +
 &
 \\
&
+
\frac{\gamma^2 \dot \rho_0  \ddot \rho_0 \dot\varphi }{\omega^2 (1-\dot \rho_0^2)^2}
\Big)
d_{\varphi} U_B
- \frac{\gamma^2 \dot \varphi^2}{\omega^2 (1-\dot\rho_0^2)}d_{\varphi}^2 U_B
+
\frac{ \gamma^2}{2} W_B - \frac{  \gamma^2 \dot \varphi^2}{2 \omega^2 (1-\dot \rho_0^2)}Z_B
\Big)_{\varphi_0(x)}
&
\end{aligned}
\end{equation}

\begin{equation}
\begin{aligned}
&
\Big[n^a d_{a} \varphi \Big]_{UV}^{IR} =
\Big[\frac{\varphi' +\dot \rho_0 \dot \varphi}{\omega \sqrt{1-\dot \rho_0^2}}\Big]_{UV}^{IR} =
\Big(
6 \Big
(\frac{\gamma  \omega  \ddot \gamma-\gamma  \dot \gamma \dot \omega+\omega  \dot \gamma^2}{\gamma ^2 \omega ^3 (\dot \rho_0^2-1)}-\frac{\dot \gamma \dot \rho_0 \ddot \rho_0}{\gamma  \omega ^2 (\dot \rho_0^2-1)^2}
\Big)
d_{\varphi} U_B
 + &\\
 &
+ d_{\varphi} W_B+
\Big(
\frac{\dot \varphi (2 \gamma  \omega ^2 \dot \rho_0 \ddot \rho_0-3 \gamma ^2 \dot \gamma)}{\gamma  \omega ^4 (1-\dot \rho_0^2)^2}+\frac{\dot \varphi (3 \omega  \dot \gamma-2 \gamma  \dot \omega)+\gamma  \omega  \ddot \varphi}{\gamma  \omega ^3
(1-\dot \rho_0^2)}
\Big) Z_B
-
\frac{\dot \varphi^2}{2 \omega^2 (1-\dot\rho_0^2)} d_{\varphi} Z_B
\Big)_{\varphi_0(x)}
&
\end{aligned}
\end{equation}
Observe that in these equations the time derivative on the RHS is a total derivative $\frac{d}{d\tau} =  \dot{}$, and the functions
$\gamma$, $\omega$ and $\varphi$ depend on $(\rho,\tau)$ on the LHS and on  $(\rho_0(\tau),\tau)$ on the RHS.
In other words, on the LHS we have the 5d objects while on the RHS we have the 4d ones.

\subsection{The Chesler-Yaffe ansatz} \label{app:ch-y}
In this appendix  we introduce the Chesler-Yaffe ansatz for the metric:
\be
ds^2 = -\frac{2}{z^2} dv dz-{\bf A} dv^2  +{\bf {\bf \Sigma}}^2 \delta_{ij} dx^i dx^j
\ee
with $x^i$ the three dimensional space like coordinates, and $z$ and $v$ time-like coordinates.
The bulk equations of motion for metric and scalar field are:
\begin{equation}
\begin{aligned}
-3\frac{ {\bf \Sigma''}}{{\bf \Sigma}}-\frac{6}{z} \frac{\bf \Sigma'}{ {\bf \Sigma}}+\frac{1}{2} \varphi'^2
&=& 0
\\
\frac{V(\varphi)}{4 z^2} -\frac{6 z {\bf A} {\bf \Sigma'}}{\bf{\Sigma}} -\frac{3 z^2 {\bf A'} {\bf \Sigma'}}{2 {\bf \Sigma}}-\frac{3 z^2 {\bf A} {\bf \Sigma'^2}}{\bf \Sigma^2}
+\frac{1}{4} z^2 {\bf A}\varphi'^2 -\frac{3 z^2 {\bf A}{\bf \Sigma''}}{\bf \Sigma}
+\frac{6 {\bf \Sigma'}{\bf \dot \Sigma}}{\bf \Sigma^2}+\frac{3 {\bf \dot \Sigma'}}{\bf \Sigma}
&=& 0 \\
\frac{1}{4} {\bf A} V(\varphi)
- \frac{6 z^3 {\bf A^2}{\bf \Sigma'}}{\bf \Sigma}
-\frac{3 z^4{\bf A A' \Sigma'}}{2 {\bf \Sigma}}
-\frac{3 z^4{\bf A^2 \Sigma'^2}}{2 {\bf \Sigma^2}}
+\frac{1}{4} z^4 {\bf A} \varphi'^2
-\frac{3 z^4 {\bf A^2 \Sigma''}}{\bf \Sigma}
&~+& \\
+\frac{3 z^2 {\bf \Sigma' \dot A}}{2 {\bf \Sigma}}
-\frac{3 z^2 {\bf A' \dot \Sigma}}{2{\bf \Sigma}}
+\frac{6 z^2{\bf A \Sigma' \dot \Sigma}}{\bf \Sigma^2}
-\frac{1}{2} z^2{\bf A} \varphi' \dot \varphi
+\frac{1}{2} \dot \varphi^2
+\frac{6 z^2 {\bf A \dot \Sigma'}}{\bf \Sigma}
-\frac{3 {\bf \ddot \Sigma}}{\bf \Sigma}
&=& 0 \\
-\frac{1}{4} V(\varphi){\bf \Sigma^2}
+ z^3 {\bf \Sigma^2 A'}
+ 4 z^3 {\bf A \Sigma \Sigma'}
+2 z^4 {\bf  \Sigma A' \Sigma'}
+z^4 {\bf A \Sigma'^2}
-\frac{1}{4} z^4 {\bf A \Sigma^2} \varphi'^2
&~+&  \\
+
\frac{1}{2} z^4 {\bf \Sigma^2 A''}
+ 2 z^4 {\bf A \Sigma \Sigma''}
-2 z^2 {\bf \Sigma' \dot \Sigma}
+\frac{1}{2} z^2 {\bf \Sigma^2}\varphi' \dot \varphi
-4 z^2 {\bf \Sigma \dot \Sigma'}
&=&0
\end{aligned}
\end{equation}
The normal vector to the brane is:
\begin{equation}
n_a
=
\frac{1}{z_0 \sqrt{z_0^2{\bf A} +2\dot z_0}}  (1,-\dot z_0,0,0,0)
\end{equation}
The pullback metric is given by
\begin{equation}
ds^2 = -\Big(\hat{\bf A}(z_0(v),v)+2\frac{\dot z_{0}(v)}{z_0(v)^2}\Big) dv^2 + \hat{\bf \Sigma}(z_0(v),v) dx^i dx^j \delta_{ij}
\end{equation}
with this time the $M^a_\mu$ in (\ref{M}) is
\begin{equation}
M=\left(
\begin{array}{ccccc}
\dot z_0& 1 & 0 & 0 & 0 \\
 0 & 0 & 1 & 0 & 0 \\
 0 & 0 & 0 & 1 & 0 \\
 0 & 0 & 0 & 0 & 1 \\
\end{array}
\right)
\end{equation}
The non vanishing components of $K_{\mu \nu}$ are
\begin{equation}
K_{vv} =
\frac{4 \dot z_0^2 - z_0^3 \dot{\hat{{\bf A}}}-2  z_0 \ddot z_0}{2 z_0^2\sqrt{z_0^2 \hat {\bf A}+2 \dot z_0}}
\quad
K_{ij} =
-\frac{z_0 \hat {\bf \Sigma} \dot{\hat{\bf \Sigma}}}{\sqrt{z_0^2 \hat {\bf A}+2 \dot z_0}}
\delta_{ij}
\end{equation}
One can verify that $K = K_{ab} g^{ab} = K_{\mu \nu} \hat{g}^{\mu \nu}$. In this case we have
\begin{equation}
K=
-\frac{3 z_0 \dot{\hat{\bf\Sigma}}}{\hat{\bf{\Sigma}} \sqrt{z_0^2 \hat {\bf A}+2 \dot z_0}}
-
\frac{4 \dot z_0^2+z_0^3 \dot{\hat{ {\bf A}}}-2z_0 \ddot z_0}{(z_0^2 \hat {\bf A}-2 \dot z_0)^{\frac{3}{2}}}
\end{equation}
The matching conditions can be written as
\begin{equation}
\begin{aligned}
&
\Big[ K_{\tau \tau} - \hat{g}_{\tau \tau} K\Big]_{UV}^{IR} =
\Big[
\frac{3\sqrt{z_0^2 \hat{\bf A}+2 \dot z_0}
(\dot z_0 \hat{\bf \Sigma}'-\dot{\hat{\bf \Sigma}})}{z_0 \hat{\bf \Sigma}}
\Big]_{UV}^{IR}
=
&
\\
&
\Big(
\frac{3 {\bf \dot {\hat {\bf \Sigma}}^2}}{{\hat {\bf \Sigma}}^2}U_B
+\frac{3 {\bf \dot {\hat {\bf \Sigma}}} \dot \varphi}{\hat {\bf \Sigma}}  d_\varphi U_B
-\frac{1}{2} {\hat{\bf A}} W_B
-\frac{\dot z_0  }{z_0^2} W_B
-\frac{1}{2}  \dot \varphi^2 Z_B
\Big)_{\varphi_0(x)}
&
\end{aligned}
\end{equation}

\begin{equation}
\begin{aligned}
&
\Big[K_{ij} - \hat{g}_{ij} K \Big]_{UV}^{IR}
= \Big[
-\frac{2 z_0 \hat{\bf\Sigma}
(\dot z_0 \hat{\bf \Sigma}'-\dot{\hat{\bf \Sigma}})
}{\sqrt{z_0^2 \hat {\bf A}+2 \dot z_0}}
+
\frac{
\hat{\bf\Sigma}^2(4 \dot z_0^2-z_0^3 \dot{\hat{{\bf A}}}-2 z_0 \ddot z_0)}{(z_0^2 \hat {\bf A}+2 \dot z_0)^{\frac{3}{2}}}
\Big]_{UV}^{IR}
\\
&
=
\Big(
\frac{z_0^2  ({\hat{\bf \Sigma}}  (\dot{\hat{{\bf \Sigma}}} (z_0^2  \dot{\hat{{\bf A}}} +2 \ddot z_0)-2
\ddot{\hat {{\bf \Sigma}}}
 (z_0^2 \hat{\bf  A}+2 \dot z_0))-
 \dot{\hat{{\bf \Sigma}}}^2
  (z_0^2 \hat {\bf  A}+2 \dot z_0))U_B}{(z_0^2 \hat {\bf  A}+2 \dot z_0)^2}
+
\\
&
+
\frac{z_0^2 \hat{\bf \Sigma}  (\dot \varphi (\hat{\bf \Sigma}  (z_0^2
\dot{\hat{{\bf A}}}
+2 \ddot z_0)-4
 \dot{\hat{{\bf \Sigma}}}
(z_0^2 \hat{\bf  A}+2 \dot z_0))-2
\hat{\bf \Sigma}  \ddot \varphi (z_0^2 \hat{\bf  A}+2 \dot z_0))d_\varphi U_B }
{2 (z_0^2 \hat{\bf  A}+2 \dot z_0)^2}
-
\\
&
-\frac{z_0^2 \hat {\bf \Sigma}^2  \dot \varphi^2 d_\varphi^2 U_B }{z_0^2 \hat {\bf  A}+2 \dot z_0}
-
\frac{z_0^2  \hat{\bf \Sigma}^2 \dot \varphi^2Z_B}{2 z_0^2 \hat{\bf  A}+4 \dot z_0}
+
\frac{\hat{\bf \Sigma}^2 W_B }{2}
\Big)_{\varphi_0(x)}
\end{aligned}
\end{equation}

\begin{equation}
\begin{aligned}
\Big[n^a d_{a} \varphi \Big]_{UV}^{IR} =
&
\Big[\frac{z_0(\varphi'(z_0^2 \hat{\bf A} +z_0')- \dot \varphi)}{\sqrt{z_0^2 \hat {\bf A} +2 \dot z_0}}
\Big]_{UV}^{IR} =
\frac{z_0^2 d_\varphi Z_B \dot \varphi^2}{2 z_0^2 \hat{\bf A}
+4 \dot z_0}+d_\varphi  W_B+
\\+&
\frac{3 z_0^2d_\varphi  U_B (\hat{\bf \Sigma}
(
\dot{\hat{{\bf \Sigma}}}
(z_0^2 \dot{\hat{{\bf A}}}+2 \ddot z_0)-2
\ddot{\hat{{\bf \Sigma}}}
(z_0^2 \hat{\bf A}+2 \dot z_0))-2
\dot{\hat{{\bf \Sigma}}}^2
(z_0^2 \hat{\bf A}+2 \dot z_0))}{
\hat{\bf \Sigma}^2
(z_0^2 \hat{\bf A}+2 \dot z_0^2)}
\\
+
&
\frac{z_0^2 Z_B (\hat{\bf \Sigma}\ddot \varphi (z_0^2 \hat {\bf A}+2 \dot z_0)-\dot \varphi (\hat{\bf \Sigma}(z^2
\dot{\hat{{\bf A}}}
+2 \ddot z_0)-3
\dot{\hat{{\bf \Sigma}}}
(z_0^2 \hat {\bf A}+2 \dot z_0)+3 z_0^2 \hat{\bf \Sigma }^2
\dot{\hat{{\bf \Sigma}}}
))}{\hat {\bf \Sigma}(z_0^2 \hat{\bf A}+2 \dot z_0^2)}
\end{aligned}
\end{equation}

\subsection{The coupled wave equation ansatz} \label{app:xristos}

We finally present in this appendix the coupled wave equation  ansatz for the metric:
\begin{equation}
ds^2 = e^{2\nu} B^{-2/3} (dr^2-dt^2) + B^{2/3} \delta^{ij} dx_i dx_j
\end{equation}
with $t$ the time coordinate, $r$ the radial coordinate, and $x^i$ the three dimensional space-like coordinate.
In this system of coordinates, the bulk equations of motion take a simple form and they are:
\begin{eqnarray}
 \ddot B- B'' &=&e^{2 \nu}   B^{\frac{1}{3}} V(\varphi)
\\
 \ddot \nu- \nu'' &=& \frac{e^{2 \nu}}{6} B^{-\frac{2}{3}} V(\varphi) -\frac{\dot \varphi^2}{4}+ \frac{\varphi'^2}{4}
\\
\ddot \varphi-\varphi'' &=& -e^{2 \nu} B^{-\frac{2}{3}} d_{\varphi}V-\frac{\dot B \dot \varphi}{B} + \frac{B' \varphi'}{B}
\\
 \dot \nu B' +  \nu' \dot B -\dot B'&=& \frac{B}{2} \dot \varphi \varphi'
\\
2 B' \nu' +2  \dot B \dot \nu -\ddot B- B'' &=& \frac{B}{2} B (\dot \varphi^2+ \varphi'^2)
\end{eqnarray}
Note the coupled two dimensional wave form of the first three equations for each of the variables while the latter two couple the variables non-linearly. This Anzatz was first used by Taub to analyse planar symmetric metrics in vacuum \cite{taub}. The bulk equations can be further simplified upon going to light-like coordinates \cite{exact}.
While the matching conditions are:
\begin{equation}
\begin{aligned}
&
\Big[ K_{tt} - \hat{g}_{tt} K\Big]_{UV}^{IR} =
-
\Big[
\frac{e^{\hat \nu}}{\hat B^{\frac{4}{3}}}
\sqrt{1-\dot r_0^2}
( \dot r_0 \dot {\hat B} + (1+\dot r_0^2) \hat B')
\Big]_{UV}^{IR}
=
&
\\
&
\Big(
\frac{\dot{\hat B} \dot \varphi  d_{\varphi}U_B}{\hat{B}}+\frac{\dot{\hat B}^2 U_B}{3 \hat{B}^2}+\frac{e^{2 \hat{\nu}} \dot r_0^2 W_B}{2 \hat{B}^{2/3}}
-\frac{e^{2 \hat{\nu} } W_B }{2 \hat{B}^{2/3}}+\frac{Z_B \dot \varphi^2}{2}
\Big)_{\varphi_0(x)}
&
\end{aligned}
\end{equation}
\\
\begin{equation}
\begin{aligned}
&
\Big[ K_{ij} - \hat{g}_{ij} K\Big]_{UV}^{IR} =
\Big[
\frac{
(\dot {\hat B} + 3 \hat B \dot {\hat \nu}) \dot r_0 (1-\dot r_0^2)
+
(\hat B' +3 \hat B \hat \nu' )(1-\dot r_0^4)
+
3 \hat B \ddot r_0
}{3 e^{\hat \nu}(1-\dot r_0^2)^{\frac{3}{2}}}
\Big]_{UV}^{IR}
=
&
\\
&
\Big(
\frac{{\hat{B}}^{\frac{2}{3}} W_B}{2}
-
\Big(\frac{\hat{B}^{\frac{1}{3}}  (\dot \varphi  (\dot{\hat B}-\hat B \dot{\hat \nu})+\hat{B} \ddot \varphi )}{1-\dot r_0^2}+\frac{\hat{B}^{\frac{4}{3}}  \dot r_0 \ddot r_0
\dot\varphi }{(1-\dot r_0^2)^2} \Big) \frac{d_{\varphi} U_B}{e^{2 \hat \nu }}
+
\frac{\hat{B}^{\frac{4}{3}}  \dot \varphi^2 Z_B}{2e^{2 \hat \nu } (1-\dot r_0^2)}
&
\\
&
-
\Big(\frac{2 \hat{B}^{\frac{1}{3}} \dot {\hat B} \dot r_0 \ddot r_0}{3 (1-\dot r_0^2)^2}+\frac{ 6 \hat{B} \ddot{\hat B}-6 \hat{B} \dot {\hat{B}} \dot{\hat \nu }-\dot{\hat B}^2}{9 \hat{B}^{\frac{2}{3}} (1-\dot r_0^2)}\Big)\frac{U_B}{e^{2 \hat \nu }}
-
\frac{\hat{B}^{\frac{4}{3}}  \dot \varphi^2}{1-\dot r_0^2} \frac{d_{\varphi}^2 U_B }{e^{2 \hat \nu} }
\Big)_{\varphi_0(x)}
&
\end{aligned}
\end{equation}
\\
\begin{equation}
\begin{aligned}
&
\Big[n^a d_{a} \varphi \Big]_{UV}^{IR} =
-
\Big[
\frac{e^{\hat \nu}}{\hat B^{\frac{1}{3}}}
\frac
{(1-\dot r_0^2)\varphi'  - \dot r_0 \dot \varphi}
{\sqrt{1-\dot r_0^2}}
\Big]_{UV}^{IR}
= \Big(
\Big(
\frac{\dot{\hat B} \dot {\hat \nu} - \ddot {\hat B}}{1-\dot r_0^2}-\frac{\dot {\hat{B}} \dot r_0 \ddot r_0}{(1-\dot r_0^2)^2}
\Big)
\frac{2d_\varphi U_B}
{e^{2 \hat \nu} \hat{B}^{\frac{1}{3}}}
+ d_\varphi  W_B
&
\\
&
+\frac{\hat{B}^{\frac{2}{3}} \dot \varphi^2  d_\varphi Z_B}{2 e^{2 \hat \nu } (1-2 \dot r_0 ^2)}
-\Big(
\frac{3 \hat{B} \dot \varphi (e^{-2\hat \nu} {\hat B}^{\frac{1}{3}} \dot{\hat B} - 2 \dot r_0 \ddot r_0)}{(1-\dot r_0^2)^2}
-\frac{(5 \dot{\hat B}-6 \hat{B}\dot{\hat \nu}) \dot \varphi+3 \hat{B} \ddot \varphi}{1-\dot r_0^2}
\Big)\frac{Z_B}{3 \hat{B}^{\frac{1}{3} }e^{2 \hat \nu }}
\Big)_{\varphi_0(x)}
\end{aligned}
\end{equation}

\section{The time-dependent domain-wall  ansatz}
\label{simplifiedApp}

In this appendix we make a first attempt in solving the time dependent
equations for the probe limit.
To do that we focus on the simple ansatz
\be
\label{simplestansatz}
ds^2 =  du^2 +e^{2 \mathcal{A}(u,t)}(- dt^2 + \delta_{ij} dx^i dx^j )
\ee
that can be seen as a simplification of  the various cases analyzed in appendix \ref{equations} to a time dependent metric specified by a single scalar function.
To simplify we first assess the existence of a solution performing a linearised analysis.
This has indeed a double objective: first of all it allows us to assess if a solution in this form exists. Indeed in general, a non-linear solution will have a linear limit as well. Hence, if a linear
solution is not found, we can conclude that, quite generically, also a full non-linear solution does not exist. The reverse is of course not true: i.e. a linear solution will not be in general a solution for the full non-linear system.
Secondly, if a linear solution is found, this can put the basis of the quest for the full non-linear solution based on solving the equations order by order.
As a result of our analysis, we show that a generic linear solution does not exist, and  therefore very probably also a generic non-linear solution of the form (\ref{simplestansatz}), does not exist.

To show this we start by linearizing the metric ansatz in (\ref{simplestansatz}) as
\begin{equation}
\label{linans}
ds^2 = du^2+\alpha(u,t) \eta_{\mu \nu} dx^\mu dx^\nu
\end{equation}
Denoting time-derivatives by a dot and a $u$-derivatives by a prime,
the bulk equations of motion (\ref{eom-app})  take the form
\begin{eqnarray}
-V' -2 \frac{\dot \alpha \dot \varphi}{\alpha^3}-\frac{\ddot \varphi}{\alpha^2} +
4 \frac{\alpha' \varphi'}{\alpha}+\varphi'' &=& 0
\nonumber \\
4 \frac{(\dot \alpha)^2}{\alpha^2}-\frac{1}{2} (\dot \varphi)^2 -2 \frac{\ddot \alpha}{\alpha}
 &=&0
\nonumber \\
3\frac{\dot \alpha \alpha'}{\alpha^2}-\frac{1}{2} \dot \varphi \varphi' -3 \frac{\dot \alpha'}{\alpha}
 &=&0
\nonumber \\
-\frac{(\dot \alpha)^2}{\alpha^4}-\frac{\ddot \alpha}{\alpha^3}+3 \frac{(\alpha')^2}{\alpha^2}-\frac{1}{2} (\varphi')^2 - 3 \frac{\alpha''}{\alpha}
 &=&0
\nonumber \\
V-3\frac{(\dot \alpha)^2}{\alpha^4}-3\frac{\ddot \alpha}{\alpha^3}+9 \frac{(\alpha')^2}{\alpha^2}+3 \frac{\alpha''}{\alpha}
 &=&0
\end{eqnarray}
By evaluating $\alpha(u,t)$ on the brane trajectory $u=u_0(t)$, we can define
$\hat \alpha(u_0(t),t)$. Using this notation, the matching conditions on the brane become
(to simplify notation, we set, $u_0(t) \equiv u_0$)
\begin{eqnarray}
&&
[K_{tt}-\gamma_{tt} K]_{IR}^{UV} =
\Big[\frac{3 \dot u_0 \sqrt{\hat{\alpha}^2-\dot u_0^2} ({\hat \alpha}' \dot u_0+\dot {\hat \alpha})}
{\hat \alpha ^2}
\Big]_{IR}^{UV}
\nonumber \\
= &&
\big(
-\frac{1}{2} W_B \alpha^2+3 U_B \frac{(\dot \alpha)^2}{\alpha^2} + 3 U_B' \frac{\dot \alpha \dot\varphi}{\alpha}
-\frac{1}{2} Z_B (\varphi)^2\big)_{\varphi(x)}
\end{eqnarray}

\begin{eqnarray}
&&
[K_{ii}-\gamma_{ii} K]_{IR}^{UV} =
\Big[
\frac{-{\hat \alpha}^3 \ddot u_0-{\hat \alpha}^2 \dot u_0 ({\hat \alpha}' \dot u_0+
\dot {\hat \alpha})+2 \dot u_0^3 ({\hat \alpha}'  \dot u_0+ \dot {\hat \alpha})}
{({\hat \alpha}^2-\dot u_0^2)^{3/2}} \Big]_{IR}^{UV}
\nonumber
\\&&
=\big(
\frac{1}{2} W_B \alpha^2+ U_B \frac{(\dot \alpha)^2}{\alpha^2} - U_B' \frac{\dot \alpha \dot\varphi}{\alpha}
+\frac{1}{2} Z_B (\dot \varphi)^2
-U_B'' (\dot \varphi)^2 -2 U_B\frac{\ddot \alpha}{\alpha} - U_B'\ddot \varphi
\big)_{\varphi(x)}
\end{eqnarray}
\begin{equation}
[n^a \partial_a \varphi]_{IR}^{UV} = \Big[
\frac{\varphi'}{\sqrt {1-(\dot u_0/\alpha)^2}}
\Big]_{IR}^{UV}
=\big(
W_B'-2Z_B\frac{\dot \alpha \dot \varphi}{\alpha^3} + Z_B' \frac{(\dot \varphi)^2}{2 \alpha^2}
-6 U_B' \frac{\ddot \alpha}{\alpha^3} + Z_B \frac{\ddot \varphi}{\alpha^2}
\big)_{\varphi(x)}
\end{equation}
To explicitly solve the equations we  need, in principle, to specify the form of the potential. 
Nevertheless, we observe in the following that it is not needed to specify the potential to assess the existence or not of the generic solution.  Hence, in the following, we  keep the potential generic, without restricting to any functional form.
To proceed, expand the scalar functions $\varphi$ and $\alpha$ as:
\begin{equation}
\alpha(u,t) = a(u) +\phi_1(u,t),\quad
\varphi(u,t)=\varphi_0(u)+\phi_2(u,t)
\end{equation}
The equations of motion at zeroth order are:
\begin{eqnarray}
-V'(\varphi_0)+\frac{4 a'}{a} \varphi_0' + \varphi_0'' &=& 0  \\
\frac{3a'^2}{a^2}-\frac{1}{2}\varphi_0'^2-\frac{3 a''}{a}&=& 0   \\
V(\varphi_0) + \frac{12 a'^2}{a^2}-\frac{1}{2}\varphi_0'^2 &=& 0
\end{eqnarray}
while at first order we obtain:
\begin{eqnarray}
&&
-\frac{\ddot \phi _2}{a^2}+\frac{4 a' \phi _2'}{a}-\frac{4 \varphi _0 \phi _1 a'}{a^2}+\frac{4 \varphi _0' \phi _1'}{a}-\phi _2 V''(\varphi _0)+\phi _2''
= 0    \\
&&
\ddot{\phi_1}= 0  \\
&&
\frac{3 a' \dot \phi_1}{a^2}-\frac{1}{2} \varphi_0' \dot \phi_2-\frac{3 \dot \phi_1'}{a}
= 0  \\
&&
-\frac{6 \phi_1 a'^2}{a^3}
+\frac{3 \phi_1 a''}{a^2}
-\frac{\ddot \phi_1}{a^3}
+\frac{6 a' \phi_1'}{a^2}
-\varphi_0' \phi_2'
-\frac{3\phi_1''}{a}
= 0   \\
&&
-18 \frac{\phi_1 a'^2}{a^3}+ \phi_2 V'(\varphi_0) -\frac{3 \phi_1 a''}{a^2} - \frac{3 \ddot \phi_1}{a^3} + \frac{18 a' \phi_1'}{a^2} + \frac{3 \phi_1''}{a}
= 0
\end{eqnarray}
A time derivative on the third equation gives:
\begin{equation}
\ddot \phi_2=0
\end{equation}
The matching condition for the continuity of the metric and of the scalar
are, at order zero:
\begin{equation}
[a]_{IR}^{UV} =0,
\quad
[\varphi_0]_{IR}^{UV} =0
\end{equation}
and at first order:
\begin{equation}
[\rho a'+\phi_1]_{IR}^{UV} =0,
\quad
[\rho \varphi_0'+\phi_2]_{IR}^{UV} =0
\end{equation}
The second matching condition, related to the
discontinuity of the extrinsic curvature and normal derivative,
 (with $u \rightarrow  u_0 + \rho(t)$) are, at zero order:
\begin{eqnarray}
\label{v8}
[K_{tt} - g_{tt} K ]_{IR}^{UV} =-[K_{ii} - g_{ii} K ]_{IR}^{UV} =
\Big[
3 a a'
\Big]_{IR}^{UV} =
\left(
-\frac{1}{2} a^2 W_B
\right)_{\varphi_0(r)} & &
\end{eqnarray}
\begin{eqnarray}
[n^a \partial_a \varphi ]_{IR}^{UV}  = [\varphi_0']_{IR}^{UV}
=
\left( W_B'
\right)_{\varphi_0(r)}
&&
\end{eqnarray}
while at first order we have:
\begin{eqnarray}
\label{eq1}
[K_{tt} - g_{tt} K ]_{IR}^{UV} =
\Big[
 3 a'(\rho a'+\phi_1) + 3 a (\rho a'' + \phi_1'  )
\Big]_{IR}^{UV} =
&& \nonumber
\\ =
\left(
-\frac{a^2}{2}  (\rho  \varphi _0' +\phi _2) W_B'-a (\rho  a' + \phi _1) W_B
\right)_{\varphi_0(r)} & &
\end{eqnarray}
\begin{eqnarray}
\label{eq2}
[K_{ii} - g_{ii} K ]_{IR}^{UV} &=&
\Big[
 -3 a'(\rho a'+\phi_1) - 3 a (\rho a'' + \phi_1'  )+ \ddot \rho
\Big]_{IR}^{UV} =
\Bigg(
\frac{a^2}{2}  (\rho  \varphi _0' +\phi _2) W_B'
+
 \nonumber
 \\
&+&
a (\rho  a' + \phi _1) W_B
-
\frac{2}{a}  (\ddot \phi_1+a' \ddot \rho)U_B- (\ddot \rho \varphi _0'+\ddot \phi_2) U_B'
\Bigg)_{\varphi_0(r)}
 \end{eqnarray}

\begin{eqnarray}
[n^a \partial_a \varphi ]_{IR}^{UV}  &=& [\rho \varphi_0''+\phi_2']_{IR}^{UV}
=
\nonumber \\
&=&
\left( (\rho  \varphi _0' +\phi _2)W_B''+
  \frac{1}{~a^2} (\ddot \rho  \varphi _0'+\ddot \phi_2)Z_B -\frac{6}{~a^3}( \ddot \rho  a' +\ddot \phi_1)U_B'
\right)_{\varphi_0(r)}
\end{eqnarray}
From the relations $\ddot \phi_1 = 0 $ and $\ddot \phi_2 = 0 $
we can expand $\phi_1(u,t) = \lambda_1(u)+\eta_1(u) t$
and $\phi_2(u,t)=\lambda_1(u)+\eta_2(u) t$.
Actually we can absorb in the background the terms in $\lambda_1(u)$ and $\lambda_2(u)$ and
study the equations for $\eta_1(u)$ and  $\eta_2(u)$.
By separating the contributions at order $t^0$ from the one at order $t$ we have the following
system of equations:
\begin{eqnarray}
\frac{4 a'}{a} \eta_2' + 4 \left(\frac{\eta_1}{a}\right)'  \varphi_0' - \eta_2 V'' + \eta_2'' &=& 0 \\
\left(\frac{\eta_1}{a}\right)' +\frac{1}{6} \eta_2 \varphi_0' &=& 0 \\
\eta_2'\varphi_0'+3 \left(\frac{\eta_1}{a}\right)''&=& 0 \\
\frac{1}{3} \eta_2 V' + 8\frac{a'} {a} \left(\frac{\eta_1}{a}\right)'+ \left(\frac{\eta_1}{a}\right)''&=& 0
\end{eqnarray}
Combining the second and the third equations we can solve for $\eta_2$ as follows
\begin{equation}
3
\left(\left(\frac{\eta_1}{a}\right)'+\frac{1}{6} \eta_2 \varphi_0' \right)'
-
\left( \eta_2'\varphi_0'+3 \left(\frac{\eta_1}{a}\right)'' \right)
\propto
\eta_2' \varphi_0' - \eta_2 \varphi_0 '' =0
\rightarrow
\eta_2 = c_1 \varphi_0'(u)
\end{equation}
Substituting this solution in the second equations  we
have the following equation
\begin{equation}
\left(\frac{\eta_1}{a}\right)' =- \frac{1}{6} c_1 (\varphi_0')^2
=-c_1 \frac{a'^2-a a''}{a^2}
\rightarrow
\eta_1 =c_1 a'(u) + c_2 a(u)
\end{equation}
Observe that the equations of motions at first order are solved by these values of $\eta_2$
and $\eta_1$ together wth the solutions of the equations at order zero.
By imposing the equations of motion
and  the fact that $[\rho a'+\phi_1]_{IR}^{UV} = 0$ the matching conditions are equivalent to
\be
\label{matchigfinal}
\left\{
\begin{array}{l}
\Big[
 \rho a'' + \phi_1'
\Big]_{IR}^{UV} = \frac{1}{6}
\left(
a W_B'(\rho  \varphi _0' + \phi _2 )+W_B (\rho  a' + \phi _1)
\right)_{\varphi_0(u)}\\
\Big[
\ddot \rho
\Big]_{IR}^{UV} =
\left(
-\frac{2 U_B a' \ddot \rho}{a}-U_B' \ddot \rho \varphi _0'
\right)_{\varphi_0(u)}
\\
\Big[
\rho \varphi_0''+ \phi_2' \Big]_{IR}^{UV}
=
\left( (\rho  \varphi _0' +\phi _2)W_B''+
  \frac{1}{~a^2}\ddot \rho \,  \varphi _0'Z_B -\frac{6}{~a^3}a' \ddot \rho \, U_B'
\right)_{\varphi_0(u)}
\end{array}
\right.
\ee
If we now define $\rho(t)  = \hat \rho \; t$, i.e. we consider a linear time response,
the linearized expansions for $\alpha$ and $\varphi$ become

\begin{equation}
\alpha(u+\rho(t),t) \rightarrow a(u) + (\hat \rho+c_1)a'(u) t + c_2 a(u) t
\end{equation}
\begin{equation}
\varphi(u+\rho(t),t) \rightarrow \varphi_0(u) +  (\hat \rho + c_1) \varphi_0'(u) t
\end{equation}
The various matching conditions are summarized as
\begin{eqnarray}
&&
[a]_{IR}^{UV} =0, \quad
[\varphi_0]_{IR}^{UV} =0
,\quad
[(\hat \rho +c_1)a'+c_2 a]_{IR}^{UV} =0
,
\quad
[(\hat \rho+c_1) \varphi_0']_{IR}^{UV} =0
\nonumber \\
&&
\Big[
  (\hat \rho   +c_1)a''\Big]_{IR}^{UV} +
 \Big[a' c_2
\Big]_{IR}^{UV} = \frac{1}{6}
\left(
  (\hat \rho   +c_1) (a \varphi _0' W_B' + a' W_B)+  c_2 a W_B
\right)_{\varphi_0(u)}
 \\
 &&
\Big[ a' \Big]_{IR}^{UV} = \left(-\frac{1}{6} a W_B\right)_{\varphi_0(u)}
\hspace{-.5cm},\quad
[(\hat \rho+c_1) \varphi_0'']_{IR}^{UV}
=
\left( (\hat \rho +c_1) W_B''
\right)_{\varphi_0(u)},
\,\,
 [\varphi_0']_{IR}^{UV} =\left( W_B'
\right)_{\varphi_0(u)}
 \nonumber
\end{eqnarray}
In order to solve these equations we fix the boundary conditions as $c_1^{(UV)} = c_2^{(UV)} = 0$.
The solutions of the matching equations are
\begin{eqnarray}
a^{(IR)} &=& a^{(UV)}
  \\
\varphi_0^{(IR)}&=&\varphi_0^{(UV)}
 \\
a'^{(IR)}&=& a'^{(UV)} + \frac{1}{6}a^{(UV)} (W_B)_{\varphi_0(u)}
 \\
\varphi_0'^{(IR)} &=& \varphi_0'^{(UV)}-(W_B')_{\varphi_0(u)}
 \\
c_2^{(IR)} &=& 0
  \\
\rho^{(UV)} &=& 0
  \\
\rho^{(IR)} &=& -c_1^{(IR)}
\end{eqnarray}
\\
\\
To assure that the set of equation above has a generic solution the matrix
\begin{tiny}
\begin{equation}
\left(
\begin{array}{cccc}
 a^{\prime (UV)}
 &
 -\frac{1}{6}  (W_B)_{\varphi_0(u)} a^{(IR)}-a^{\prime (UV)}
 &
 -\frac{1}{6}  (W_B)_{\varphi_0(u)} a^{(IR)}-a^{\prime (UV)}
 &
 -a^{(UV)}
  \\
 \varphi _0^{\prime (UV)}
 &
  (W_B)_{\varphi_0(u)}'-\varphi _0^{\prime (UV)}
 &
  (W_B)_{\varphi_0(u)}'-\varphi _0^{\prime (UV)} & 0
 \\
 a^{\prime \prime (\text{UV})}
 &
K
 &
K
  &
 -\frac{1}{6}  (W_B)_{\varphi_0(u)} a^{(IR)}-a^{\prime (UV)}
 \\
 \varphi _0^{\prime \prime(UV)}
 &
 -\varphi _0^{\prime \prime (IR)}- (W_B'')_{\varphi_0(u)}
 & -\varphi _0^{\prime \prime(IR)}- (W_B'')_{\varphi_0(u)}
 &
 0
 \\
\end{array}
\right)
\end{equation}
\end{tiny}
needs to have rank 2, where $$K \equiv
-\frac{\varphi _0^{\prime (IR)}  (W_B')_{\varphi_0(u)}' a^{(IR)}+(W_B)_{\varphi_0(u)} a^{\prime (IR)}}{6} -a^{\prime \prime(IR)}.$$
 This correspond to the fact that the
following expression is  vanishing
\footnotesize
\be
\label{finalex}
\begin{array}{c}
(W_B'')_{\varphi_0(u)} \Big(a^{'(IR)}a^{'(UV)} - a^{UV} (a^{''(UV)}+\varphi_0^{'(UV)})\Big)
+
a^{'(IR)} a^{'(UV)} (\varphi_0^{''(IR)} - \varphi_0^{''(UV)} )
+\frac{1}{6} a^{(UV)}(
\\
6 \varphi_0^{''(UV)}(a^{''(IR)}+\varphi_0^{'(IR)})
-6 \varphi_0^{''(IR)}(a^{''(UV)}+\varphi_0^{'(UV)})
+a^{(UV)} \varphi_0^{'(IR)}
(\varphi_0^{'(UV)} -\varphi_0^{'(IR)}) \varphi_0^{''(UV)})
=0
\end{array}
\ee
\normalsize
where we have used the relations
\begin{equation}
(W_B)_{\varphi_0(u)} =\frac{6( a^{'(IR)} -  a^{'(UV)})}{a^{(UV)}},
\quad
(W_B')_{\varphi_0(u)} = \varphi_0^{'(IR)} -  \varphi_0^{'(UV)}
\end{equation}

Nevertheless the vanishing of (\ref{finalex})  is possible only for tuned choices of UV and IR conditions, and hence we conclude that a generic solution does not exist.
We leave to the interested reader the assessment of existence of specific solutions of the constraint above to determine non-generic linear solutions.

\section{Derivation of the probe brane action\label{approbe}}

In this appendix, we   provide a derivation of the probe brane action, equation
(\ref{c12}). In the probe limit, the  bulk metric
and bulk scalar field  take  the form 
\be \label{c01}
ds^2 = du^2 + e^{A(u)}\eta_{\mu\nu}dx^\mu dx^\nu, \qquad \f=\f(u). 
\ee
We choose world-volume coordinates adapted to the brane coordinates,
i.e.$\xi^\mu = (t,x^i)$. Then the brane embedding is $u=u(t)$, and the induced world-volume metric is then given by
\be\label{c02}
d\hat{s}^2 =  -\left(e^{2A(u(t))} - \dot{u}^2\right)dt^2 +
e^{2A(u(t))} \delta_{ij} dx^i dx^j , \quad i,j=1\ldots 3. 
\ee
To compute the action for $u(t)$  we have to compute separately the three terms
appearing in the brane action, equation (\ref{A3}).

Consider first the induced Einstein-Hilbert term. From equation (\ref{c02}), one obtains
\be
\sqrt{-\gamma}\hat{R} = 6 {\dot{A}^2 +\ddot{A} \over \left(e^{2A} -
    \dot{u}^2\right)^{1/2}} + 6
{\dot{u}\ddot{u} \dot{A} - \dot{u}^2 \dot{A}^2 \over \left(e^{2A} -
  \dot{u}^2\right)^{3/2}}. 
\ee
Adding an appropriate total derivative to eliminate  $\ddot{A}$, and
using the definition of the superpotential (\ref{c7}) to write 
\be
\dot{A} =  -{W \over 6}\dot{u}, 
\ee
we arrive at the expression: 
\be\label{c8}
\sqrt{\hat g}\hat R=
-{1\over 6}{W^2e^{2A}\dot u^2\over \sqrt{1-e^{-2A}\dot u^2}}-\pa_t\left({ {W}e^{2A}\dot u\over \sqrt{ 1-e^{-2A}\dot u^2}}\right)
\ee
Inserting this expression in the second term of the brane action
(\ref{A3}), we obtain:
\be
\int \sqrt{-\hat g}U_B(\varphi)\hat R=-{1\over 6}\int {U_BW^2e^{2A}\dot u^2\over \sqrt{1-e^{-2A}\dot u^2}}-\int U_B\pa_t\left({ {W}e^{2A}\dot u\over \sqrt{ 1-e^{-2A}\dot u^2}}\right)
\label{c9}\ee
Integrating the second term by parts and using
\be
\pa_t U_B=\pa_{\f} U_B  {\pa\phi\over \partial u}{\partial u\over \partial t}=\pa_{\f} U_B \pa_{\f}W \dot u\equiv U_B'W'\dot u
\label{d1}\ee
we finally obtain
\be\label{d2}
\int  d^3x \,dt\, \sqrt{-\hat g}U_B(\varphi)\hat R
=V_3\int dt~ {e^{2A}\dot u^2\over \sqrt{ 1-e^{-2A}\dot u^2}}\left[-{U_BW^2\over 6}+WW'U_B'\right]
\ee
where we separated the spatial volume $V_3$.

For the brane potential term, i.e. the first in equation (\ref{A3}),
we find
\be
-\int  d^3x \,dt\, \sqrt{-\hat g} W_B=-V_3\int dt~W_Be^{4A}\sqrt{ 1-e^{-2A}\dot u^2}.
\label{c10}\ee

Finally, we consider  the brane-induced scalar field kinetic
term. From the second of equations (\ref{c7}), we have
\be
\dot{\f} = {dW \over d\f} \dot{u}.
\ee
Then, the third term in equation  (\ref{A3}) takes the form
\be
-{1\over 2}\int  d^3x \,dt\, \sqrt{-\hat g} Z_B(\pa\varphi)^2={V_3\over 2}\int
dt~ {e^{2A} \dot u^2\over \sqrt{1-e^{-2A}\dot u^2}} Z_B\, W'^2
\label{c11}\ee
It is convenient to rewrite the prefactor in equations (\ref{d2}) and
(\ref{c11}) in the form 
\be\label{c11-i}
e^{4A}{e^{-2A} \dot u^2\over \sqrt{1-e^{-2A}\dot u^2}} =e^{4A}\left( {1\over
  \sqrt{1-e^{-2A}\dot u^2}} -  \sqrt{1-e^{-2A}\dot u^2}\right). 
\ee
Defining the quantity
\be\label{c11-ii}
F = -{U_BW^2\over 6}+W {d W\over d\f} {d U_B \over d\f}+{1\over
  2}Z_B\left({d W \over d\f}\right)^2
\ee
and collecting together the expressions (\ref{d2}), (\ref{c10}) and
(\ref{c11}),  we obtain 
\be\label{c11-iii}
S_b = M^3 V_3 \int dt e^{4A}\left[ {F \over  \sqrt{1-e^{-2A}\dot u^2}}
  -  \sqrt{1-e^{-2A}\dot u^2}\left(W_B + F\right)\right]
\ee
i.e. equation (\ref{c12}) as claimed. 

The equations of motion derived from this action by varying $u(t)$ are
\be
{\pa_u(e^{4A}F)\over \sqrt{ 1-e^{-2A}\dot u^2} }-\pa_u(e^{4A}(W_B+F))\sqrt{ 1-e^{-2A}\dot u^2}-{A'e^{2A}\dot u^2F \over ( 1-e^{-2A}\dot u^2)^{3\over 2}}-{A'(W_B+F)e^{2A}\dot u^2\over ( 1-e^{-2A}\dot u^2)^{1\over 2}}=
\label{c16}\ee
$$
=\pa_t\left[e^{2A}\dot u\left({F\over  ( 1-e^{-2A}\dot u^2)^{3\over 2}}+{W_B+F\over  ( 1-e^{-2A}\dot u^2)^{1\over 2}}\right)\right]
$$

\section{Solving the cubic equation}
\label{AppCubic}
In this appendix we provide a discussion of the solutions of the cubic equation (\ref{cc3}) used in the main text to study the brane cosmology.

Observe that the  cubic equation in (\ref{cc3})  is  of the form
\begin{equation}
\label{form}
y^3 + b y + c = 0,
\quad
\text{with}
\quad
b = \frac{W_B}{F}-1
\quad
\text{and}
\quad
c  = -  \frac{E}{e^{4A}F}
\end{equation}
Substituting $y= z-\frac{b}{3 z}$ the equations simplifies to
\begin{equation}
z^6 + c z^3- \frac{b^3}{27} = 0
\rightarrow
z^3=\pm \sqrt{\frac{b^3}{27}+\frac{c^2}{4}}-\frac{c}{2}
=\frac{E}{e^{4A} F}
\pm \sqrt{\left(\frac{W_B-F}{3 F}\right)^3+\frac{E^2}{e^{8A} F^2}}
\end{equation}
The argument in the square root is the discriminant $\Delta_3$
\begin{equation}
\label{disc}
\Delta_3 \equiv \frac{b^3}{27}+\frac{c^2}{4}
\end{equation}
of the cubic equation (\ref{form}).
If $\Delta_3>0$ there is only one real root,
if $\Delta_3 =0$ there are three real roots, where two are coincident,
while if $\Delta_3 <0$ there are three different real roots.

Indeed, it is possible to find that the three solutions
of the cubic equation for $y$ are:
\begin{eqnarray}
y_A= S-\frac{ b}{3S},
\quad
y_B=
\frac{e^{-\frac{2i \pi}{3} } b}{3 S}-e^{\frac{2i \pi }{3}} S,
\quad
y_C = \frac{e^{\frac{2i \pi }{3}} b}{3 S}-e^{-\frac{2i \pi}{3} } S
\end{eqnarray}
with
\begin{equation}
S\equiv \left(\sqrt{\Delta_3}-\frac{c}{2}\right)^\frac{1}{3},
\end{equation}
The real solutions of the cubic equation can hence be classified as follows:
\begin{itemize}
\item $\Delta_3>0$: in this case there is a single real solution, that corresponds to
$y= y_A$ for $c<0$ and
$y= y_B$ for $c>0$.
Only in the first case, $c<0$,  we can have  $y>1$, and this is possible  if
$-3\left(\frac{c^2}{4}\right)^{\frac{1}{3}}<b< -1-c$.
Observe that this condition can be satisfied as long as
$c<-\frac{1}{4}$.
\item
$\Delta_3= 0 $: in this case there are three real solutions, two of those are coincident. They are
\begin{eqnarray}
c>0: &&
y_1=-2 \left(~~\frac{c}{2}\right)^{\frac{1}{3}},
\quad
y_{2,3} =  ~~\left(~~\frac{c}{2}\right)^{\frac{1}{3}}
\\
 c<0: &&
y_1=~~2 \left(-\frac{c}{2}\right)^{\frac{1}{3}},
\quad
y_{2,3} =  -\left(-\frac{c}{2}\right)^{\frac{1}{3}}
\end{eqnarray}
In the first case $y_{2,3}>1$ if $c>2$
while in the second case $y_1>1$ if $c<-\frac{1}{4}$.
\item
$\Delta_3< 0 $: in this case the three solutions above are real. They can be reformulated as
\begin{eqnarray}
\label{various}
y_1&=& ~~2\sqrt{-\frac{b}{3}}
\sin \left(\frac{1}{3} \arcsin \
\left(\frac{c}{2}
\left(-\frac{3}{b}\right)^\frac{3}{2}
 \right)\right)\nonumber \\
y_2&=& - 2\sqrt{-\frac{b}{3}}
\sin \left(\frac{1}{3} \arcsin \
\left(\frac{c}{2}
\left(-\frac{3}{b}\right)^\frac{3}{2}
 \right)+\frac{\pi}{3}\right)\nonumber \\
y_3 &=& ~~ 2\sqrt{-\frac{b}{3}}
 \cos \left(\frac{1}{3} \arcsin \
\left(\frac{c}{2}
\left(-\frac{3}{b}\right)^\frac{3}{2}
 \right)+\frac{\pi}{6}\right) \nonumber  \\
\end{eqnarray}
For $y_i$ to be an acceptable  solution, it is required that $y_i >1$.
If $b<-(c+1)$ there is only one solutions satisfying this requirement.
If $b>-(c+1)$ and $c>2$ a second solutions with $y>1$ appears.
\end{itemize}
The various possibilities are summarized in the table here below:
\begin{center}
\begin{tabular} {|c|c|}
\hline
& Number of solutions with $y>1$ \\
\hline
$\Delta_3>0$ & One, if $b<-(c+1)$         \\
\hline
$\Delta_3=0$ &
Two (coincident), if  $c>2$
\\
\hline
$\Delta_3=0$ &One, if $c<-\frac{1}{4}$   \\
\hline
$\Delta_3<0$ & One, if  $b<-(c+1)$      \\
\hline
$\Delta_3<0$ & Two, if $c>2$ and $b>-(c+1)$   \\
\hline
\end{tabular}
\end{center}

\section{Computational details on the asymptotic cosmologies}
\label{appdet}

In this appendix we provide some computational details on the
asymptotic cosmologies discussed in the body of the paper.
Moreover we discuss the case with a power law bulk superpotential in the IR.

\subsection{Near-AdS}
\label{appd1}
Here we study the cubic equation (\ref{cubicUVIR}) both in the UV and in the IR case.
\begin{itemize}
\item UV case:  in this case the discriminant is, at lowest order
in $\varphi$
\begin{equation}
\Delta_3 =
-\frac{\left(h_U+h_W\right){}^3}{27 h_U^3} + \mathcal{O} \left(\varphi^{\frac{4}{\Delta_+}}\right)
<0
\end{equation}
And it follows that there are three real solutions of the form
\begin{equation}
y_1 =\frac{ e^{-4 A_0} E \ell^2 \varphi ^{4/\Delta_+ }}{(h_U+h_W)}
+ \mathcal{O} \left(\varphi^{\frac{8}{\Delta_+}}\right),
\quad
y_{2,3} =
\pm \sqrt{\frac{h_W}{h_U}+1}
-\frac{ e^{-4 A_0} E \ell^2 \varphi ^{4/\Delta_+ }}{2(h_U+h_W)}
+ \mathcal{O} \left(\varphi^{\frac{8}{\Delta_+}}\right)
\end{equation}
At small $\varphi$ we keep only the solution $y_3$,
because we require $y>1$. This solution
can be approximated as
\begin{equation}
y = y_3 = \sqrt{\frac{h_W}{h_U}+1}
+ \mathcal{O} \left(\varphi^{\frac{4}{\Delta_+}}\right),
\end{equation}
\item
IR case: in this case the discriminant is, at lowest order
\begin{equation}
\Delta_3 =
\left(\frac{E \ell^2}{2 e^{4 A_0} \varphi ^{\frac{4}{\Delta _-}} h_U}\right)^2
 > 0
\end{equation}
And it follows that there is only one real solutions of the form
\begin{equation}
y=
\varphi ^{\frac{4}{3 \Delta _-}} \left(\frac{E \ell^2}{e^{4 A_0} h_U}
\right)^{1/3}
\end{equation}
\end{itemize}

\subsection{The exponential bulk superpotential}

\subsubsection{The ultra-relativistic regime}
\label{appIRexpUR}

The cubic equation (\ref{cc3}) is solved as
\begin{eqnarray}
\text{If} \,\,\gamma_U< \gamma_Z
\quad
&\rightarrow&
\quad
y =\frac{1}{\epsilon^{\frac{1}{3} \left(\frac{2}{3 \kappa^2}-\frac{\gamma_Z}{\kappa}-2 \right)}}
\left(\frac{2 E}{e^{4 A_0} W_\infty^2 \kappa ^2 h_Z}\right)^{1/3}
\\
\text{If} \,\, \gamma_U> \gamma_Z
\quad
&\rightarrow&
\quad
y =\frac{1}{\epsilon^{\frac{1}{3} \left(\frac{2}{3 \kappa^2}-\frac{\gamma_U}{\kappa}-2 \right)}}
\left(\frac{6 E}{W_\infty^2 e^{4 A_0} h_U \left(6 \kappa  \gamma _U-1\right)}\right)^{1/3}
\end{eqnarray}
The equation for the brane dynamics in term of the variable $\epsilon(\tau)$ is
\begin{eqnarray}
\text{If} \,\,  \gamma_U< \gamma_Z
\quad
&\rightarrow&
\quad
\frac{d \epsilon}{d \tau} =\pm
W_\infty \kappa ^2 \epsilon ^{\frac{1}{3} \left(2-\frac{2}{3 \kappa ^2}+\frac{\gamma _Z}{\kappa }\right)} \left(\frac{2 E}{W_\infty^2 e^{2 A_0} \kappa ^2 h_Z}\right)^{2/3}
\\
\text{If} \,\,  \gamma_U> \gamma_Z
\quad
&\rightarrow&
\quad
\frac{d \epsilon}{d \tau} = \pm W_\infty \kappa ^2 \epsilon ^{\frac{1}{3} \left(2-\frac{2}{3 \kappa ^2}+\frac{\gamma _U}{\kappa }\right)} \left(\frac{6 E}
{W_\infty^2 e^{2A_0} h_U \left(6 \kappa  \gamma _U-1\right)}\right)^{2/3}
\end{eqnarray}
Solving these equations we arrive  at
\begin{eqnarray}
\text{If} \,\,  \gamma_U< \gamma_Z
\quad
&\rightarrow&
\quad
a(\tau)
=
\left(\frac{2 W_\infty E \left(\tau -\tau _0\right){}^3 \left(3 \kappa ^2-3 \kappa  \gamma _U+2\right)^3 }{243
e^{6 A_0 \kappa  (\gamma _U-\kappa)} h_U \left(6 \kappa  \gamma _U-1\right)}\right)^{\frac{1}{6 \kappa ^2-6 \kappa  \gamma _U+4}}
\\
\text{If} \,\,  \gamma_U> \gamma_Z
\quad
&\rightarrow&
\quad
a(\tau)
=
\left(\frac{2 W_\infty E \left(\tau -\tau _0\right){}^3 \left(3 \kappa ^2-3 \kappa  \gamma _Z+2\right)^3}{729 \kappa ^2 h_Z e^{6 A_0 \kappa  \left(\gamma _Z-\kappa \right)}}\right)^{\frac{1}{6 \kappa ^2-6 \kappa  \gamma _Z+4}}
\end{eqnarray}

\subsubsection{The  non-relativistic regime}
\label{appIRexpNR}

The cubic equation (\ref{cc3}) can be studied in this regime by expanding $y$ in (\ref{cc3}) as $y = 1 + \delta_x$,  for small $\delta_x$.
At lowest order the equation is solved by
\begin{eqnarray}
\text{If} \,\,  \gamma_U> \gamma_Z
\quad
&\rightarrow&
\quad
y= 1+ \delta_y = 1+
\frac{3 E \epsilon^{2-\frac{2}{3 \kappa ^2}+\frac{\gamma _U}{\kappa }}}{W_\infty^2 e^{4 A_0} h_U \left(6 \kappa  \gamma _U-1\right)}
\\
\text{If} \,\,  \gamma_U< \gamma_Z
\quad
&\rightarrow&
\quad
y= 1+ \delta_y = 1+
\frac{E \epsilon^{2-\frac{2}{3 \kappa ^2}+\frac{\gamma _Z}{\kappa }}}{W_\infty^2 e^{4 A_0} \kappa ^2 h_Z}
\end{eqnarray}
and the expansion is consistent only if $\delta_y \rightarrow 0^+$.
The differential equation for the dynamics $\epsilon(\tau)$    is solved by
\begin{eqnarray}
\text{If} \,\,  \gamma_Z< \gamma_U
\quad
&\rightarrow&
\quad
\epsilon(\tau) = \left(\frac{\left(\left(\tau -\tau _0\right) \left(2-3 \kappa  \gamma _U\right)\right)
}{e^{2 A_0}}
\sqrt{\frac{E}{6 h_U \left(6 \kappa  \gamma _U-1\right)}}
\right)^{\frac{6 \kappa ^2}{2-3 \kappa  \gamma _U}}
\\
\text{If} \,\, \gamma_U< \gamma_Z
\quad
&\rightarrow&
\quad
\epsilon(\tau)  =
\left(\frac{ \left(\left(\tau -\tau _0\right) \left(2-3 \kappa  \gamma _Z\right)\right)}{3 e^{2 A_0} \kappa \sqrt{\frac{E}{2 h_Z}} }\right)^{\frac{6 \kappa ^2}{2-3 \kappa  \gamma _Z}}
\end{eqnarray}
By plugging this solution in $A(\epsilon)$ we obtain the scale factor $a(\tau)$ as
\begin{eqnarray}
\text{If} \,\, \gamma_Z< \gamma_U
\quad
&\rightarrow&
\quad
a(\tau) =
e^{A_0} \left(\frac{\left(\tau -\tau _0\right) \left(2-3 \kappa  \gamma _U\right)}{e^{2 A_0}}
\sqrt{\frac{E}{6 h_U \left(6 \kappa  \gamma _U-1\right)}}
\right)^{\frac{1}{2-3 \kappa  \gamma _U}}
\\
\text{If} \,\, \gamma_U< \gamma_Z
\quad
&\rightarrow&
\quad
a(\tau) =
e^{A_0} \left( \frac{(\tau -\tau _0) (2-3 \kappa  \gamma _Z)}{3 e^{2 A_0} \kappa }
\sqrt{\frac{E}{2 h_Z}}\right)^{\frac{1}{2-3 \kappa  \gamma _Z}}
\end{eqnarray}

\subsection{Power law bulk superpotential in the IR}
\label{PL}

We now consider the case of a power law bulk superpotential as follows:
\begin{equation}
W = a  \varphi^p
\end{equation}
with $p$ a positive integer number. For the brane superpotentials we consider:
\begin{equation}
U_B = h_U \varphi^{q_U}, \quad
W_B = h_W \varphi^{q_W}, \quad
Z_B = h_Z \varphi^{q_Z}
\end{equation}
with $ q_I $ positive integers and $0< q_I < p$ for $I=U,W,Z$.
Solving the equations  $A'=-\frac{W}{6}$ for $A$ we have
\begin{equation}
\frac{d A}{d u}
=\frac{d A}{d \varphi} \, \frac{d \varphi}{d u}
=
\frac{d A}{d \varphi} \, \frac{d W}{d \varphi}
= -\frac{W}{6}
\quad
\rightarrow
\quad
A = A_0 -\frac{\varphi^2}{12 p}
\end{equation}
The power $p>0$ has to be chosen such that $\varphi$
explodes in the IR. Solving $W' = \varphi'$ we hence obtain:
\begin{equation}
\varphi= (W_\infty p (p-2) (u_0-u))^\frac{1}{2-p}
\equiv \epsilon^{\frac{1}{2-p}}
\end{equation}
where we defined $\epsilon$ as
\begin{equation}
\epsilon = W_\infty p (p-2) \left(u_0-u\right)
\end{equation}
In the IR, corresponding to $\epsilon \rightarrow 0$, we have to require $p>2$.

We want to solve the equation for the brane dynamics, that in this case corresponds to
(\ref{dynphi}),
where the cubic equation for $y$ is
\begin{equation}
\label{cubicultra}
E = \frac{e^{4 A_0} y \left(6 h_W \varphi ^{q_W+2}-W_\infty^2 \left(y^2-1\right) \varphi ^{2 p} \left(h_U \varphi ^{q_U} \left(\varphi ^2-6 p q_U\right)-3 p^2 h_Z \varphi ^{q_Z}\right)\right)}{6 \varphi ^2 e^{\frac{\varphi ^2}{3 p}}}
\end{equation}
that can be approximated as
\begin{equation}
\label{cubicultra2}
E = \frac{e^{4 A_0} y \left(W_\infty^2 \left(y^2-1\right) \varphi ^{2 p} \left(3 p^2 h_Z \varphi ^{q_Z-2}-h_U \varphi ^{q_U}\right)\right)}{6  e^{\frac{\varphi ^2}{3 p}}}
\end{equation}
In the IR the non-relativistic regime is not allowed.
This case can be studied in the ultra-relativistic regime.
In this case, depending on the hierarchy between $q_Z$ and $q_U$: i.e.
$q_Z > q_U+2$ or $q_Z < q_U+2$,
the cubic equation (\ref{cubicultra2}) is solved, at leading order for large $\varphi$, by
\begin{equation}
y = \left(
\frac{6 E e^{\frac{\varphi ^2}{3 p}}}{W_\infty^2 h_U \varphi ^{2 p+q_U}e^{4 A_0}}
\right)^\frac{1}{3}
\quad
\text{or}
\quad
y = \left(
\frac{2 E e^{\frac{\varphi ^2}{3 p}}}{W_\infty^2 e^{4 A_0} p^2 h_Z \varphi ^{2 (p-1)+q_Z}}
\right)^\frac{1}{3}
\end{equation}
Equation  (\ref{dynphi}), at large $\varphi$ becomes
\begin{equation}
\frac{d \varphi}{d \tau}
=
\pm
W_\infty p  \varphi^{p-1} y
\end{equation}
Solving equation  (\ref{dynphi}) for such a $y$  boils down to compute the integral

\begin{equation}
\label{lamb}
\begin{array}{c}
\tau \propto
\int_{\varphi}^{\infty} e^{-\frac{\varphi ^2}{9 p}} \varphi ^{\frac{1}{3} (q_U-p)+1} d \varphi
\simeq
-\frac{9p }{2}   e^{-\frac{\varphi ^2}{9 p}} \varphi ^{\frac{1}{3} \left(q_Z-p-2\right)}
=
-\frac{9 p}{2} a^{\frac{4}{3}} \left(- 12 p \log a\right)^{\frac{1}{6} \left(q_U-p\right)}
\\
\text{or}
\\
\tau \propto
\int_{\varphi}^{\infty} e^{-\frac{\varphi ^2}{9 p}} \varphi ^{\frac{1}{3} (q_Z-p+1)} d \varphi
=
-\frac{9 p}{2}  e^{-\frac{\varphi ^2}{9 p}} \varphi ^{\frac{1}{3} \left(q_U-p\right)}
=
-\frac{9 p}{2} a^{\frac{4}{3}} \left(- 12 p \log a\right)^{\frac{1}{6} \left(q_Z-p+1\right)}
\quad
\end{array}
\end{equation}
The integral corresponds to the expansion at large $\varphi$ of the incomplete Gamma
function.
The relation (\ref{lamb}) can be inverted in order to obtain $a(\tau)$. We observe that the leading behavior is
$a(\tau) = (\tau-\tau_0)^{3/4} (\dots)$
where $\dots$ represent a logarithmic correction.
This is consistent with the fact that this is a limiting case
of the exponential one discussed above.

\addcontentsline{toc}{section}{References}


\end{document}